\documentclass[english]{article}
\usepackage[T1]{fontenc}
\usepackage[latin9]{inputenc}
\usepackage{amstext}
\usepackage{graphicx}
\usepackage{float}

\makeatletter

\providecommand{\tabularnewline}{\\}

\makeatother

\usepackage{babel}
\begin{document}

\title{Optical effects related to Keplerian discs orbiting Kehagias-Sfetsos naked singularities}

\author{Zden\v{e}k Stuchlík and Jan Schee}
\maketitle
\begin{abstract}
We demonstrate possible optical signatures of the Kehagias-Sfetsos naked singularity spacetimes representing spherically symmetric vacuum solution of the modified Ho\v{r}ava gravity. In such spacetimes, accretion structures significantly different from those present in the standard black hole spacetimes occur due to the "antigravity" effect causing existence of an internal static sphere surrounded by Keplerian discs. We focus our attention on the optical effects related to the Keplerian accretion discs, constructing the optical appearance of the Keplerian discs, the spectral continuum due to their thermal radiation, and spectral profiled lines generated in the innermost parts of such discs. The KS naked singularity signature is strongly encoded in the characteristics of predicted optical effects, especially in the case of the spectral continuum and spectral lines profiled by the strong gravity of the spacetimes, due to the region of the vanishing of the angular velocity gradient influencing the effectivity of the viscosity mechanism. We can conclude that optical signatures of the Kehagias-Sfetsos naked singularities can be well distinguished from the signatures of the standard black holes.  
\end{abstract}

\section{Introduction}

The Ho\v{r}ava (or Ho\v{r}ava-Lifshitz) gravity \cite{Hor:2009:PHYSR4:} is one of the promising approaches to the quantum gravity that recently attracts strong attention. The Ho\v{r}ava quantum gravity is a field-theory that is based on the ideas of the solid state physics \cite{Lif:1941:ZhEkspTheorPhys:} and uses an anisotropic scaling of space and time. Its Lagrangian demonstrates the Lorentz invariance at low energies, while the Lorentz invariance is broken at high energies. In the Ho\v{r}ava gravity the space dimensions are scaled as $x \rightarrow bx$, while the time is scaled as $t \rightarrow b^{z}t$ with $z$ being an integer; the Lorentz invariance is recovered for $z=1$ \cite{Hor:2009:PHYSR4:,Hor:2009:PHYSRL:,Hor-MelT:2010:PHYSR4:,And-etal:2012:PHYSR4:,Gri-Hor-MelT:2012:JHEP:,Gri-Hor-MelT:2013:PHYSRL:,Ver-Sot:2012:PHYSR4:,Lib-Mac-Sot:2012:PHYSRL:}.

The solutions of the Ho\v{r}ava-Lifshitz effective gravitational equations have been found in \cite{Bar-Sot:2012:PHYSRL:,Bar-Sot:2013:PHYSRL:,Bar-Sot:2013:PHYSR4:}. The spherically symmetric solution, compatible with the Minkowski vacuum and having asymptotically the Schwarzschildian character, has been found in the framework of the modified Ho\v{r}ava model and is given by the so called Kehagias-Sfetsos (KS) metric \cite{Keh-Sfe:2009:PhysLetB:,Park Mu-In:2009:JHEP:}. The KS metric involves two parameters - the gravitational mass $M$ determining the distance scales in the metric, and the Ho\v{r}ava parameter $\omega$ reflecting the influence of the quantum effects. The character of the KS spacetime is governed by the dimensionless product $\omega M^2$. Considering the Ho\v{r}ava parameter to be an universal constant, the character of the spacetime is governed by the mass parameter $M$. If the product $\omega M^2 \geq 1/2$, the KS metric describes a black hole, while for mass parameter small enough, when $\omega M^2 < 1/2$, there are no event horizons, and the metric describes a naked singularity. The observational limits on $\omega$ presented in \cite{Ior-Rug:2010:IJMPA:,Liu-etal:2011:GenRelGrav:,Ior-Rug:2011:IJMPD:} do not exclude the existence of stellar-mass KS naked singularities. Therefore, it is of crucial interest to look for signatures of KS naked singularity spacetimes that enable to clearly distinguish them from the black hole backgrounds due to the appearance of accretion phenomena.

For the accretion phenomena, the KS black hole spacetimes has been extensively studied in a series of works related both to the particle motion \cite{Hak-etal:2010:MPLA:,Ali-Sen:2010:PHYSR4:,Abd-Ahm-Hak:2011:PHYSR4:,Ior-Rug:2011:IJMPD:,Eno-etal:2011:PHYSR4:,Hor-Ger-Ker:2011:PHYSR4:,Hak-Abd-Ahm:2013:PHYSR4:} and optical phenomena \cite{Ama-Eir:2012:PHYSR4:,Eir-Sen:2012:PHYSR4:,Ata-Abd-Ahm:2013:AstrSpaScie:} that can be relevant for tests of validity of the Ho\v{r}ava gravity. Quite recently, some works have been related to the KS naked singularity spacetimes. The nature of the geometrical properties of the KS spacetimes has been represented by the embedding diagrams in \cite{Gol-etal:2013:submitted:}, the ultra-high-energy particle-collisions in the deep gravitational potential of KS naked singularity spacetimes have been addressed in \cite{Stu-Sche-Abd:2013:submitted:}, and the circular geodesics were studied in \cite{Vie-etal:2013:submitted:}, where close similarity to the properties of the circular geodesics of the Reissner-Nordstrom naked singularity spacetimes \cite{Stu-Hle:2002:ActaPhysSlov:,Hac-etal:2008:PHRVD:,Pug-etal:2011:PHYSR4:} has been explicitly demonstrated. In all of the considered phenomena, an "antigravity" effect \cite{Vie-etal:2013:submitted:} plays a fundamental role. The situations is similar also in the braneworld naked singularity spacetimes \cite{Stu-Kot:2009:GenRelGrav:,Sche-Stu:2009:IJMPD:,Stu-Kol:2012:JCAP:,Ali-etal:2013:CLAQG:}, but it seems to be different from the Kerr naked singularity spacetimes \cite{Stu:1980:BAC:, Vir-Ell:2002:PhRvD:,Stu-Sche:2010:CLAQG:,Stu-Hle-Tru:2011:CLAQG:,Pat-Jos:2011:CLAQG:,Stu-Sche:2012a:CLAQG:,Stu-Sche:2013:CLAQG:,Sche-Stu:2013:JCAP:,Kol-Stu:2013:PHYSR4:,Bam-Mal:2013:PhRvD:}. 

Due to the "antigravity" effect, all the KS naked singularity spacetimes contain a "static" (or "antigravity") sphere where test particles can remain in stable equilibrium positions. The static sphere represents the innermost limit on existence of circular geodesics (corresponding to orbits with zero angular momentum). However, the properties of the circular motion strongly depend on the dimensionless parameter $\omega M^2$ \cite{Vie-etal:2013:submitted:,Stu-Sche-Abd:2013:submitted:}. For $0 < \omega M^2 < (\omega M^2)_{ms} = 0.2811$, stable circular orbits exist starting at the static sphere of radius $r_{stat}$ up to infinity. For $(\omega M^2)_{ms} < \omega M^2 < (\omega M^2)_{ph} = 0.3849$, two distinct regions of stable circular geodesics exist - the inner one between the static radius $r_{stat}$ and the outer marginally stable orbit at $r_{OSCO}$, and the outer one extends between the inner marginally stable orbit $r_{ISCO}$ and infinity; at the region $r_{OSCO} < r < r_{ISCO}$ unstable circular geodesics are located. For $(\omega M^2)_{ph} < \omega M^2 < 1/2$, the inner region of stable circular orbits terminates at a stable photon circular orbit at radius $r_{ph/o}$, while an unstable photon circular orbit at $r_{ph/i}$ represents a limit of unstable circular geodesics located at $r_{ph/i} < r < r_{ISCO}$, located under the outer region of stable circular geodesics; at the region $r_{ph/o} < r < r_{ph/i}$, no circular geodesics are allowed \cite{Vie-etal:2013:submitted:}. 

The standard Keplerian accretion discs are allowed in the regions of the outer stable circular geodesics; in the case of $0 < \omega M^2 < (\omega M^2)_{ms} = 0.2811$, this is the whole region of the stable circular geodesics, in the other cases of the KS naked singularity spacetimes, such discs are limited by $r_{ISCO}$ from below. The inner Keplerian discs related to the inner region of the stable circular geodesics cannot be created by the standard accretion processes, but we can consider some other processes, as gravitational radiation of orbiting matter supplied by the standard (outer) Keplerian discs, with the angular momentum limited by $L=L_{ISCO}$ corresponding to the inner marginally stable circular orbit, or with larger values of angular momentum, $L > L_{ISCO}$, corresponding to matter inflowing from thick, toroidal accretion discs. The orbiting matter can succesively settle down to a nearly circular geodesic motion, forming and inner Keplerian disc.  An alternative possibility of creating an inner Keplerian disc occurs when particles falling freely from infinity enter high-energy collisions slightly above the static radius \cite{Stu-Sche-Abd:2013:submitted:}. Some of the new particles resulting from the high-energy collisions can follow circular or nearly circular geodesics, being trapped in the "antigravity" region; particles on such circular geodesics can even have very large (covariant) specific energy (and specific angular momentum) since the high-energy particle collisions near the static radius can convert the rest energy of heavy infalling particles into kinetic energy of created light particles \cite{Stu-Sche-Abd:2013:submitted:}.

Here we focus our attention on the optical phenomena related to the outer (standard) Keplerian discs orbiting the KS naked singularity spacetimes, and compare them to those related to the KS black hole spacetimes or Schwarzschild spacetimes. We discuss their appearance, their spectral continuum created by the thermaly radiating Keplerian discs, and, finally, we construct the spectral profiled lines generated by the innermost parts of the outer Keplerian accretion discs. For comparison, we take into account in some cases also the effects generated by the inner Keplerian discs or some parts of these discs. In our study, we assume the motion of photons and matter governed by the General Relativity laws, i.e., the Infra-Red end of the Ho\v{r}ava theory, and possible effects of the Lorentz invariance violation, expected at the Ultra-Violet end of the theory \cite{Hor:2009:PHYSR4:,Hor:2009:PHYSRL:,Keh-Sfe:2009:PhysLetB:,Park Mu-In:2009:JHEP:}, are not considered.

The KS geometry and the equations of motion of test particles and photons are given in Section 2. In Section 3, circular geodesics are discussed and properties of the photon motion in the KS naked singularity spacetimes are determined and compared to those corresponding to the KS black hole spacetimes (and Schwarzschild spacetimes). In Section 4, properties of the geodetical circular orbits in the KS naked singularity spacetimes are discussed and related to the Keplerian discs and the general accretion phenomena in these spacetimes. Then we present in Section 5 the appearance of the innermost parts of the Keplerian accretion discs under assumption of discs radiating at a given frequency, in order to demonstrate the effects of the gravitational lensing and the Doppler and gravitational frequency shift. In Section 6, we give the spectral continuum under assumption of thermaly radiating Keplerian discs (of the Page-Thorne type), and in Section 7 we construct the profiled spectral lines generated by the Keplerian discs if their radiation is assumed at a fixed frequency related, e.g., to some of the Fe spectral lines. The results of the calculations of the spectral effects are summarized and discussed in Section 8. Concluding remarks are presented in Section 9.

\section{Kehagias-Sfetsos geometry and its geodesics}

\subsection{Geometry}

The spacetime interval of the Kehagias-Sfetsos solution of the modified Ho\v{r}ava gravity in the standard Schwarzschild coordinates and the geometric units ($c=G=1$) reads \cite{Keh-Sfe:2009:PhysLetB:}
\begin{equation}
ds^{2}=-f(r)dt^{2}+\frac{1}{f(r)}dr^{2}+r^{2}d\theta^{2}+r^{2}\sin^{2}\theta d\phi^{2} .
\end{equation}
The metric coefficients $g_{tt}$ and $g_{rr}$ are determined by the function  
\begin{equation}
f(r)=1+r^{2}\omega\left(1-\sqrt{1+\frac{4M}{\omega r^{3}}}\right) , 
\end{equation}
where the free parameter $\omega$ governs the role of the modified Ho\v{r}ava gravity and the parameter $M$ gives the gravitational mass, and the dimensional scale of the solution. Note that although the KS solution of the modified Ho\v{r}ava gravity is a vacuum solution, it is not a Ricci flat solution - for details see \cite{Gol-etal:2013:submitted:}. Its physical singularity, located at $r=0$, has a special character, since the metric coefficients $g_{tt}(r=0)=g_{rr}(r=0)=1$ are finite there, but their radial derivatives are divergent - this property implies that particles freely radially falling from infinity can reach the physical singularity, if they do not loose energy during their fall; however, any particle with non-zero angular momentum is repulsed by the centrifugal repulsive barriere acting in accord with the "antigravity" effect and cannot reach the physical singularity \cite{Vie-etal:2013:submitted:}. 

Properties of the KS spacetimes, including their geodetical structure, are governed by the dimensionless parameter $\omega M^2$ \cite{Vie-etal:2013:submitted:,Stu-Sche-Abd:2013:submitted:}. If the Ho\v{r}ava parameter $\omega$, reflecting the role of the quantum effects, is assumed to be fixed, then the gravitational mass parameter $M$ governs the character of the KS spacetimes. The horizons of the KS spacetimes are given by the equation
\begin{equation}
1+r^{2}\omega\left(1-\sqrt{1+\frac{4M}{\omega r^{3}}}\right)=0
\end{equation}
that implies the loci of the horizon at 
\begin{equation}
 r_{\pm}=M\pm\sqrt{M^{2}-\frac{1}{2\omega}}.
\end{equation}
Two horizons of the KS black hole spacetimes exist, if 
\begin{equation}
\omega M^2 \geq \omega M^{2}_{h}=\frac{1}{2}.
\end{equation}
In the case when equality holds, the horizons coincide, giving an extreme black hole KS spacetime. The KS naked singularity spacetimes occur for 
\begin{equation}
\omega M^2 < \omega M^{2}_{h}
\end{equation}
Therefore, considering the parameter $\omega$ fixed, the gravitational mass $M$ small enough guarantees existence of a KS naked singularity. Notice that recent works on the influence of the dimensionless Ho\v{r}ava parameter in astrophysical systems \cite{Ior-Rug:2010:OAJ:,Har-Kov-Lob:2011:CLAQG:,Dwo-Hor-Ger:2013:AstrNach:} put very weak limit 
\begin{equation}
         \omega M^2 > 8 \times 10^{-10}
\end{equation}
that is very far from the limit corresponding to the KS black hole spacetimes. 

\subsection{Geodesic equations}

The motion of test particles and photons is governed by geodesics of the spacetime. Because of the spacetime symmetries, the geodesic equations can be easily given in a separated and integrated form. 
Due to the axial symmetry and stationarity of the KS spacetimes two constants of motion arise related to the covariant components of the particle (photon) 4-momentum  
\begin{equation} 
 P_{\phi}=L,  P_{t}=-E ,
\end{equation} 
that are identified with the axial angular momentum and energy measured by distant static observers. In spherically symmetric spacetimes, the motion occurs in the central planes; for a single particle motion, the plane can be chosen to be the equatorial plane. Considering motion in a general central plane, an additional motion constant, $Q^{2}$, can be introduced by the formula
\begin{equation}
P_{\theta}^{2}+\frac{L^{2}}{\sin^{2}\theta}=L^{2}+Q^{2}.\label{eq_Q_def}
\end{equation}
The total angular momentum $\mathcal{L}$ of the particle is then given by 
\begin{equation}
        \mathcal{L}^2 = L^2 + Q^2 .
\end{equation}
The latitudinal component of test particle 4-momentum is given by the formula (\ref{eq_Q_def}) and it takes the form \cite{Stu:1983:BAC}
\begin{equation}
\left[P^{\theta}\right]^{2}=\frac{1}{r^{4}}\left[Q^{2}+L^{2}\left(1-\frac{1}{\sin^{2}\theta}\right)\right] \label{eq_latitudunal}
\end{equation}
that holds for both test particles and photons. For the motion in the equatorial plane ($\theta=\pi/2$), there is $Q=0$ and $\mathcal{L} = L$. 

Using the norm of the 4-momentum $-m^2 = g_{\mu\nu}P^{\mu}P^{\nu}$, where $m$ denotes the rest energy of the particle ($m=0$ for photons), the equation for the radial component of the 4-momentum of test particles and photons can be expressed in the form 
\begin{equation}
\left[P^{r}\right]^{2}=E^{2}-V_{eff}
\end{equation}
where the effective potential is introduced that takes the form 
\begin{equation}
V_{eff}=f(r)\left(m^2 + \frac{L^{2}+Q^{2}}{r^{2}}\right) . 
\end{equation}

\section{Circular geodesics}

We give a short overview of the character of the circular geodesics of the KS spacetimes. We make comparison of the behavior of circular geodesics as studied in the KS black hole spacetimes \cite{Chen-Wang:2010:IJMPA:,Abd-Ahm-Hak:2011:PHYSR4:} and in the KS naked singularity spacetimes \cite{Vie-etal:2013:submitted:,Stu-Sche-Abd:2013:submitted:} and extend in some aspects the discussion of their properties in the naked singularity spacetimes. 

\subsection{Photon circular geodesics}

\begin{figure}
\begin{centering}
\includegraphics[scale=0.8]{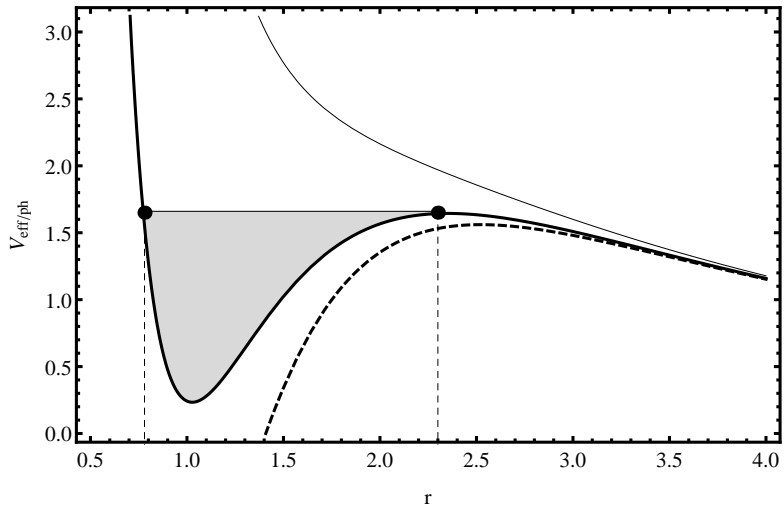}
\caption{The effective potential $V_{ff/ph}(r,\omega)$ of the phton motion is given for three representative values of Ho\v{r}ava parameter: $\omega=0.3<\omega_{ph}$ (full light line), $0.49$
(full thick line) and $0.6>\omega_{h}$ (dashed line). Shaded is the region that represents the trapped photons in the KS naked singularity spacetime with $\omega = 0.1$.}
\end{centering}
\end{figure}

For motion in the equatorial plane ($\theta = \pi/2$) there is $q=0$; then the turning points of the radial photon motion are given by using an appropriate effective potential related to the impact parameter $l$ of photons  
\begin{equation}
             l^2 = V_{eff/ph} \equiv \frac{r^2}{f(r)}. 
\end{equation}
We illustrate the radial profile of the effective potential $V_{eff/ph}(r;\omega)$ for representative choices of the parameter $\omega$ in Figure 1. The photon circular orbits, i.e., their radii $r_{ph}$ and their impact parameters $l_{ph}$, can be found from the condition
\begin{equation}
\frac{dV_{eff/ph}}{dr}=\frac{2}{f(r)^{2}} \left(r-\frac{3M}{\sqrt{1+\frac{4M}{r^{3}\omega}}}\right)=0 . \label{eq_radil_null}
\end{equation}
The photon circular orbits are located at radii satisfying the cubic equation \cite{Stu-Sche-Abd:2013:submitted:,Vie-etal:2013:submitted:}
\begin{equation}
r^{3}-9r+\frac{4}{\omega M^2}=0.\label{eq_cubic}
\end{equation}
Its solutions exist, if 
\begin{equation}
 \omega M^2 > \omega M^{2}_{ph}=\frac{2}{3\sqrt{3}}.
\end{equation}
There are four cases of the occurence of the photon circular orbits, one for the KS black holes, three for the KS naked singularities.
\begin{enumerate}
\item $\omega M^2 \geq \omega M^2_{h}$- in the black hole spacetimes, only one photon orbit exists at the radius
\begin{equation}
r_{ph/i}=2\sqrt{3}M\cos\left(\frac{1}{3}\cos^{-1}\left(-\frac{2}{3\sqrt{3}\omega M^2}\right)-\frac{2\pi}{3}\right)\label{eq_rph1}
\end{equation}
This orbit is always unstable relative to radial perturbations and represents an inner boundary for existence of circular geodetical orbits.
\item $\omega M^2_{h} > \omega M^2 > \omega M^2_{ph}$ - in these naked singularity spacetimes, two photon circular orbits exist - the outer (unstable) one at the radius given by formula (\ref{eq_rph1}) and the inner (stable) one at the radius given by 
\begin{equation}
r_{ph/o}=2\sqrt{3}M\cos\left(\frac{1}{3}\cos^{-1}\left(-\frac{2}{3\sqrt{3}\omega M^2}\right)\right).
\end{equation}
Between the stable and unstable circular orbits no circular geodesics are possible. This stable circular photon orbit is the outer boundary of the inner region of the circular geodetical orbits in some of these KS naked singularity spacetimes. Note that there is $r_{ph/o}(\omega) < r_{ph/i}(\omega)$. 
\item $\omega M^2 = \omega M^2_{ph}$- only one photon circular orbit at the radius $r_{ph}=\sqrt{3}M$ remains,
\item $\omega M^2 < \omega M^2_{ph}$- no photon circular orbits exist in such naked singularity spacetimes.
\end{enumerate}
The corresponding value of the impact parameter $l$ follows from the equation (\ref{eq_radil_null}) and reads 
\begin{equation}
\frac{l_{ph}^{2}}{M^2}=\frac{r^{2}}{1+r^{2}\omega\left(1-\sqrt{1+\frac{4M}{r^{3}\omega}}\right)}.
\end{equation}
The impact parameters of the stable and unstable circular photon orbits are given in Figure 2. For the limiting values of the dimensionless parameter $\omega M^2 = \omega M^2_{h}$ ($\omega M^2 = \omega M^2_{ph}$), the photon circular orbit is has the radius and impact parameter $r_{ph(h)}=2.37M$, 
$l_{ph(h)}\doteq4.69M$ ($r_{ph(c)}=\sqrt{3}M\doteq1.73M$, $l_{ph(c)}\doteq4.40M$).

\begin{figure}
\begin{centering}
\includegraphics[scale=0.8]{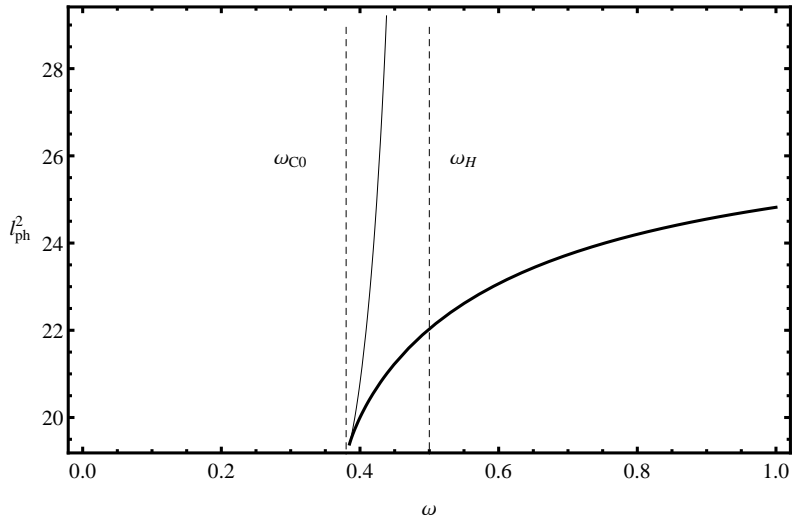}\caption{Impact parameters of photons corresponding to the stable and unstable photon circular orbits are given as function of Ho\v{r}ava parameter $\omega$. The thin line corresponds to the unstable photon orbit at $r_{ph/i}(\omega)$ and the thick line corresponds to the stable photon orbits at $r_{ph/o}(\omega)$.}
\par\end{centering}
\end{figure}

In the KS naked singularity spacetimes with $\omega M^2_{h} > \omega M^2 > \omega M^2_{ph}$, allowing existence of the stable and unstable photon circular geodesics, a region of trapped photon orbits exists in the vicinity of the stable photon circular geodesic as demonstrated in Figure 1. This situation is analogoues to those discovered in the field of Kerr naked singularity spacetimes \cite{Stu-Sche:2010:CLAQG:}. Such trapped photons can strongly influence the accretion phenomena, especially in the near-extreme Kerr naked singularity states \cite{Stu:1980:BAC:,Stu-Hle-Tru:2011:CLAQG:,Stu-Sche:2012a:CLAQG:,Stu-Sche:2012b:CLAQG:} and we can expect a similar strong influence in the KS naked singularity spacetimes too. Of course, in such situations, only photons with the impact parameter smaller than those corresponding to the unstable circular geodesic could reach distant observers and give the information on the accretion processes in the region of the trapped photons. 

In the following, we shall simplify our calculations putting $M=1$; then both the radial coordinate and the Ho\v{r}ava parameter will be dimensionless (in other words, they are expressed in units of the mass parameter). \footnote{Notice that in this case also the time coordinate becomes dimensionless.}
The resulting formulae can be easily transformed to expressions containing the mass parameter $M$ by  transformations $r \rightarrow r/M$ and $\omega \rightarrow \omega M^2$. 

\subsection{Keplerian circular orbits}

The equatorial circular orbits of test particles are determined by the condition for the local extrema of the effective potential, $\frac{dV_{eff}}{dr}=0$, that implies the radial profile of the specific angular momentum (related to the unit rest mass of the particle) in the form 
\begin{equation}
\frac{L_{K}^{2}(r;\omega)}{m^2}=\frac{r^{2}}{rA-3}\left[r^{3}\omega A-(1+r^{3}\omega)\right]  \label{eq_L2circ1}
\end{equation}
and the specific energy (related to the unit rest mass of the particle) in the form 
\begin{equation}
\frac{E_{K}^{2}(r;\omega)}{m^2}=\left[1+r^{2}\omega\left(1-\sqrt{1+\frac{4}{r^{3}\omega}}\right)\right]\left(1+\frac{1}{rA-3}\left[r^{3}\omega A-(1+r^{3}\omega)\right]\right)
\end{equation}
where we have introduced the function  
\begin{equation}
A(r;\omega)=\sqrt{1+\frac{4}{r^{3}\omega}}.
\end{equation}
In the following, we use the simplified notation $L_{K}/m \to L_{K}$ and $E_{K}/m \to E_{K}$. The radial profile of the angular frequency of the test particle motion on the Keplerian (geodesic) circular orbits, $\Omega_{K}=\frac{U^{\phi}}{U^{t}}=\frac{f(r)}{r^{2}}\frac{L_{K}}{E_{K}}$,  then takes the form

\begin{equation}
 \Omega_{K}(r;\omega)=\sqrt{\frac{r^3\omega [A(r;\omega)-1]-1}{r^3A(r;\omega)}}.\label{eq30}
\end{equation} 
We can now summarize properties of the radial profiles of the specific energy, specific angular momentum and angular frequency of the Keplerian orbits in dependence on the parameter $\omega$. 

\begin{figure}
\begin{center}
\includegraphics[scale=0.7]{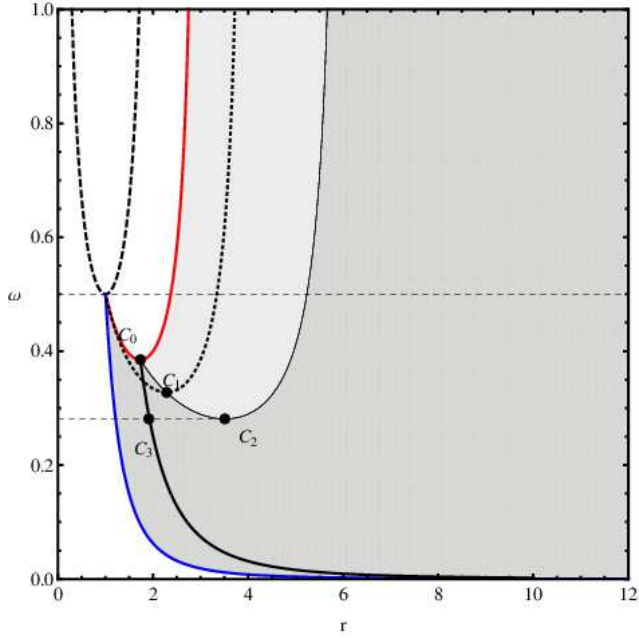}
\caption{Regions of the stable and unstable circular geodetical orbits are given in dependence on the Ho\v{r}ava parameter $\omega$ along with the critical radii of the geodesic motion. The radii of stable and unstable circular orbits $r_{ph/o}$ and $r_{ph/i}$ are given by the red curve, the static radius $r_{stat}$ is given by the blue curve, the radius $r_{\Omega/max}$ of the local maximum of the angular velocity is given by the green curve. Radii of the marginally stable circular orbits $r_{OSCO}$ and $r_{ISCO}$ are given by the black curve. Radii of the circular orbits with $E_{K}=1$ are given by the black dotted line (its right wing determines the marginally bound orbits). Radii of the black hole horizons $r_{h\pm}$ are given by the black dashed line. The critical points $C_0, C_1, C_2, C_3$ are defined in the text. The region of the stable circular geodesics is dark shaded, while the region of the unstable circular geodesics is light shaded. The standard MRI viscosity mechanism can be at work in the region of stability up to the radius $r_{\Omega/max}$ where the angular velocity gradient governing the viscosity vanishes.}
\end{center}
\end{figure}

\begin{figure}
\begin{center}
\begin{tabular}{c}
\includegraphics[scale=0.8]{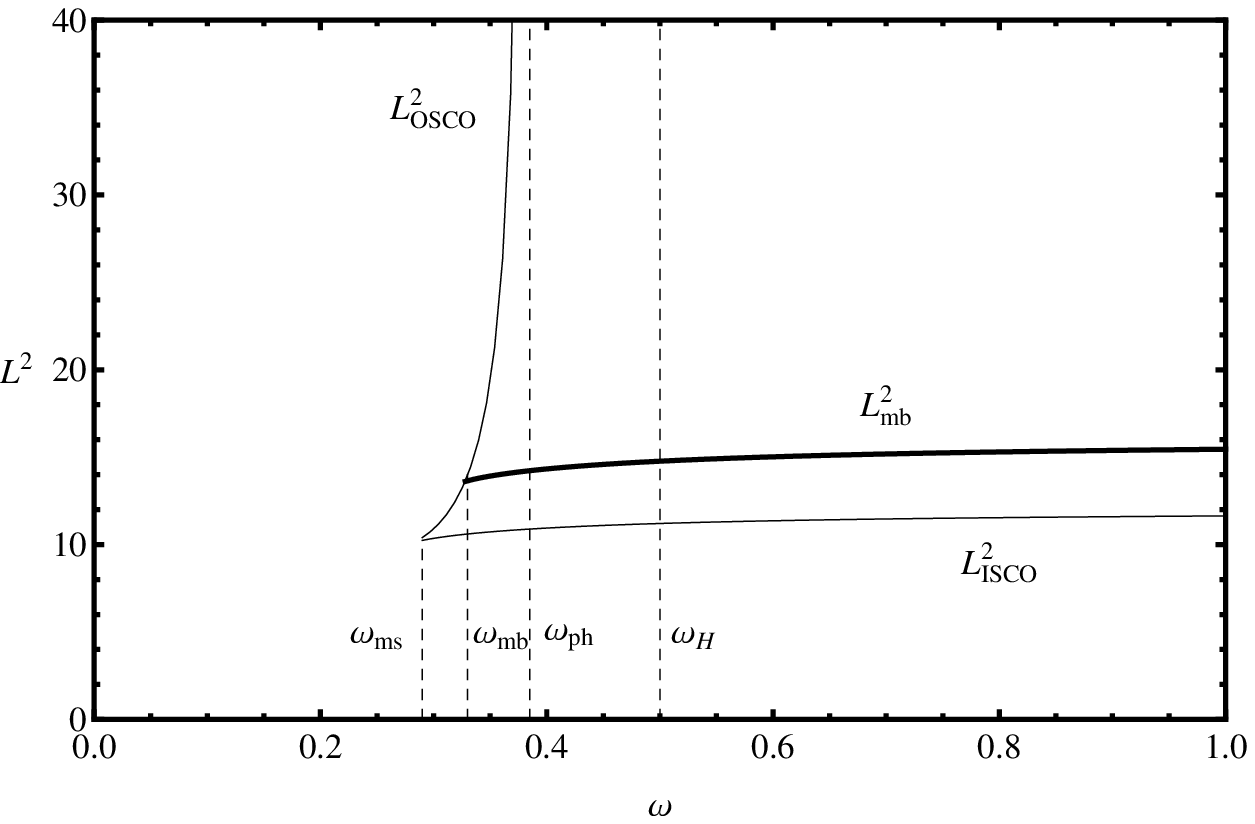}\\
\includegraphics[scale=0.8]{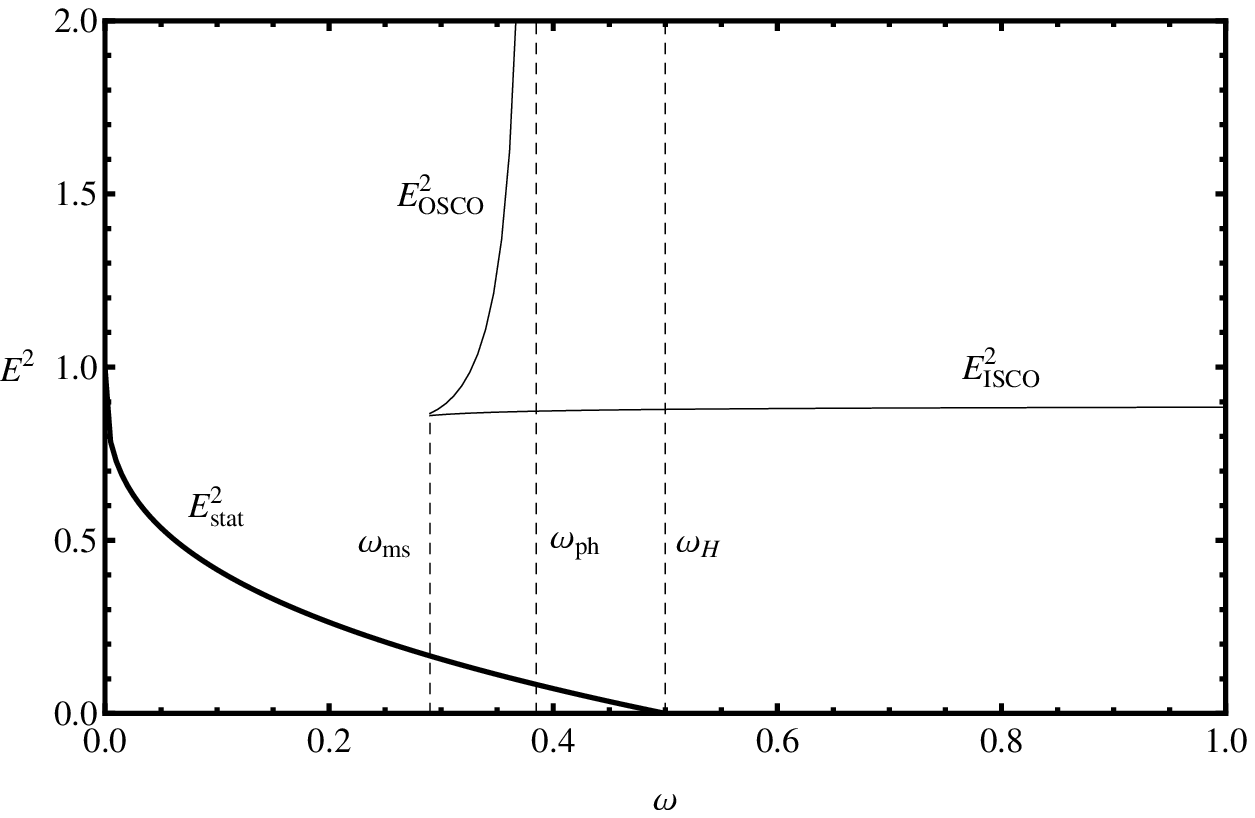}\tabularnewline
\end{tabular}\caption{The extremal values of the squared specific angular momentum, related to the marginally stable orbits and the marginally bound orbits ($L_{OSCO(ISCO)}^{2}$, $L_{mb}^{2}$) (the upper figure), and the specific energy, related to the marginally stable orbits and the equilibrium points at the static radius ($E_{OSCO(ISCO)}^{2}$, $E_{stat}^{2}$) (the lower figure), given as functions of the Ho\v{r}ava parameter $\omega$. There is $L_{OSCO}^{2}\rightarrow \infty$ and $E_{OSCO}^{2} \rightarrow \infty$ as $\omega\rightarrow\omega_{ph}$.}
\end{center}
\end{figure}

\begin{table}[H]
\begin{center}
\begin{tabular}{|c|c|c|c|c|c|}
\hline
	& $\omega_{ms}$&$\omega_{mb}$&$\omega_{ph}$&$\omega_h$ & Schw\\
\hline
$E^2_{stat}$&$0.1747$ & $0.1313$ &$0.0835$ & $0$ & -\\
$E^2_{OSCO}$&$0.8565$ & $1.0002$ &$\infty$ & - & -\\
$E^2_{ISCO}$&$0.8565$ & $0.8674$ &$0.8728$ & $0.8779$ & $0.8889$\\
$L^2_{OSCO}$&$10.101$ & $13.6106$ & $\infty$ & - & -\\
$L^2_{ISCO}$&$10.101$ & $10.5908$ & $10.8881$ & $11.2072$ & $12.0$\\
\hline
\end{tabular}
\caption{The critical values of the energy and angular momentum, $E^2_{stat}$, $E^2_{OSCO}$, $E^2_{ISCO}$, $L^2_{OSCO}$, and $L^2_{ISCO}$, given for the critical values of the parameter $\omega$, and for the case of Schwarzchild spacetime.}
\end{center}
\end{table}

\begin{figure}
	\begin{center}
	\begin{tabular}{cc}
	\includegraphics[scale=0.7]{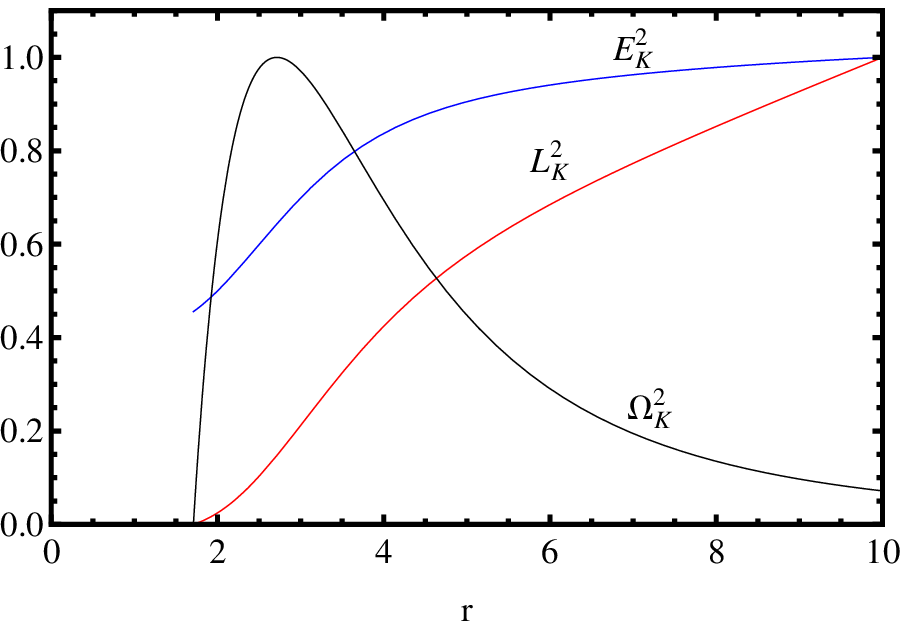}&\includegraphics[scale=0.7]{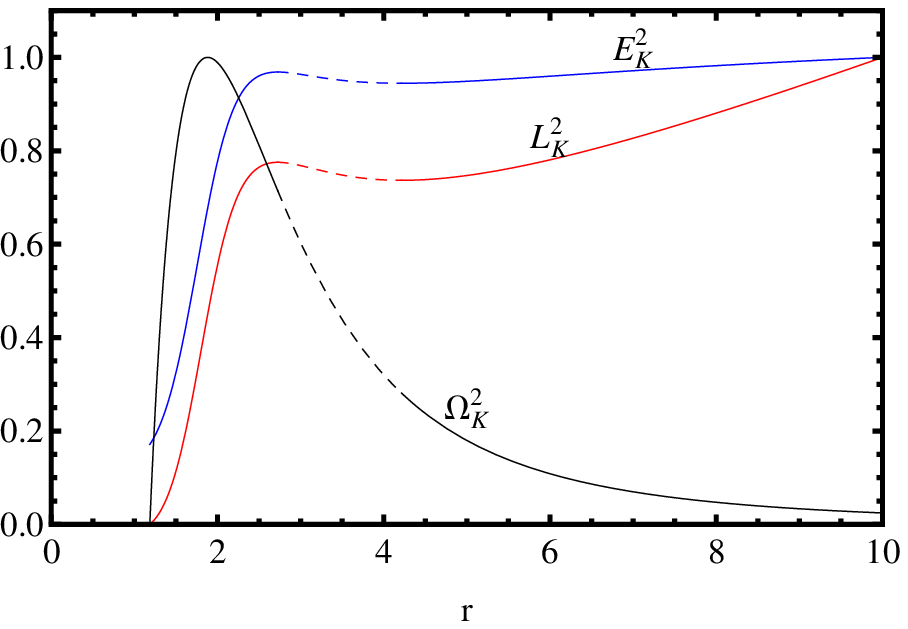}\\
	\includegraphics[scale=0.7]{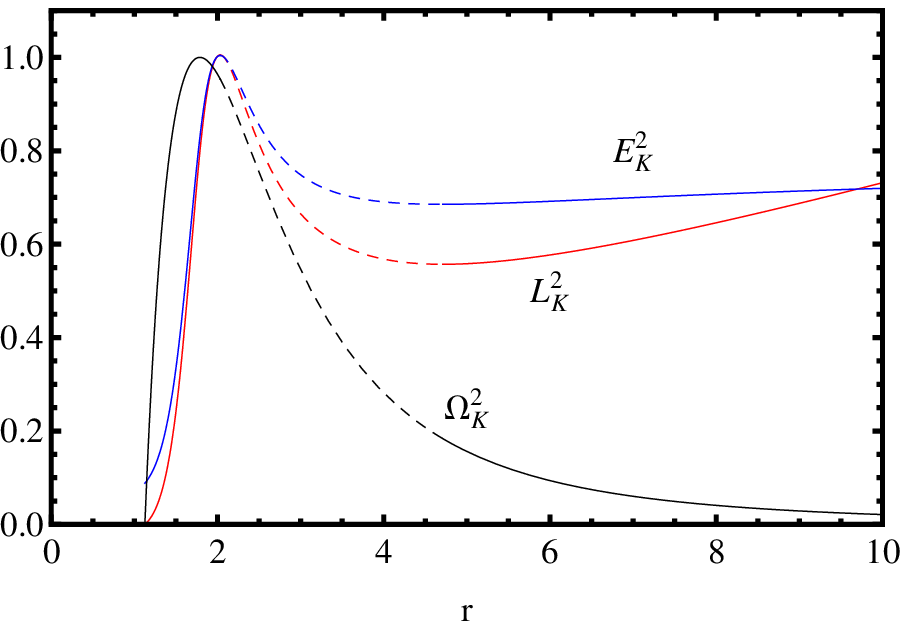}&\includegraphics[scale=0.7]{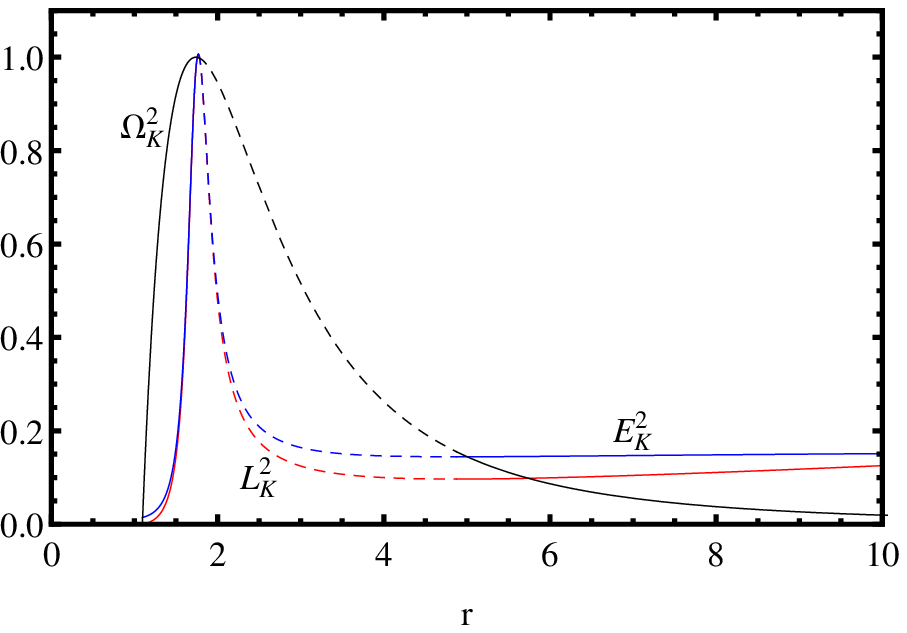}\\
	\includegraphics[scale=0.7]{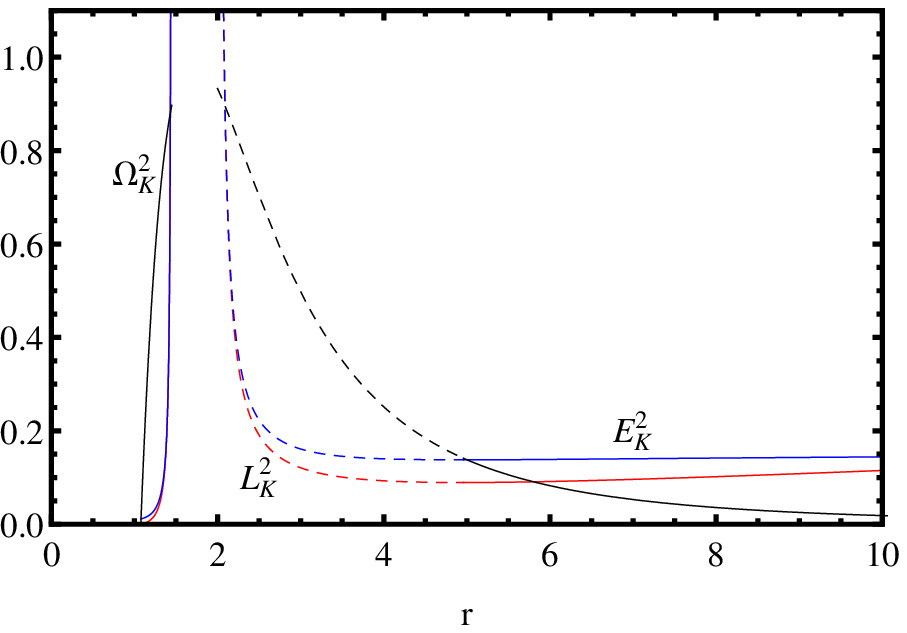}&\includegraphics[scale=0.7]{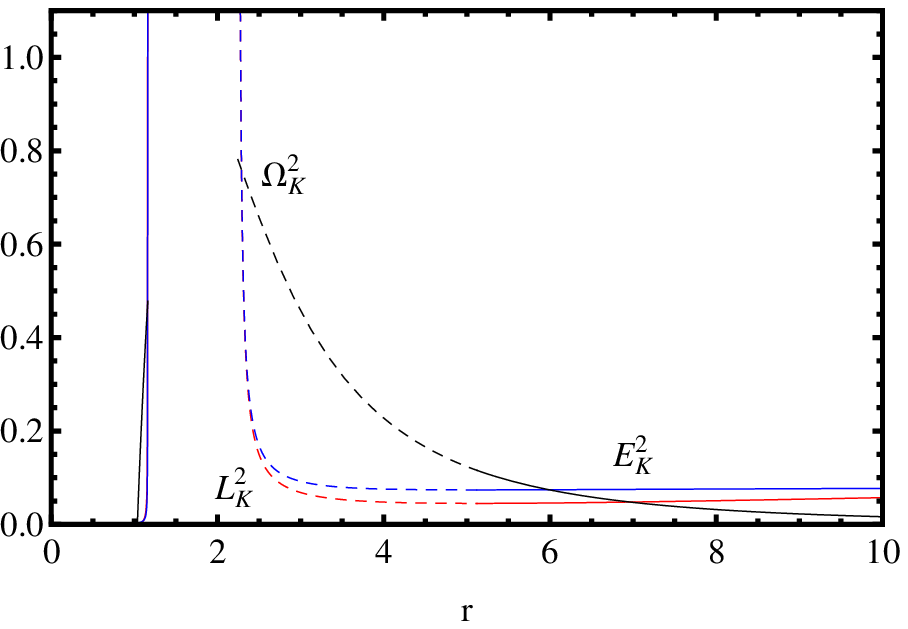}
	\end{tabular}
	\caption{Radial profiles of the specific angular momentum $L^2_K$ (red), the specific energy $E^2_K$ (blue), and the angular velocity $\Omega^2_K$ (black) of the Keplerian (geodetical) circular orbits in the equatorial plane of the typical KS naked singularity. Each panel is plotted for a particular value of the parameter $\omega=0.1$, $0.3$, $0.35$, $0.38$, $0.4$, and $0.45$ (from top left to right bottom). The radial profiles are finished at the static radius. The dashed part of the radial profiles correspond to the unstable circular geodesics. Only the stable parts of the radial profiles related to the decreasing parts of the $\Omega_{K}(r;\omega)$ function of the radius can be atributed to the Keplerian thermaly radiating discs with the standard MRI viscosity mechanism of generation of the heat energy.}
	\end{center}
\end{figure}

The divergence of both the specific energy and the specific angular momentum is given by the condition 
\begin{equation}
     rA(r;\omega) - 3 = 0 
\end{equation}
that is identical with the condition for the circular photon orbits discussed in the previous subsection. 

The zero points of the radial profiles of the specific angular momentum ($L^2_{K}(r;\omega)=0$) give the so called static (or "antigravity") radii corresponding to stable equilibrium points where the particle is at rest relative to distant static observers. The static radius is given by \cite{Vie-etal:2013:submitted:} 
\begin{equation}
                    r_{stat}(\omega) = \frac{1}{(2\omega)^{1/3}} . 
\end{equation}
The static radius $r_{stat}$ gives the lower limit for existence of circular orbits of test particles in the field of KS naked singularity spacetimes. No zero points of the specific energy radial profile exist, since $E^{2}_{K} = f(r,\omega)(1 + \frac{L^{2}_{K}}{r^2})$.  At the static radius, the specific energy of the Keplerian circular orbits approaches its minimum 
\begin{equation}
                    E_{K}(r=r_{stat},\omega) \equiv E_{stat} = 1 - (2\omega)^{1/3} . 
\end{equation}
There is $E_{stat}(\omega \rightarrow 1/2) \rightarrow 0$, and $E_{stat}(\omega \rightarrow 0) \rightarrow 1$. 

Since $L_{K}=0$ at the static radius, there is naturally also $\Omega_{K}(r_{stat},\omega)=0$ in all KS naked singularity spacetimes. This implies that a maximum of the radial profile of the Keplerian angular velocity has to exist. The condition $\frac{d\Omega_{K}}{dr}=0$, giving the local extrema of the angular velocity radial profile, implies the location of the local maximum at 
\begin{equation}
                    r_{\Omega/max}(\omega) = (\frac{2}{\omega})^{1/3} = 4^{1/3} r_{stat}(\omega), 
\end{equation}
i.e., close to the static radius. The function $r_{\Omega/max}(\omega)$ terminates at 
\begin{equation}
                    r_{\Omega/max}(\omega = \omega_{ph}) = \sqrt{3} = r_{ph}(\omega = \omega_{ph}).
\end{equation}
The circular geodetical orbits with zero gradient of the angular velocity exist only in the KS naked singularity spacetimes with $\omega < \omega_{ph}$. In the spacetimes with $\omega_{h} > \omega > \omega_{ph}$, the radius is located in the region forbidden for the circular geodesics. Then the angular velocity increases with decreasing radius in the external region of the circular geodesics located above the $r_{ph/i}$, and it decreases with decreasing radius in the internal region of the circular geodesics, located between the static radius $r_{stat}$ and the photon stable circular orbit at $r_{ph/o}$ - see \cite{Vie-etal:2013:submitted:}. 

The marginally stable circular orbits have to satisfy simultaneously the conditions
\begin{equation}
\frac{dV_{eff}}{dr}=0\quad\textrm{and}\quad\frac{d^{2}V_{eff}}{dr^{2}}=0.
\end{equation}
The first of these equations implies the radial profile of the specific angular momentum $L_{K}^2(r;\omega)$ given by (\ref{eq_L2circ1}). The second one implies the relation 
\begin{equation}
L_{ms}^{2}(r;\omega)=\frac{r^{2}\left[2-4r^{3}\omega(2-A(r))-r^{6}\omega^{2}(1-A(r))\right]}{30-3A(r)r(4+r^{3}\omega)+12r^{3}\omega}.\label{eq_L2circ2}
\end{equation}
The marginally stable orbits are then given as solutions of the equation 
\begin{equation}
        L_{K}^2(r;\omega) = L_{ms}^2(r;\omega).          
\end{equation}
The loci of the marginally stable circular orbits are determined by numerical calculations. They depend significantly on the value of the parameter $\omega$. There exists a critical value of the Ho\v{r}ava parameter, $\omega_{ms}$, separating spacetimes admitting two marginally stable radii $r_{ISCO}$ and $r_{OSCO}$, and no marginally stable radii. The critical value of the Ho\v{r}ava parameter for the stable orbits reads  
\begin{equation}
              \omega_{ms} = 0.281100.
\end{equation}

Similarly, we can look for radii of the marginally bound orbits, $r_{mb}$, i.e., the unstable circular geodesic orbits having the specific energy $E_{K} = 1$. They can be found again by numerical calculations from the condition
\begin{equation}
                  E_{mb} \equiv E_{K}(r_{mb},L_{mb}) = 1 
\end{equation}
that governs both the radius $r_{mb}$ and the angular momentum $L_{mb}$ of the Keplerian orbit. Note that this condition determines both the unstable outer orbits with $E_{K}=1$ and $L_{K}=L_{mb}$ and the stable inner orbits with the same energy $E_{K}=1$, but different angular momentum $L > L_{mb}$, if they exist. Clearly, these orbits can exist only in the spacetimes allowing for existence of two stable circular geodesics for a given angular momentum, i.e. the spacetimes with $\omega > \omega_{ms}$. There exists a critical value of the Ho\v{r}ava parameter, $\omega_{mb}$, separating spacetimes where the marginally bound unstable circular orbits are allowed, and where it is not possible - in such spacetimes the inner marginally stable orbits have $E_{OSCO} < 1$. The critical value of the Ho\v{r}ava parameter for the spacetimes with marginally bound orbits reads  
\begin{equation}
              \omega_{mb} = 0.327764.
\end{equation}

The marginally stable orbits radii $r_{ms}$ ($r_{OSCO}$, $r_{ISCO}$), and the marginally bound orbit radii $r_{mb}$ are given as functions of the parameter $\omega$ in Figure 3, along with the radii of the photon circular geodesics $r_{ph}$ ($r_{ph/o}, r_{ph/i}$), the static radius, $r_{stat}$, and the radius where the gradient of the angular velocity of the circular geodesics vanishes, $r_{\Omega/max}$. The critical points where the curves $r_{ph}(\omega)$, $r_{mb}(\omega)$ and $r_{ms}(\omega)$ have the local extremum are given in the following way: 
\begin{eqnarray}
                C_{0}& =& (r_{phc}, \omega_{ph}) = (1.7321, 0.3849), \\
                C_{1}& = &(r_{mbc}, \omega_{mb}) = (2.2818, 0.3278),  \\
                C_{2}& =& (r_{msc}, \omega_{ms}) = (3.5140, 0.2811),\\
                C_{3}&=&(r_{\Omega max},\omega_{ms})=(1.921,0.2811).
\end{eqnarray}
The point $C_{3}$ determining the radius $r_{\Omega/max}(\omega)$ for the critical value $\omega_{ms}$ is also included as this radius can be important for the Keplerian disc structure during the transition accross the critical point in the evolution of a KS naked singularity due to the accretion of matter. 

We can see that there is always $r_{\Omega/max}(\omega) < r_{ms}(\omega)$. For completeness, we have included also the values of the parameter $\omega > 1/2$ corresponding to the KS black holes -- notice that in this case, there are no circular orbits present under the inner horizon, in constrast to the situation in the standard general relativistic Reissner-Nordstrom or Kerr black holes. The points $C_0$ and $C_2$ (or $C_1$) give the classification of the KS naked singularity spacetimes according to the properties of circular geodesics governing the Keplerian accretion discs. The points $C_0$, $C_1$ and $C_2$ also determine the maximal extension of the inner Keplerian discs in the corresponding classes of the KS naked singularity spacetimes. In the KS naked singularity spacetimes with $\omega > \omega_{ms}$, the inner Keplerian discs, constituted by the separated inner parts of the stable circular geodesics, are located at small distance from the naked singularity at $r=0$ and cannot exceed the radius $r_{msc}=3.514$ determined by the point $C_2$. On the other hand, in the spacetimes with $\omega < \omega_{ms}$, we can alternatively consider as the inner Keplerian discs the parts of the Keplerian disc located under the radius $r_{\Omega/max}$. Extension of such a disc can be much larger that $r_{msc}$ - of course, such an inner Keplerian disc can evolve into a toroidal, geometrically thick disc, if viscosity mechanisms are not efficient in the region $r < r_{\Omega/max}$. 
    
In Figure 4, we give the values of the specific energy $E_{K}(\omega)$ of the marginally stable orbits, along with the specific energy at the static radius and at the radius of vanishing angular-velocity gradient that can be relevant for the efficiency of accretion of matter in the Keplerian discs. We give there also the specific angular momentum $L_{K}(\omega)$ of the marginally stable and marginally bound Keplerian orbits. Note that the marginally bound circular geodesics are given only by the the right branch of the function $r_{mb}(\omega)$, since the left branch governs the stable circular geodesics with $E_{K}=1$ (see Figure 3). The values of the energy and the angular momentum for the ISCO and OSCO orbits at the critical values of the parameter $\omega$ are given in Table 1, along with the energy at the static radius.

\section{Properties of Keplerian accretion discs in KS naked singularity spacetimes}

The Keplerian discs orbiting the KS black holes were extensively discussed in the literature -- see, e.g., \cite{Chen-Wang:2010:IJMPA:,Har-Kov-Lob:2011:CLAQG:,Abd-Ahm-Hak:2011:PHYSR4:}. Therefore, we focus here our attention to the Keplerian discs in the KS naked singularity spacetimes, using the classification of these spacetimes according to the properties of the circular geodesics presented above and in \cite{Vie-etal:2013:submitted:,Stu-Sche-Abd:2013:submitted:}; for each class of the KS naked singularity spacetimes the astrophysically relevant Keplerian discs can have quite different properties. 

In dependence on the (dimensionless) parameter $\omega M^2$, we can distinguish four types of the KS spacetimes. For each class of the KS spacetimes, we give the typical radial profiles of the specific energy, specific angular momentum, and angular velocity of the circular geodesics governing behavior of the Keplerian discs in Figure 5. The corresponding cases of the behavior of the effective potential governing the Keplerian discs are presented in \cite{Vie-etal:2013:submitted:} and will not be repeated here. 

Note that in all the classes of the KS naked singularity spacetimes a sphere of particles at stable equilibrium positions can be created at the static radius. Such a sphere could be a final state of the accreted matter, if dissipative mechanisms (viscosity, gravitational radiation, etc.) could work at time scales long enough. The physical singularity at $r=0$ can be reached only by particles falling freely from infinity with angular momentum $L=0$ and $E \geq 1$. If their specific energy $E < 1$, they will oscillate in the potential well, loosing successively energy because of gravitational (or electromagnetic) radiation and declining to the equilibrium position at the static radius $r_{stat}$  \cite{Vie-etal:2013:submitted:}. Matter falling freely from infinity with a non-zero angular momentum cannot reach the physical singularity at $r=0$ being reflected by the combined effect of the "antigravity" and the centrifugal repulsion \cite{Vie-etal:2013:submitted:}. If spherical shells radially falling from large distance (with zero or small angular momentum) collide successively at the deep gravitational field near KS naked singularities, large amount of energy can be released in very short time scales \cite{Stu-Sche-Abd:2013:submitted:}, and observationally interesting effects could appear that we plan to study in some future work. 

Here, we focus our attention to the accretion processes with large angular momentum, typical for binary systems with compact objects, when Keplerian discs governed by the structure of the circular geodesics can be established. We separate our discussion into four cases, in dependence on the magnitude of the Ho\v{r}ava dimensionless parameter $\omega$. 

\subsection{Naked singularities with $\omega < \omega_{ms}$}

The KS naked singularity spacetimes containing no photon circular orbits and no unstable circular orbits. Only stable circular orbits are allowed for the specific angular momentum from the interval $0 < L^2 < \infty$. The radial profiles of the specific angular momentum, the specific energy, and the angular velocity of the circular geodetical orbits are given in Figure 5. (The behaviour of the effective potential can be found in \cite{Vie-etal:2013:submitted:}.) Since all the circular geodesic orbits are stable, with energy and angular momentum decreasing with decreasing radius, we can consider the Keplerian discs in the whole range of $r_{stat} < r < \infty$. 

However, a critical point of the Keplerian discs in these KS naked singularity spacetimes is related to the radius $r_{\Omega/max}$ where the angular-velocity gradient vanishes. At this radius and its vicinity, the standard viscosity mechanism based on the non-zero gradient of the angular velocity of the accreting matter ceases its relevance. It can be substituted there by the mechanism based on the gravitational (or electromagnetic) radiation of the orbitting matter -- the corresponding looses of energy and angular momentum can transmit the accreting matter to the region where the viscosity mechanism causing heat production and thermaly radiating Keplerian discs could continue its work.   

As an alternative, we have to consider the Keplerian discs limited by the condition $\frac{d\Omega_{K}}{dr} < 0$ (and having the inner edge at the critical radius $r_{\Omega/max}$) that assumes the Magneto-Rotational Instability (MRI) mechanism \cite{Bal-Haw:1998:RevModPhys:} related to this condition to be the only mechanism of generating the viscosity effects that enable accretion of matter in the Keplerian discs \cite{Vie-etal:2013:submitted:}. If some other viscosity mechanism related to the oppositely oriented gradient of the angular velocity ($\frac{d\Omega_{K}}{dr} > 0$) works, the internal parts of the Keplerian discs can also radiate giving thus the complete thermaly radiating discs discussed above. \footnote{A different scenario related to the Keplerian discs is possible due to the long time scales of the gravitational radiation of orbiting matter. In the region where $\frac{d\Omega_{K}}{dr} \sim 0$, matter has to be accumulated, creating a limited toroidal structure of non-Keplerian character where pressure gradients will be relevant. If such a toroid represents an inner boundary of the accreting and efficiently radiating structure, it means that only the MRI viscosity is at work, while a continuation of the accretion structure by an internal part of the Keplerian thermaly radiating disc indicates some other efficient viscosity mechanism at work. If such a viscosity mechanism is not available in the inner part of the Keplerian disc under $r_{\Omega/max}$, the accretion can continue because of the gravitational radiation, but such Keplerian discs will be cold and their radiation cannot be of thermal origin. We plan to study the toroidal accretion discs around KS naked singularities in a future work.} 

\subsection{Naked singularities with $\omega_{ph} > \omega > \omega_{ms}$}

The KS naked singularity spacetimes containing no photon circular orbits, but two marginally stable circular orbits at $r_{OSCO}$ and $r_{ISCO}$. The dependence of these radii on the parameter $\omega$ is presented in Figure 3. The stable circular orbits, potentially corresponding to Keplerian discs, are located in the region $r_{stat} < r < r_{OSCO}$ (the inner Keplerian disc) and $r_{ISCO} < r < \infty$ (the outer Keplerian disc). For the marginally stable circular orbits, the energy and angular momentum satisfy the conditions $E_{ISCO} < E_{OSCO}$ and $L_{ISCO} < L_{OSCO}$. The unstable circular geodesics are located at $r_{OSCO} < r < r_{ISCO}$, and their energy and angular momentum belong to the intervals $E_{ISCO} < E < E_{OSCO}$ and $L_{ISCO} < L < L_{OSCO}$. The radial profiles of the specific angular momentum, the specific energy and the angular velocity of the circular geodetical orbits are demonstrated in Figure 5. There exists one stable circular orbit for $L^{2}\in[0,L_{OSCO}^{2})\cup\left(L_{ISCO}^{2},\infty\right]$, while for $L$ from the interval $L^{2}\in\left[L_{ISCO}^{2},L_{OSCO}^{2}\right]$, there are two stable circular orbits and one unstable. The behaviour of the effective potential can be found in \cite{Vie-etal:2013:submitted:}. 

Since the critical radius $r_{\Omega/max} < r_{OSCO}$, we can conclude that the complete outer Keplerian disc is quite regular relative to the MRI viscosity mechanism; the critical point of the vanishing angular-velocity gradient is related to the inner Keplerian disc only. Therefore, we can conclude that the accretion Keplerian discs work quite well down to the ISCO circular orbit and the complexities related to the vanishing of the angular-velocity gradient are relevant only in the inner Keplerian discs, if such discs will be created. At the ISCO the accreting matter starts to move freely, with energy $E_{ISCO}$ and angular momentum $L_{ISCO}$, in the deep well of the effective potential, loosing slowly its energy and angular momentum due to gravitational (or electromagnetic) radiation and declining successively to the stable equilibrium point at the static radius. We can expect that matter  accumulates under the ISCO orbit radius, forming successively a perfect-fluid toroidal (or spherical-like) structure above the static radius (or around the static radius). We can expect that under some properly established conditions, such a toroidal structure could evolve to a Keplerian-like structure with matter in quasi-circular geodesic motion. 

In order to create an inner Keplerian disc by the accretion processes, we have to consider a non-Keplerian, perfect fluid accretion with relevance of the pressure gradients \cite{Koz-Jar-Abr:1978:ASTRA:,Stu-Sla-Hle:2000:ASTRA:}. Then Keplerian orbits corresponding to the unstable circular geodesics appear at the inner edge of equilibrium toroidal configurations of accreting matter, with energy and angular momentum that are larger than those corresponding to the ISCO enabling thus a successive creation of an inner Keplerian disc or an inner toroidal perfect fluid accretion structure, as demonstrated, e.g., in \cite{Sla-Stu:2005:CLAQG:,Stu:2005:MPLA:,Kuc-Sla-Stu:2011:JCAP:}. \footnote{We plan to study the toroidal structures in the KS naked singularity spacetimes in a future paper. We expect that their behavior will be similar to those discovered for the Reissner-Nordstrom naked singularity spacetimes \cite{Kuc-Sla-Stu:2011:JCAP:}}. Another possibility enabling the existence of inner Keplerian discs is based on creation of matter structures in near-circular motion due to particle collisions \cite{Stu-Sche-Abd:2013:submitted:} or due to an efficient circularization of originally strongly elliptical motion of matter infalling from the inner edge of the outer Keplerian discs. 

In the KS naked singularity spacetimes allowing for existence of the unstable circular geodesics, the marginally bound (unstable) circular geodesics with specific energy given by the condition $E_{mb} = 1$ play a crucial role. Such orbits can be approached approximately by particles starting at rest at infinity with specific energy $E=1$ and angular momentum $L \sim L_{mb}$ and are significant in theory of toroidal configurations of perfect fluid orbiting black holes or naked singularities, giving the upper limit on the existence of marginally stable toroidal configurations \cite{Koz-Jar-Abr:1978:ASTRA:,Stu-Sla-Hle:2000:ASTRA:,Stu:2005:MPLA:,Stu-Sla-Kov:2009:CLAQG:}. The marginally bound orbits can exist only in the KS naked singularity spacetimes having the Ho\v{r}ava parameter  
\begin{equation}
                 \omega_h > \omega > \omega_{mb}.
\end{equation}
In the KS spacetimes with $\omega < \omega_{b}$ no unstable circular orbits with $E>1$ exist - then even particles following unstable circular orbits can remain on a bounded orbit, if perturbed radially,  and the equilibrium toroidal configurations of perfect fluid cannot extend up to infinity in such spacetimes \cite{Koz-Jar-Abr:1978:ASTRA:,Stu-Sla-Hle:2000:ASTRA:}. 

In the inner Keplerian disc, the critical point of the vanishing angular velocity gradient and its relation to the standard MRI viscosity mechanism is relevant, influencing thus the properties of the disc along the lines discussed in the previous case of the KS naked singularity spacetimes. The creation of some toroidal structures can be thus relevant also for this reason. Nevertheless, we can again assume that under some special conditions an inner Keplerian disc can be created at radii $r_{stat} < r < r_{\Omega/max}$ that has the specific angular momentum in the range $0 < L < L_{mb}$ and specific energy in the range $E_{stat} < E < E_{\Omega/max}$, being probably radiatively inefficient, if a viscosity mechanism cannot be at work. 

\subsection{Naked singularities with $\omega_h > \omega > \omega_{ph}$}

The KS naked singularity spacetimes containing two photon circular orbits, the inner stable one at $r_{ph/o}$, the outer unstable one at $r_{ph/i}>r_{ph/o}$, and one marginally stable orbit at $r_{ISCO} > r_{ph/i}$. The stable circular orbits are located at $r_{stat} < r < r_{ph/o}$, and at $r > r_{ISCO}$. Unstable circular orbits are located at $r_{ph/i} < r < r_{ISCO}$. At the region $r_{ph/o}<r<r_{ph/i}$, no circular geodesics are allowed. There is no critical point related to the vanishing of the angular-velocity gradient. The angular velocity increases (decreases) with decreasing radius at the outer (inner) Keplerian disc. The radial profiles of the specific angular momentum, the specific energy and the angular velocity of the circular geodetical orbits are given in Figure 5. The behaviour of the related effective potential can be found in \cite{Vie-etal:2013:submitted:}. 

In the outer Keplerian disc the viscosity MRI mechanism can work quite well and we can expect the standard Keplerian thermaly radiating disc in the whole region above $r_{ISCO}$. In the inner Keplerian disc some other viscosity mechanism could be working, however, its structure is quite non-standard, as the energy and angular momentum diverge and change extremely fastly near the stable  photon circular orbit, especially in the case of near-extreme KS naked singularity spacetimes. Creation of such a Keplerian structure by natural astrophysical processes seems to be very improbable. Accretion of matter from the edge of the outer Keplerian disc at the radius $r_{ISCO}$, with energy $E_{ISCO}$ and angular momentum $L_{ISCO}$, implies possibility of creation of the inner Keplerian at the low energy end near the static radius. Matter following the high-energy end of the inner Keplerian disc near the photon circular orbit could be created by particles incoming from infinity due to their high-energy collisions near the radius of the stable photon circular orbit \cite{Stu-Sche-Abd:2013:submitted:}. However, such possibilities are not astrophysically realistic, although they are not quite excluded. We can expect that in the innermost region of near-extreme KS naked singularities a toroidal accretion disc can occur, or a spherical configuration of mixture of matter and trapped photons can be created around the stable circular photon orbit, and the static radius. In the following we shall consider for simplicity only the inner Keplerian discs limited at the energy level corresponding to the ISCO energy. 

\subsection{Black holes -- $\omega > \omega_h$}

The KS black hole spacetimes have the circular geodesics above the outer horizon only. The inner limit is given by the photon circular orbit $r_{ph}$. The unstable circular orbits are located between the photon circular orbit and the marginally stable orbit, $r_{ISCO}$, above which the stable circular orbits are located, similarly to the case of the Schwarzschild spacetimes. The radial profiles of the specific angular momentum, the specific energy and the angular velocity of the circular orbits are given in Figure 5. Of course, the standard Keplerian thermaly radiating disc with the MRI viscosity mechanism extends in whole the region above the ISCO.

\subsection{Efficiency of the Keplerian accretion}

The Keplerian-accretion efficiency is given by the energy difference of the initial and final state of the matter in the disc. Of course, we have to consider a reasonable initial state. Usually, it is the rest state at infinity, with energy $E=1$. This condition is relevant for the outer Keplerian discs around KS naked singularities. In the case of the inner Keplerian discs the situation is more complex and depends on the value of the Ho\v{r}ava parameter of the KS spacetime. In realistic astrophysical situations, the initial state in the inner disc has to be related somehow to the final state of the accretion in the outer Keplerian or toroidal accretion disc. Therefore, for the Keplerian discs in the KS spacetimes with $\omega > \omega_{ms}$, it cannot be higher than $E_{ISCO}$, while for the toroidal perfect fluid configurations in cannot exceed $E=1$ if $\omega > \omega_{mb}$, and $E_{OSCO}$ if $\omega < \omega_{mb}$. The final state is given by $E_{ISCO}$ in the outer Keplerian discs, and by $E_{stat}$ in the inner Keplerian discs. 

For $\omega > \omega_{ms}$, the efficiency of the outer Keplerian discs is given by the formula 
\begin{equation}
                 \eta^{o}(\omega) = 1 - E_{ISCO}(\omega) ,
\end{equation}
while for the inner Keplerian discs it is limited from above by  
\begin{equation}
                 \eta^{i}(\omega) = E_{ISCO}(\omega) - E_{stat}(\omega) ,
\end{equation}
if inflow from the outer Keplerian discs is considered only. Higher values are allowed, if the initial state of the inner accretion is related to toroidal configurations, when the efficiency can go up to 
\begin{equation}
                 \eta^{i}(\omega) = 1 - E_{stat}(\omega) = (2\omega)^{1/3}.
\end{equation}
In the case of Keplerian discs in the spacetimes with stable photon circular orbits, when energy of matter in the inner Keplerian discs can grow without limit, we have to input somehow (e.g. by the collisional processes) this energy into the disc -- therefore, it cannot increase the efficiency of the accretion process relative to the distant observers. 

In the KS naked singularity spacetimes allowing existence of stable circular orbits only (with $\omega < \omega_{ms}$), the efficiency of the complete Keplerian accretion is again given by 
\begin{equation}
                 \eta(\omega) = 1 - E_{stat}(\omega) .
\end{equation}
However, if we restrict attention to the astrophysically realistic case of the Keplerian discs with the MRI driven viscosity mechanism, the accretion efficiency is given by 
\begin{equation}
                 \eta^{o}(\omega) = 1 - E_{\Omega/max}(\omega) .
\end{equation}

The Keplerian accretion efficiency is for all the considered cases illustrated in Figure 6. We can see that the efficiency of the Keplerian thermaly radiating accretion discs related to the MRI viscosity mechanism (at the outer Keplerian discs corresponding to the region of stable circular geodesics with the angular velocity decreasing with increasing radius) demonstrates quite interesting behavior. We observe a sharp discontinuity of the efficiency when evolution of the KS naked singularity due to the accreted matter causes crossing of the critical parameter $\omega = \omega_{ms}$. Then the efficiency jumps from the value of $\eta^{o} \sim 0.25$ down to $\eta^{o} \sim 0.072$ that is close to the efficiency of the accretion onto a Schwarzschild black hole $\eta_{Schw} \sim 0.0587$. \footnote{The maximal accretion efficiency of the Keplerian discs with the MRI governed viscosity mechanism, $\eta^{o}_{max} \sim 0.275$, is obtained for the Ho\v{r}ava parameter $\omega \sim 0.2$.} The inner edge of the Keplerian disc then jumps from the radius $r_{in} = r_{\Omega/max}(\omega_{ms}) \sim 2.1$ up to $r_{in}=r_{ms}(\omega_{ms}) \sim 3.5$. Such a dramatic change of the disc extension and the accretion efficiency should have a clear observational consequences that will be discussed in a future paper since a more detailed study of the innermost parts of the disc before the transition should take into account toroidal structures that probably will be developed in the innermost regions of the Keplerian discs, and some other physical phenomena.

If we consider, under assumption of existence of a hypothetical viscosity mechanism working also in the region where the circular geodesics have the angular velocity increasing with increasing radius,  complete Keplerian discs extending down to the static radius, the accretion efficiency can grow up substantially, going up to the value of $\eta = 1$ when $\omega \to 1/2$. Notice that in this limit the KS naked singularity evolution stops because $E=L=0$, and no energy or angular momentum are added to the KS naked singularity background being radiated during the accretion process. 

\begin{figure}
	\begin{center}
		\includegraphics[scale=0.9]{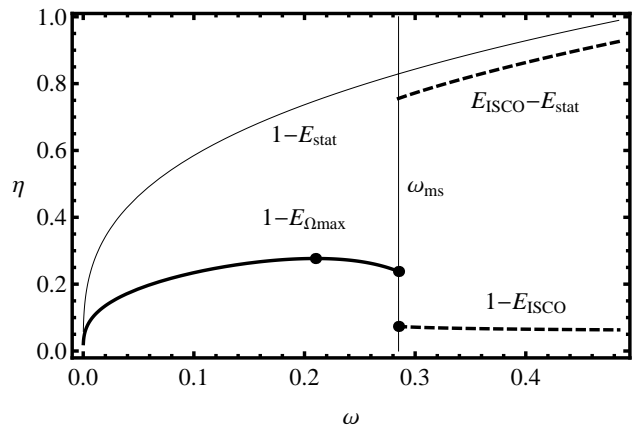}
		\caption{Efficiency of the Keplerian acretion in dependence on the Ho\v{r}ava parameter $\omega$. The efficiency is determined for the astrophysically relevant, thermaly radiating outer Keplerian discs where the standard MRI viscosity mechanism works (thick lines). The efficiency is also given for the complete theoretical model related to the final state of the accretion determined by the static radius where the accretion motion is stopped (light curves), or for the inner Keplerian disc assuming its initial state at the energy corresponding to the final state of the accretion in the outer thermaly radiating disc. The maximum of the accretion efficiency $1-E_{\Omega max}$ is at $(\omega,\eta)=(0.2099,0.2767)$. At the transition point, the efficiencies read $1-E_{\Omega max}|_{\omega _ms}=0.2379$ and  $1-E_{ISCO}|_{\omega_ms}=0.0735$.} 
	\end{center}
\end{figure}


\section{Optical phenomena and the appearance of Keplerian discs}

We study the optical phenomena occuring in all three classes of the KS naked singularity spacetimes; in the KS black hole spacetimes, the optical phenomena were discussed in \cite{Har-Kov-Lob:2011:CLAQG:,Ata-Abd-Ahm:2013:AstrSpaScie:} and will not be repeated here. We present in this section the optical appearance of the Keplerian accretion discs reflecting both their shape distortions due to the gravitational lensing, and the frequency shift due to the gravitational and Doppler effects. In order to visualize clearly the combined gravitational and Doppler shifts, we assume the Keplerian discs radiating locally at a fixed frequency (corresponding, e.g., to a Fe X-ray line). In the following section, we construct the spectral continuum of the standard thermaly radiating Keplerian discs with the thermal radiation profile given by the Page-Thorne model \cite{Pag-Tho:1974:ApJ:}. Finally, we construct profile of the spectral lines (usually the fluorescent spectral Fe lines are assumed) generated in the innermost regions of the Keplerian discs. The disc appearance can be relevant for sources close enough that could enable a detailed study of the innermost parts of the accretion structures, as can be expected in near future for the Sgr A* source \cite{Doe-etal:2009:APJ:}. Of course, the spectral continuum and the profiled spectral lines are relevant also for much more distant sources.  

We shall demonstrate the appearance of the Keplerian discs for small, mediate, and large inclination angle of the discs relative to the distant observers. We shall construct first the appearance of the Keplerian discs related to the MRI viscosity mechanism. Therefore, in the KS naked singularity spacetimes with $\omega > \omega_{ms}$, we restrict our attention to the outer Keplerian discs located at $r > r_{ISCO}$, while in the spacetimes with $\omega < \omega_{ms}$, we shall consider the part of the Keplerian discs at $r > r_{\Omega/max}$. Alternatively, we add also some inner parts of the Keplerian discs. For the KS naked singularity spacetimes with $\omega > \omega_{ms}$, we add only the part of the inner Keplerian discs having the outer edge given by the energy $E = E_{ISCO}$ and extending down to the static radius, in order to keep some connection to the accretion from the outer Keplerian discs; more complex studies are planned in future papers. For the KS spacetimes with $\omega < \omega_{ms}$, we consider the complete Keplerian discs extending down to the static radius. Note that in the alternative cases, we obtain a special internal accretion disc structure, because in the inner edge (and its vicinity) of the inner Keplerian discs the frequency shift will not be dependent on the azimuthal angle as the Doppler shift becomes irrelevant at the static radius.

We apply the same alternatives also when constructing the spectral continuum and the profiled spectral lines generated by the Keplerian discs.

\subsection{Photon motion}

The radial component of the photon 4-momentum reads  
\begin{equation}
\left[P^{r}\right]^{2}=E^{2}-f(r)\left(\frac{L^{2}+Q^{2}}{r^{2}}\right).\label{eq_radial}
\end{equation}
The trajectories of photons are independent of energy, therefore, it is convenient to relate the effective potential of photons relative to the impact parameters 
\begin{equation}
                l = \frac{L}{E} , q = \frac{q}{E}
\end{equation}
For general, non-equatorial photon motion, it is convenient to use the coordinates 
\begin{equation}
       u=\frac{1}{r} , m=\cos\theta
\end{equation}
and to reparameterize the radial motion equation by $Ew\rightarrow w$, using simultaneously also the constants of motion in the form of impact parameters $l=L/E$ and $q^{2}=Q^{2}/E^{2}$. Equations (\ref{eq_radial}) and (\ref{eq_latitudunal}) then transfrom into
\begin{equation}
\frac{du}{dw}=\pm u^{2}\sqrt{1-\widetilde{f}(u)\left(l^{2}+q^{2}\right)u^{2}}
\end{equation}
where  
\begin{equation}
\tilde{f}(u)=f(1/u)
\end{equation}
and
\begin{equation}
\frac{dm}{dw}=\pm u^{2}\sqrt{q^{2}-(l^{2}+q^{2})m^{2}} 
\end{equation}
that can be properly integrated when photons radiated by Keplerian discs are considered \cite{Rau-Bla:1994:ApJ:,Sche-Stu:2009:IJMPD:,Stu-Sche:2010:CLAQG:}. 

\subsection{Frequency shift}

The frequency shift of radiation emitted by a point source moving along a circular orbit in the equatorial plane is given in the standart manner. The frequency shift $g$ between the emitter (e) and observer (o) is defined by the formula
\begin{equation}
g=\frac{k_{\mu}U^{\mu}|_{o}}{k_{\mu}U^{\mu}|_{e}}.
\end{equation}
In case of circular orbits, the emitter four-velocity components are
\begin{equation}
U^{\mu}=\left[U^{t},0,0,U^{\phi}\right].
\end{equation}
For the static observers at infinity, the frequency shift formula $g$ then reads
\begin{equation}
g=\frac{1}{U_{e}^{t}(1-l\Omega)}, 
\end{equation}
where $l$ is the impact parameter of the photon, and $\Omega$ is the angular velocity of the emitter (the behavior of the frequency shift is illustrated in Figure 7). The temporal component of the emitter four-velocity one obtains from the norm of the four-velocity:
\begin{equation}
 \frac{1}{[U_{e}^{t}]^{2}}=f(r)-r_{e}^{2}\Omega^{2}=1+r_{e}^{2}\omega\left[1-\sqrt{1+\frac{4}{r_{e}^{3}\omega}}\right]-r_{e}^{2}\Omega^{2}.
\end{equation}
In our study, the angular velocity of the emitter is given by the Keplerian angular velocity, $\Omega = \Omega_{K}$, determined by equation (\ref{eq30}). 
\begin{figure}[H]
	\begin{center}
	\includegraphics[scale=0.8]{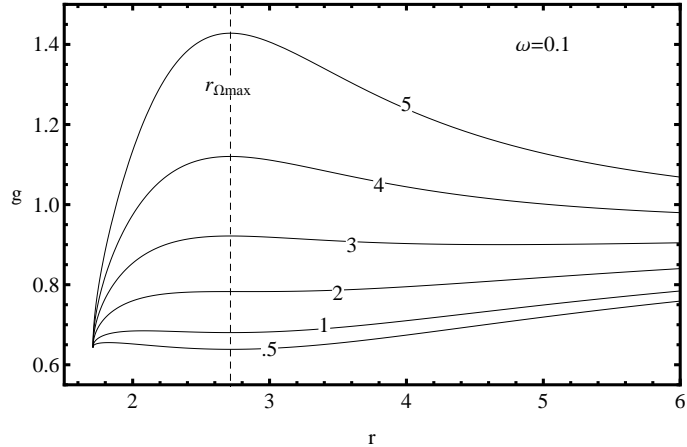}
	\caption{The frequency shift $g$ is given as function of the radial coordinate $r$ for fixed representative value of $\omega=0.1$ and some characteristic values of the impact parameter $l$.}
	\end{center}
\end{figure}
\subsubsection{Appearance of Keplerian discs}

We give the appearance of the Keplerian discs for all the three relevant classes of the KS naked singularity spacetimes. Modifications of the shape of the discs are caused by the gravitational lensing effect, while the frequency shift is caused by the combined effect of the gravitational redshift and the Doppler shift. We assume for simplicity that the disc is radiating with a fixed frequency along whole active area that is considered to be restricted to the outer Keplerian disc, i.e., to the region between the $r_{ISCO}$ and $r=20M$ in the case of KS naked singularity spacetimes with $\omega > \omega_{ms}$, and to the region between $r_{\Omega/max}$ and $r=20M$ in the case of the KS spacetimes with $\omega < \omega_{ms}$. Therefore, we give first the appearance of the Keplerian discs governed by the condition $\frac{d\Omega_{K}}{dr} < 0$ that assumes the MRI mechanism as the only mechanism of generating the viscosity effects enabling accretion in the Keplerian discs.  We give the disc appearance in dependence on the inclination angle relative to the observer, for small (Figure 8), mediate (Figure 9) and large (Figure 10) inclination angles, for typical values of the parameter $\omega$. For all the inclination angles we present for comparison also the appearance of the Keplerian disc orbiting a KS black hole ($\omega > 1/2$).

Alternatively, we construct for the KS naked singularity spacetimes with $\omega > \omega_{ms}$ the models of the disc appearance including also the inner part of the Keplerian disc located between $r_{stat}$ and $r_{E=E_{ISCO}}$ for the chosen inclination angles (Figure 11). For the KS naked singularity spacetimes with $\omega=0.1 < \omega_{ms}$ we construct alternatively in Figure 12 the appearance of the Keplerian discs including their inner part, i.e., the complete Keplerian discs extending between the static radius $r_{stat}$ and $r=20M$. For comparison, we give in Figure 13 the appearance of the complete Keplerian disc and the outer Keplelrian disc in the KS spacetime with the parameter $\omega=0.2$, when the accretion efficiency of the outer Keplerian disc is close to the maximal value related to the standard Keplerian discs governed by the MRI viscosity mechanism. In the alternate cases, we assume a different viscosity mechanism, related to the opposite gradient of the angular velocity ($\frac{d\Omega_{K}}{dr} > 0$), at work (it is not clear, if the MRI viscosity can occur also in such situations), implying that the inner parts of the Keplerian discs can also efficiently radiate giving thus completely radiating discs, since with decreasing radius of the orbiting matter both its angular momentum and energy decrease. However, such an assumption is not necessary, as we consider discs radiating with (locally) fixed frequency corresponding to some spectral line that could be induced by irradiation of the disc -- then an efficient viscosity is not necessary in the Keplerian disc.

In all the constructed images of the Keplerian discs we give the map of the relative frequency shift 
\begin{equation}
                    g*=\frac{g-g_{min}}{g_{max}-g_{min}}
\end{equation}
that is related to the range $(g_{max}-g_{min})$ reflecting the extension of the frequency shift in all the considered cases of the modelled appearance of the discs. Note that the frequency range, if observed, could serve as a strong tool for determining the naked singularity spacetime parameters, as demonstrated in the case of the Kerr naked singularities \cite{Stu-Sche:2010:CLAQG:}. The disc appearance is constructed in the observer detection plane, using the coordinates $\alpha, \beta$ defined in \cite{Bar:1973:BlaHol:,Sche-Stu:2009:IJMPD:}. 

\begin{figure}[H]
\begin{center}
\includegraphics[scale=0.8]{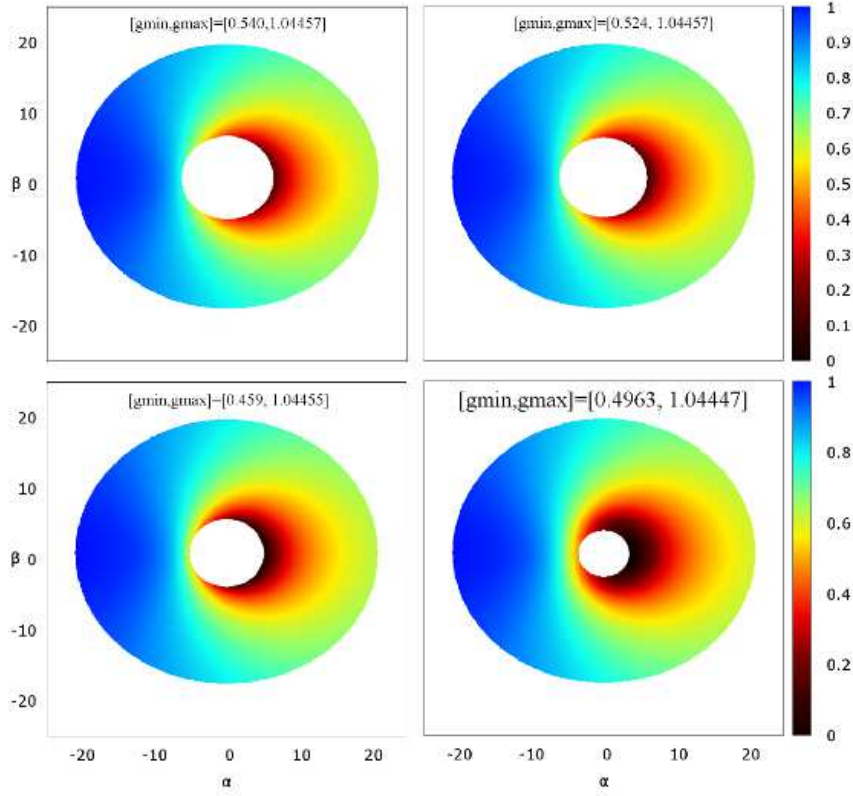}
\end{center}
\caption{Appearance of the outer Keplerian discs (where the MRI viscosity mechanism can work). The images are generated for the observer inclination angle $\theta_{0}=30^\circ$, and for four representative values of the KS spacetimes $\omega=0.6$, $0.45$, $0.3$, and $0.1$ (from left top to right bottom). The value of $\omega=0.6$ gives a KS black hole spacetime, while the other values of $\omega$ correspond to the KS naked singularity spacetimes. The inner edge of the disc corresponds to $r_{ISCO}$ for the spacetimes with $\omega=0.3$, $0.45$, $0.6$, and to $r_{\Omega/max}$ for the spacetime with $\omega=0.1$. The outer edge of the disc is located at $r=20$M in all the cases. The frequency shift is expressed by $g*=(g-g_{min})/(g_{max}-g_{min})$, where $g_{min}$ ($g_{max}$) is the minimal (maximal) frequency shift from all four models with $\theta_{0}=30^\circ$. The Keplerian discs are assumed to be rotating anticlockwise.}
\end{figure}

\begin{figure}[H]
\begin{center}
\includegraphics[scale=0.8]{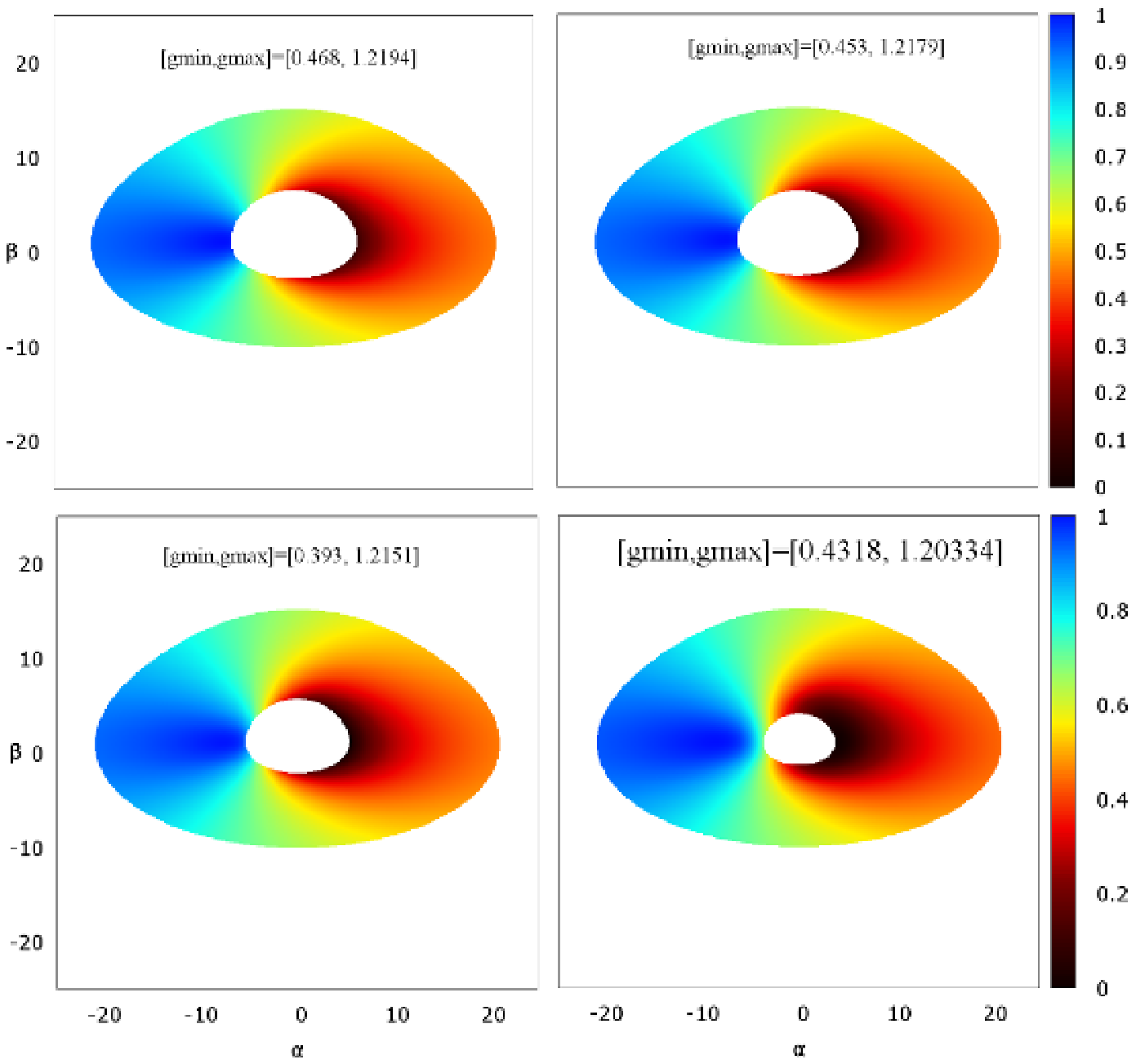}
\end{center}
\caption{Appearance of the outer Keplerian discs. The images are generated for observer inclination angle $\theta_{0}=60^\circ$, and for the four representative values of $\omega=0.6$, $0.45$, $0.3$, and $0.1$ (from left top to right bottom). The inner edge of the disc is at $r_{ISCO}$ for $\omega=0.3$, $0.45$, $0.6$, and at $r_{\Omega max}$ for $\omega=0.1$. The outer edge of the disk at $r=20$M. The frequency shift is given by $g*=(g-g_{min})/(g_{max}-g_{min})$, where $g_{min}$ ($g_{max}$) is the minimal (maximal) frequency shift from all four models with $\theta_{0}=60^\circ$.}
\end{figure}

\begin{figure}[H]
\begin{center}
\includegraphics[scale=0.8]{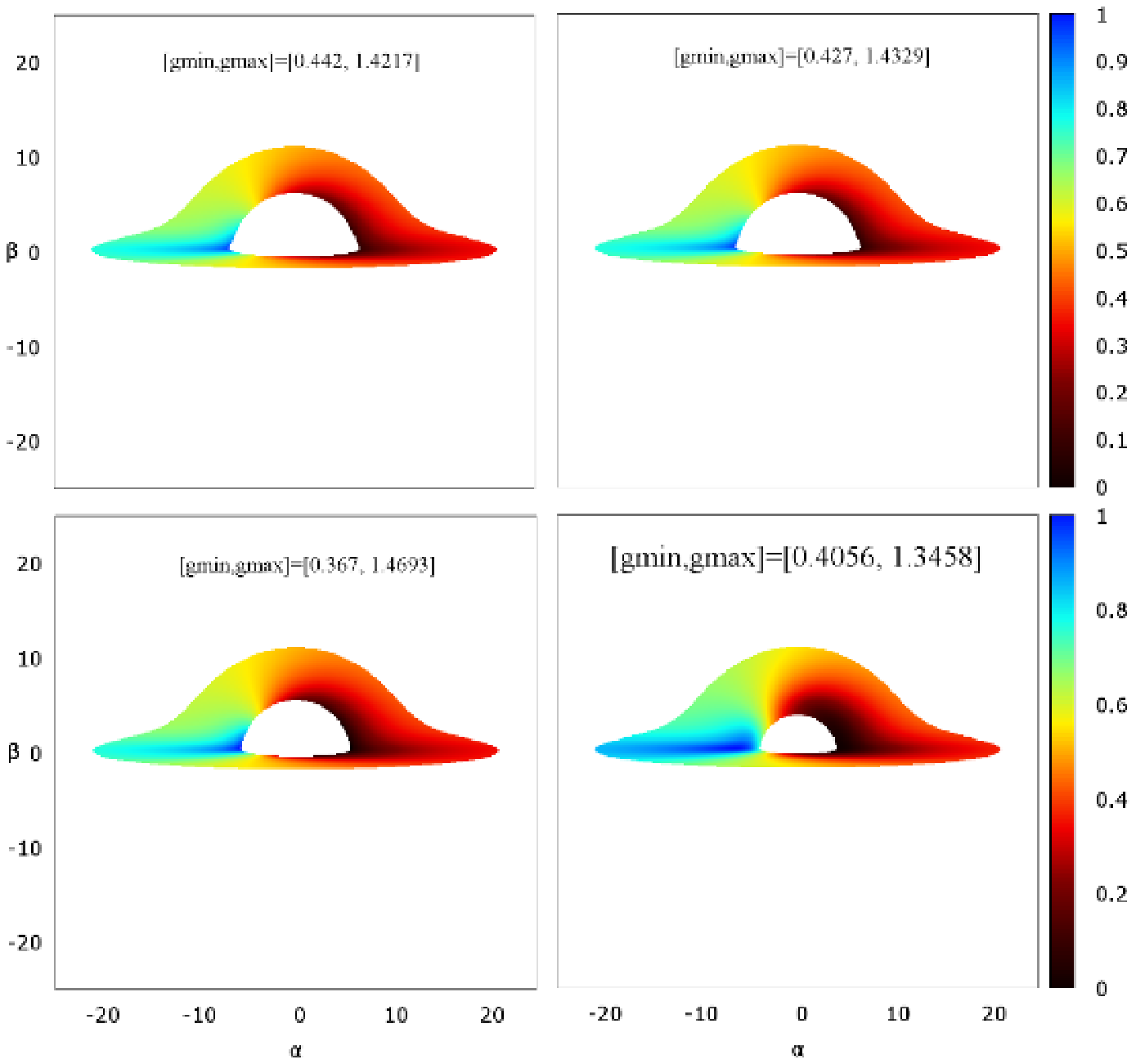}
\end{center}
\caption{Appearance of the outer Keplerian discs. The images are generated for observer inclination angle $\theta_{0}=85^\circ$, and for the four representative values of $\omega=0.6$, $0.45$, $0.3$,  $0.1$ (from left top to right bottom). The inner edge of the disc is at $r_{ISCO}$ for $\omega=0.3$, $0.45$, $0.6$, and at $r_{\Omega max}$ for $\omega=0.1$. The outer edge of the disk at $r=20$M. The frequency shift is given by $g*=(g-g_{min})/(g_{max}-g_{min})$, where $g_{min}$ ($g_{max}$) is the minimal (maximal) frequency shift from all four models with $\theta_{0}=85^\circ$.}
\end{figure}

\begin{figure}[H]
\begin{center}
\includegraphics[scale=0.8]{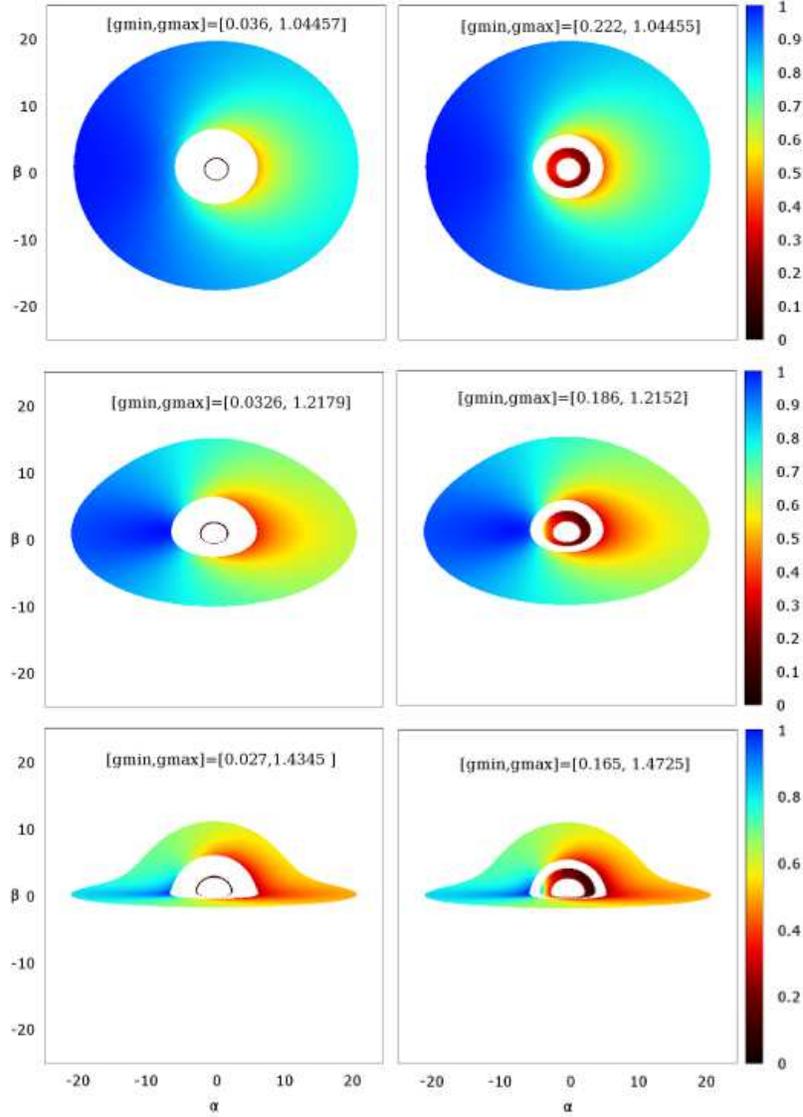}
\par\end{center}
\caption{Appearance of the inner and outer Keplerian discs in the field of KS naked singularities. The images of the discs are generated generated for the three representative values of the distant observer inclination $\theta_{0}=30^\circ$ (top), $60^\circ$(middle), $85^\circ$ (bottom), and for two representative values of the Ho\v{r}ava parameter $\omega=0.45$ (left), $0.3$ (right). The inner disc: the inner edge corresponds to $r_{stat}$, and the outer edge is located at the radius where the energy of the circular Keplerian orbit is the same as energy of marginally stable orbit of the outer disc ($E_{ISCO}$). The outer disc: the inner edge is at $r_{ISCO}$ and the outer edge of the disc is set to $r=20$M. The frequency shift is expressed by $g*=(g-g_{min})/(g_{max}-g_{min})$, where $g_{min}$ ($g_{max}$) is the minimal (maximal) frequency shift taken from all images with $\theta_o$ kept fixed.}
\end{figure}

\begin{figure}[H]
\begin{center}
\includegraphics[scale=0.8]{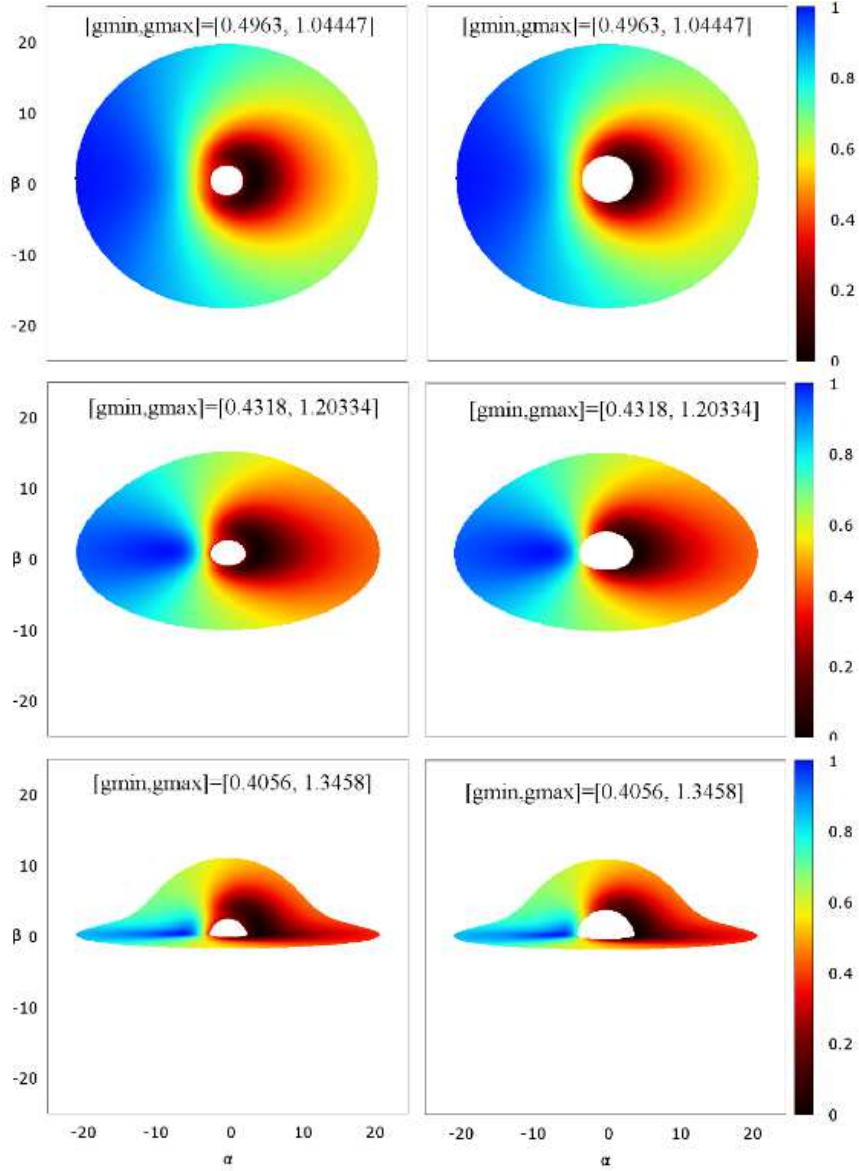}
\par\end{center}
\caption{Appearance of the complete Keplerian disc compared to the appearance of the outer Keplerian disc. The images are generated for the three representative values of the observer inclination angle  $\theta_{0}=30^\circ$ (top), $60^\circ$(middle) and $85^\circ$ (bottom), and for the Ho\v{r}ava parameter $\omega=0.1$. The inner edge of the Keplerian disc corresponds to $r_{stat}=r_{stat}(\omega)$ for the complete disc (left image), and to $r_{\Omega max}=r_{\Omega max}(\omega)=(2/\omega)^{1/3}$ for the outer disc (right image). The outer edge of the disc is located at   $r_{out}=20M$. The frequency shift is given by $g*=(g-g_{min})/(g_{max}-g_{min})$, where $g_{min}$ ($g_{max}$) is the minimal (maximal) frequency shift from all images with $\theta_{0}$ kept fixed.}
\end{figure}

\begin{figure}[H]
\begin{center}
\includegraphics[scale=0.8]{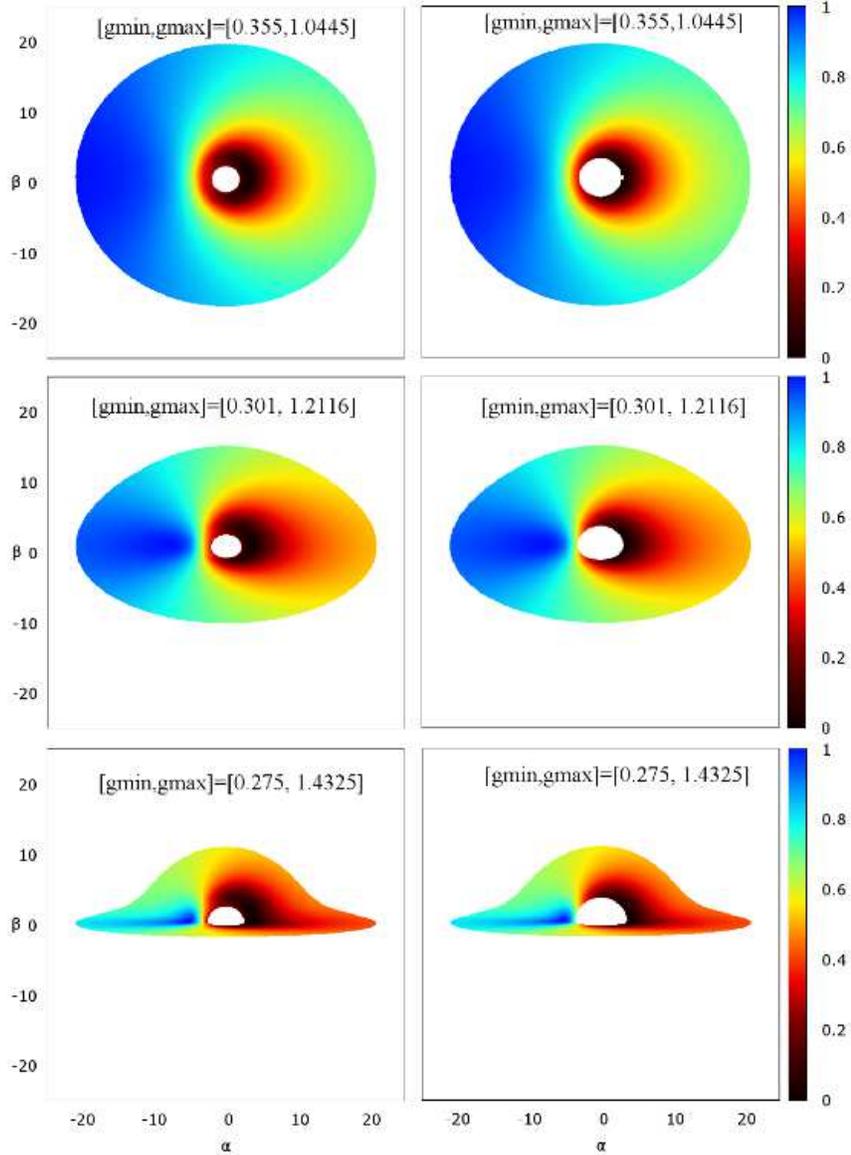}
\par\end{center}
\caption{Appearance of the complete Keplerian disc compared to the appearance of the outer Keplerian disc. The images are given for the three representative values of the observer inclination angle  $\theta_{0}=30^\circ$ (top), $60^\circ$ (middle) and $85^\circ$ (bottom), and for the parameter $\omega=0.2$ when the accretion efficiency of the outer disc reaches maximum. The inner edge of the complete dics is at $r_{stat}=r_{stat}(\omega)$ (left image) and at $r_{\Omega max}=r_{\Omega max}(\omega)=(2/\omega)^{1/3}$ for the outer disc (right image). The outer edge of the disc is at  $r_{out}=20M$. The frequency shift given by $g*=(g-g_{min})/(g_{max}-g_{min})$, where $g_{min}$ ($g_{max}$) is the minimal (maximal) frequency shift from all images with $\theta_{0}$ kept fixed.}
\end{figure}

\section{Spectral continuum of thermaly radiating Keplerian discs}

We assume the Planck spectrum of the thermaly radiating Keplerian disc, given by the standard Novikov-Thorne, or more precisely Page-Thorne model \cite{Nov-Tho:1973:BlaHolg:,Pag-Tho:1974:ApJ:}. In this model the Keplerian disc is assumed to be in a thermal equilibrium, i.e., the energy generated locally at a given radius by friction (viscosity processes) is thermalized and effectively radiated at the given radius. 

We let the Keplerian disc to have the particular temperature radial profile $T(r)$ given by the Page-Thorne model, and the locally emitted spectrum to be Planckian, i.e., given by the relation 
\begin{equation}
I_{e}=B(\nu_{e},T)=\frac{2\nu_{e}^{3}}{\exp\left[\nu_{e}/T\right]-1}.
\end{equation}

\begin{figure}[H]
\begin{center}
\includegraphics[scale=0.8]{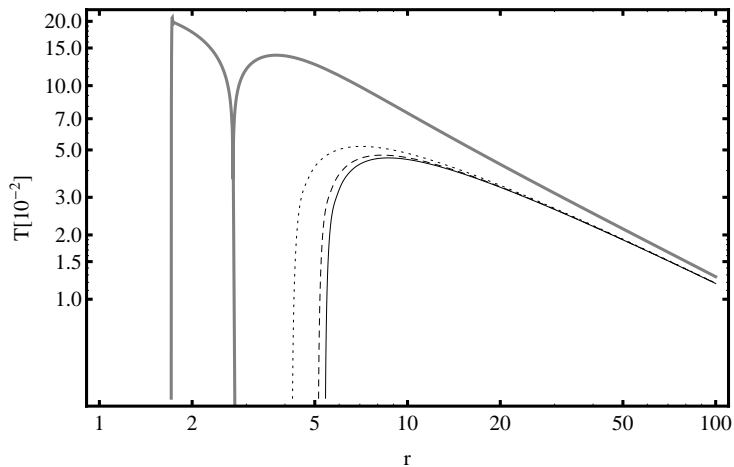}
\caption{Temperature profile of the outer Keplerian discs thermalized by an efficient viscosity mechanism are constructed for the KS spacetimes with four representative values of Ho\v{r}ava parameter $\omega=0.1$ (gray, thick), $0.3$ (black,dotted), $0.45$ (black, dashed), and $0.6$ (black).  In the case of the KS naked singularity spacetime with $\omega = 1$, the complete Keplerian disc is considered assuming an efficient viscosity mechanism and thermalization of the disc also under the critical radius where the angular velocity and temperature vanish.}
\end{center}
\end{figure}

We express both the frequency and the temperature in energy units. The radial profile of the effective temperature of the Keplerian accretion disc is governed by the standard Stefan-Boltzman flux formula
\begin{equation}
F(r)=\sigma T^{4}(r)
\end{equation}
where $\sigma$ is the Stefan-Bolzmann constant; the radiation-flux radial profile of the thermaly (as a black body) radiating, and properly averaged, standard Keplerian disc with inner edge located at the innermost stable circular geodesic is given by the standard formula related to the energy, angular momentum and angular velocity radial profiles of the geodesic circular motion, expressed by the Page-Thorne model \cite{Pag-Tho:1974:ApJ:}
\begin{equation}
F(r)=-\frac{\dot{M_{0}}}{4\pi\sqrt{-g}}\frac{\Omega_{K},_{r}}{(E_{K}-\Omega_{K} L_{K})^{2}}\int_{r_{ISCO}}^{r}\left(E_{K}-\Omega_{K} L_{K}\right)L_{K},_{r}dr.\label{eq_bb_temp}
\end{equation}
The term $\dot{M_{0}}$ denotes the accretion flow of the rest mass in the Keplerian disc that is considered to be a constant. There are some restrictions on the allowed values of the rest mass accretion flow in the thermalized Keplerian discs that are discussed, e.g., in \cite{McCli-etal:2012:CLAQG:} - here we simply assume this constraint to be fulfilled, not going into details that have to be related to concrete observed objects. We normalize our results to the accretion flow. Further, the radiation flux is scaled by the KS spacetime mass parameter as $1/M^2$, while the disc temperature is scaled as $1/\sqrt{M}$ \cite{Pag-Tho:1974:ApJ:}.  

If we consider in the case of the KS naked singularities the whole Keplerian disc to be radiating, we have to use the relation
\begin{equation}
F(r)=\frac{\dot{M_{0}}}{4\pi\sqrt{-g}}\frac{|\Omega_{K},_{r}|}{(E_{K}-\Omega_{K} L_{K})^{2}}\int_{r_{edge}}^{r}\left(E_{K}-\Omega_{K} L_{K}\right)L_{K},_{r}dr,\label{eq_bb_temp}
\end{equation}
where $r_{in}$ denotes the edge of the radiating Keplerian disc under consideration. Therefore, we can assume $r_{edge}$ corresponding to the static radius $r_{stat}$, or corresponding to $r_{\Omega/max}$, if we assume only the MRI induced viscosity requiring $\frac{d\Omega_{K}}{dr} < 0$ \cite{Bal-Haw:1998:RevModPhys:}

The observed spectral continuum is given by the redshifted black body spectrum integrated accross the radiating Keplerian disc. The luminosity $L$ given in dependence on the observed frequency $\nu$ is determined by the formula
\begin{equation}
     L(\nu) = \frac{8}{\pi c^2} cos\theta_{0} \int_{r_{i}}^{r_o}\int_{0}^{2\pi}
     \frac{\nu_{e}^{3}r dr d\phi}{\exp\left[\nu_{e}/T\right]-1}
\end{equation}
where $r_i$ ($r_o$) determines the inner (outer) edge of the radiating Keplerian disc, $\theta_{0}$ denotes the inclination angle of the disc relative to the distant observer, and the emitted frequency is related to the observed frequency by the redshift formula \cite{Lum:1979:ASTRA:,Bao-Stu:1992:ApJ:,Stu-Bao:1992:GerRelGrav:} 
\begin{equation}
         \frac{\nu_e}{\nu} = \frac{1+\Omega_{K} r \sin\phi \sin\delta}{\sqrt{f(r)-\Omega_{K}^2 r^2}}.
\end{equation}
At a given radius $r=r_e$, the temperature is determined by the radial profile $T(r)$, and the redshift formula is expressed for all photons reaching directly a distant observer fixed by the inclination angle $\theta_{0}$ and the azimuthal angle $\phi_{o} = 0$. The angle $\phi$ determines position of the emitter along the circle with radius $r_e$. 

Mixing the relations for the radiation flow $F(r)$ and the disc temperature, the radial profile of the temperature of the Keplerian discs $T(r)$ can be determined - the temperature profiles are calculated and illustrated in Figure 14 for the outer Keplerian discs in the KS naked singularity spacetimes with the parameter $\omega > \omega_{ms}$ and a black hole spacetime with $\omega > 1/2$, while the temperature profile of the complete Keplerian disc is presented for the KS naked singularity spacetime with $\omega < \omega_{ms}$. (The temperature profiles are normalized to the expression containing the mass accretion term $\frac{\dot{M_{0}}}{4\pi\sqrt{-g}}$; the same normalization is applied also in the figures giving the radiation-flux radial profiles and the spectral continuum.) The temperature profiles of the outer Keplerian discs in the KS naked singularity and black hole spacetimes with $\omega > \omega_{ms}$ vanish at the inner edge of the discs located at ISCO. In the KS spacetimes with $\omega < \omega_{ms}$, the temperature profile of the complete Keplerian disc vanishes at the inner edge of the complete disc, the static radius, and at the $r=r_{\Omega/max}$ where the angular velocity gradient vanishes, giving thus a clear qualitative signature of the naked singularity spacetimes. 

In order to have an overview of the situation related to the completely thermalized Keplerian discs, we give the radial profiles of the radiation flux of both the inner and outer Keplerian discs in the KS naked singularity spacetimes with $\omega > \omega_{ms}$ in Figure 15, assuming that also in the whole inner Keplerian discs an efficient viscosity mechanism works (although such an assumption is not realistic). For thermaly radiating complete Keplerian discs around KS naked singularity spacetimes with $\omega < \omega_{ms}$, the radiation flux is demonstrated in Figure 16.  

\begin{figure}[H]
\begin{center}
\begin{tabular}{cc}
\includegraphics[scale=0.7]{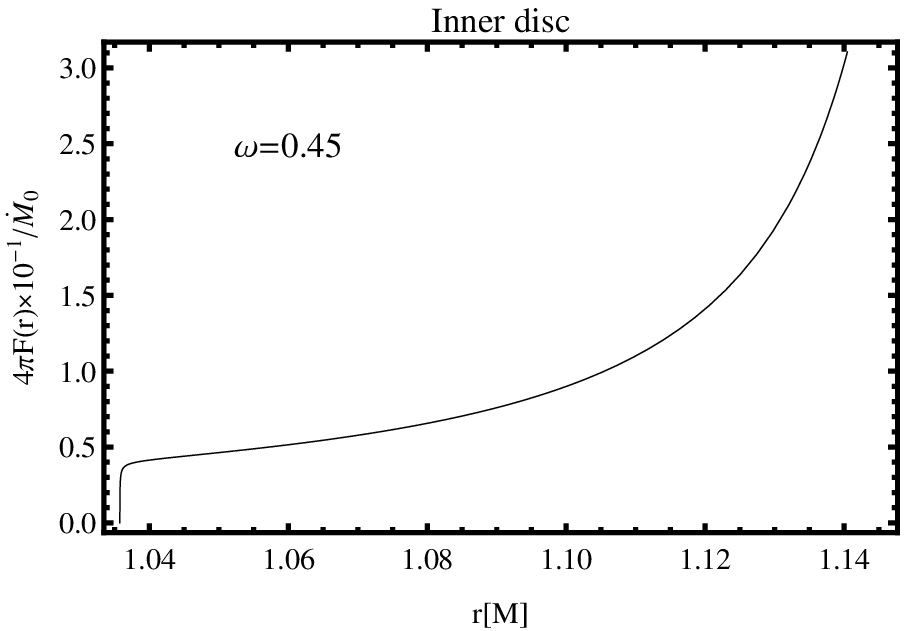}&\includegraphics[scale=0.7]{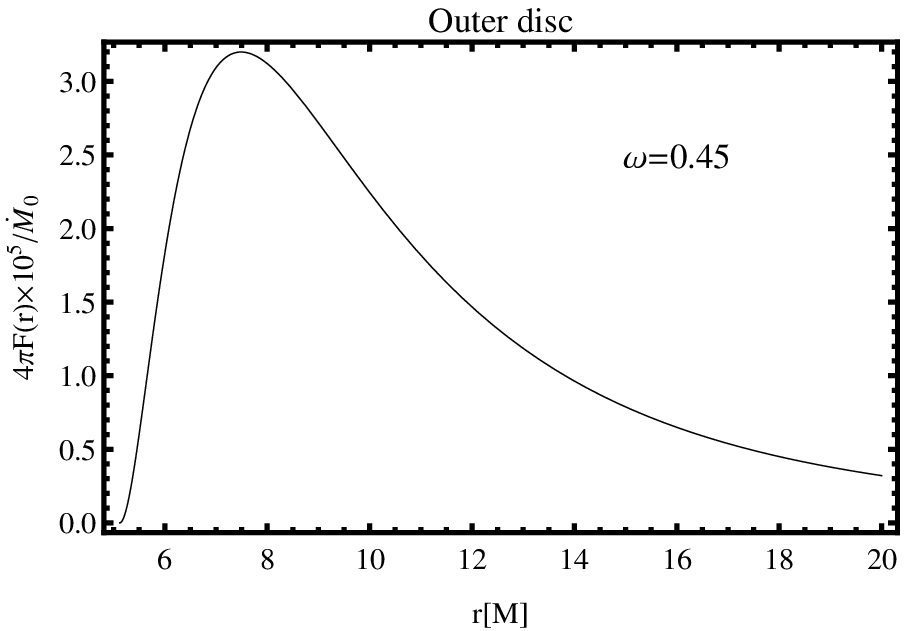}\\
\includegraphics[scale=0.7]{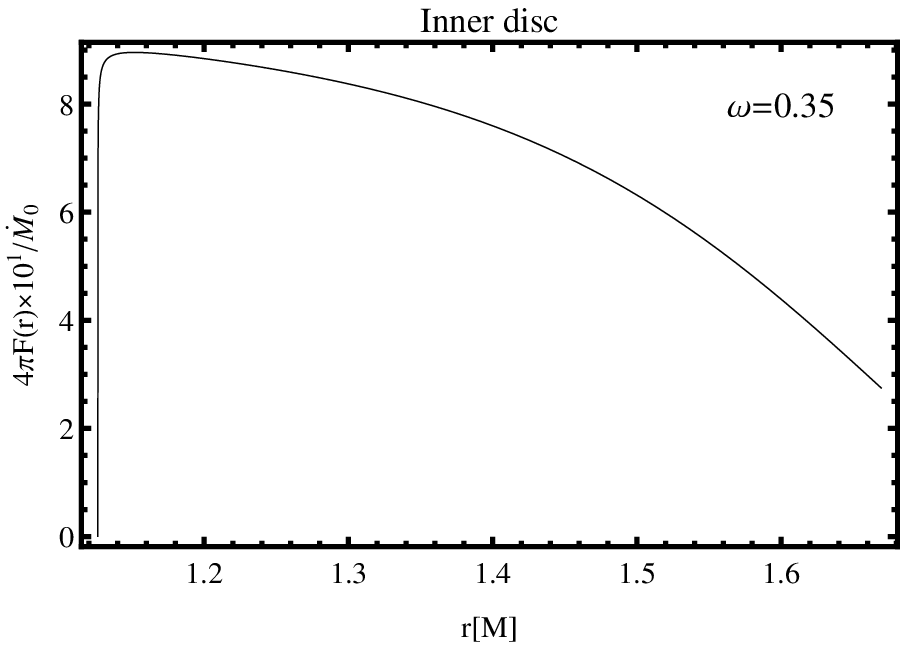}&\includegraphics[scale=0.7]{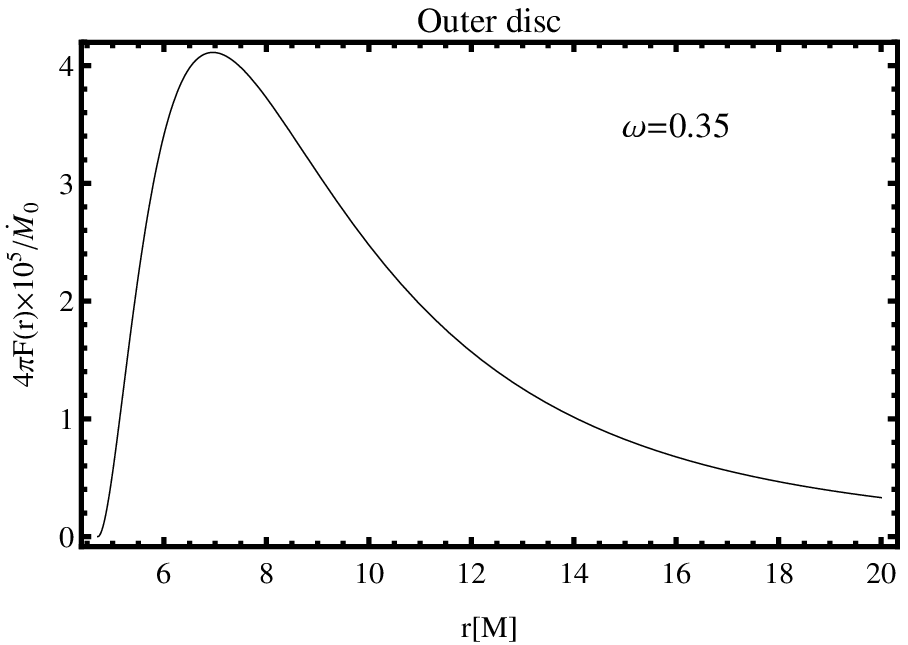}\\
\includegraphics[scale=0.7]{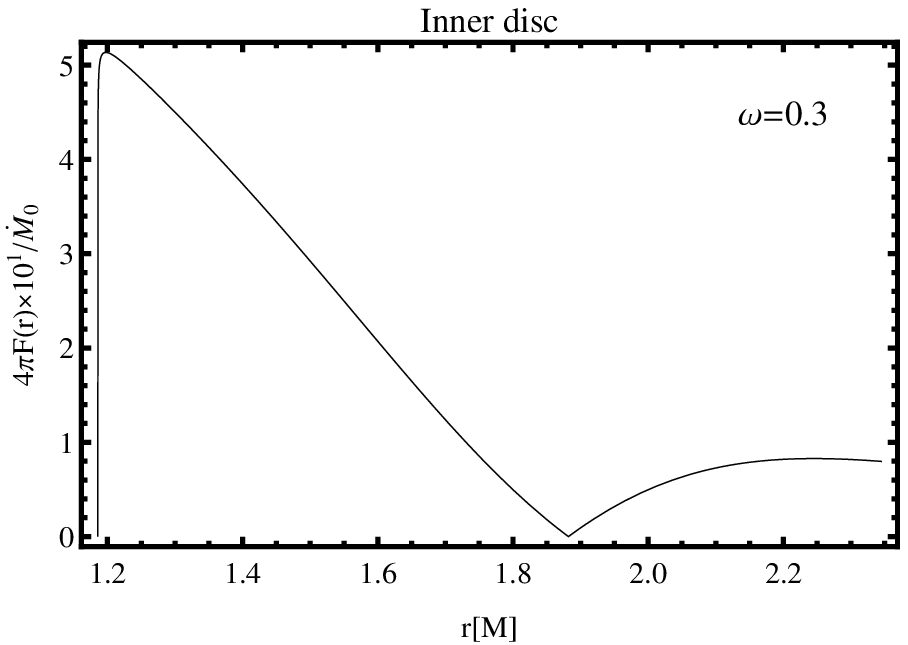}&\includegraphics[scale=0.7]{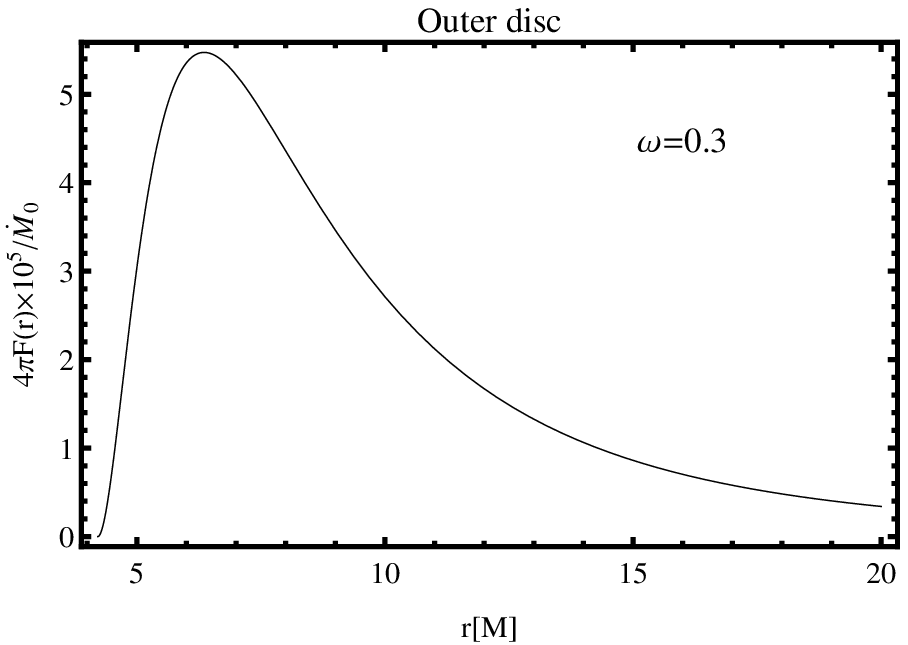}
\end{tabular}
\caption{Radial profiles of the radiative flux from the thermalized Keplerian accretion disc orbiting the KS naked singularity spacetimes with parameter $\omega > \omega_{ms}$. In the left columnn, the profile is given for the inner Keplerian discs, in the right column, for the outer Keplerian discs. The upper row corresponds to the spacetimes with $\omega > \omega_{ph}$ where the angular-velocity gradient is non-zero accross both the inner and outer discs. The two lower rows correspond to the spacetimes with $\omega_{ph} > \omega > \omega_{ms}$ where the angular-velocity gradient, and the radiation flux, vanish in the inner disc separating thus such discs in two, inner and outer, parts. Notice that for a fixed accretion flow the radiation flux from the inner discs exceeds by several orders the radiation flux from the outer discs, indicating necessity of an unrealistically efficient viscosity mechanism in the inner discs.} 
\end{center}
\end{figure} 

The spectral continuum, i.e., the $\nu L(\nu)$ dependence, is constructed for the outer Keplerian discs orbiting and thermaly radiating in the field of KS naked singularity (and black hole) spacetimes with characteristic values of the Ho\v{r}ava parameter $\omega$. We assume that the outer Keplerian discs extend between the ISCO (or the critical radius $r_{\Omega/max}$) and infinity; the resulting spectra are presented for the small, middle and large inclination angles in Figure 17. For the KS naked singularity spacetime with $\omega=0.1 < \omega_{ms}$, where the region of stable circular orbits extends from infinity down to the static radius, the spectral continuum has been constructed in two alternatives in Figure 18. In the first one, the spectral continuum is calculated for the outer disc with the inner edge at $r_{\Omega/max}$ where the model of the MRI viscosity has its limit. In the second version, we added also the radiation from the region between $r_{stat}$ and $r_{\Omega/max}$ assuming a different mechanism for the viscosity and thermalization of the Keplerian disc. 

Finally, we compare for the three typical inclination angles relative to the distant observers the spectral distribution of the radiation flux ($F(\nu_{o})$) for the thermaly radiating outer Keplerian discs and complete Keplerian discs in the KS naked singularity spacetimes with $\omega < \omega_{ms}$ in Figure 19, and of the outer Keplerian disc and the outer plus inner Keplerian discs in the KS naked singularity spacetime with $\omega_{ms} < \omega < \omega_{ph}$ in Figure 20. For the KS naked singularity spacetimes with $\omega_{ms} < \omega < \omega_{ph}$, the inner Keplerian disc in the region between the static radius and the OSCO orbit has been limited in its extension to the radius corresponding to the energy $E_{ISCO}$, at its outer edge. We thus relate the extension of the inner disc to matter inflowing from the outer Keplerian discs. 

The thermalized inner Keplerian discs are not considered here for the KS naked singularity spacetimes with $\omega_{h} > \omega > \omega_{ph}$ -- existence of the thermalized inner Keplerian discs is astrophysically highly improbable in such spacetimes because of the existence of trapped photons in the region where the inner Keplerian discs have to be located, similarly to the case of the Kerr naked singularity spacetimes \cite{Stu-Sche:2010:CLAQG:,Stu-Sche:2012:CLAQG:}.  We can expect that instead of the inner Keplerian discs, some inner toroidal or spherical accretion structures will evolve mixing the accreting matter and the trapped photons. We plan to study such structures in future. 

\begin{figure}[H]
\begin{center}
\begin{tabular}{cc}
\includegraphics[scale=0.7]{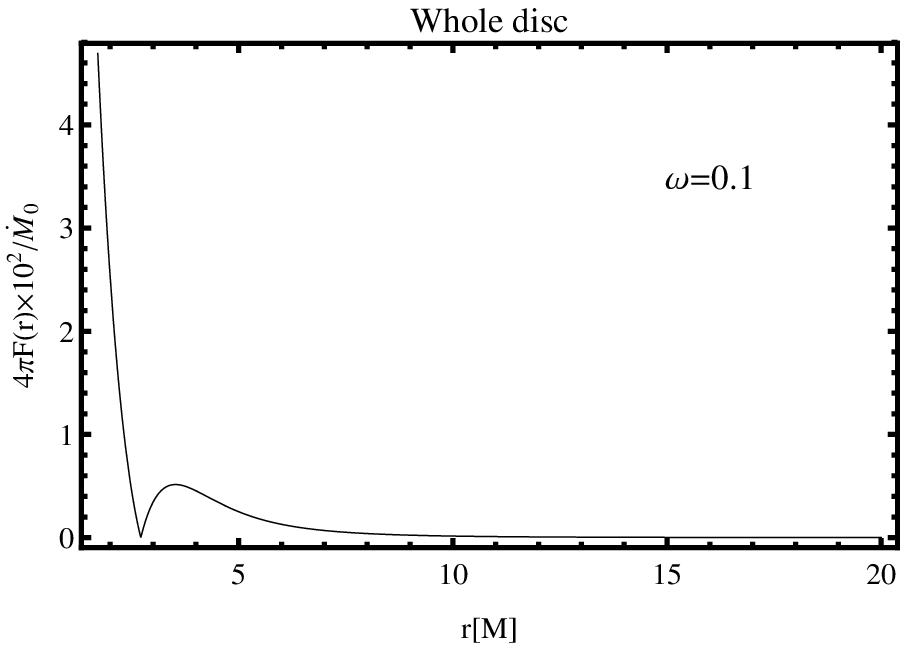}&\includegraphics[scale=0.7]{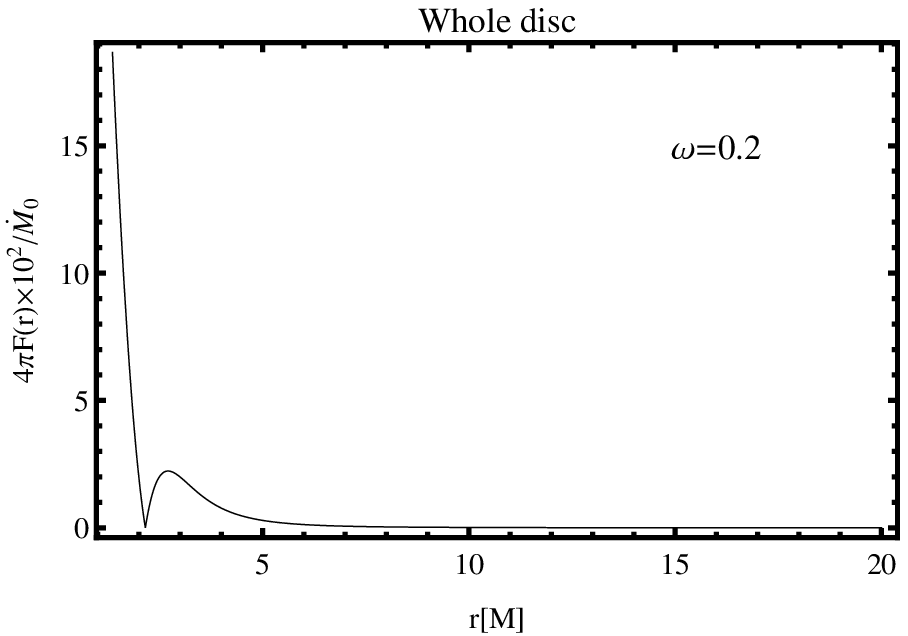}
\end{tabular}
\caption{Radial profiles of the radiative flux from the thermalized complete Keplerian accretion discs orbiting the KS naked singularity spacetimes with parameter $\omega < \omega_{ms}$ are given for two characteristic values of the parameter $\omega = 0.1, 0.2$. The complete Keplerian disc is separated into the inner and the outer part by the radius where the angular-velocity gradient, and the radiative flux, vanish. Notice that for a fixed accretion flow the radiation flux from the inner discs exceeds again the radiation flow from the outer part of the discs, but not by many orders as in the inner discs orbiting KS naked singularities with $\omega > \omega_{ms}$.}
\end{center}
\end{figure} 

\begin{figure}[H]
\begin{center}
\begin{tabular}{c}
\includegraphics[scale=0.9]{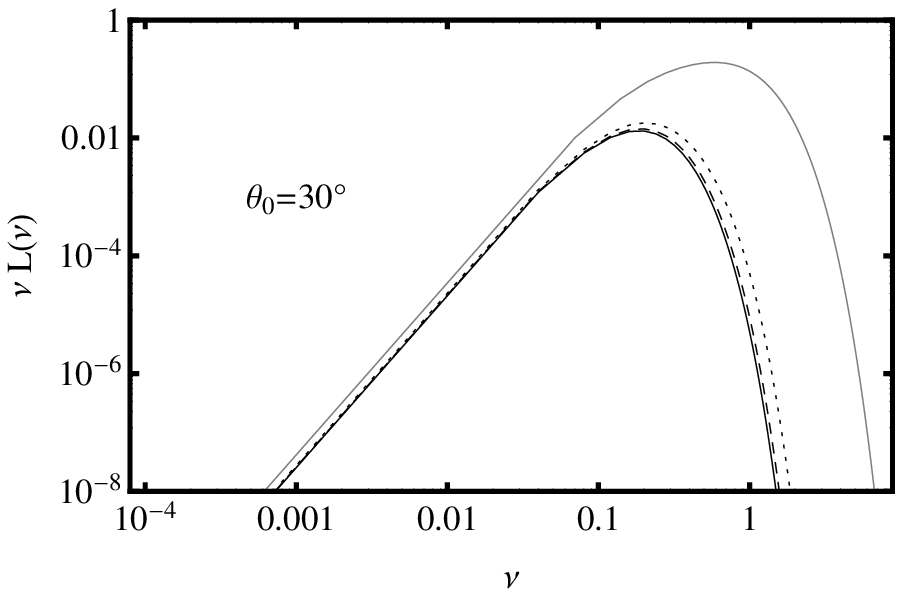}\\
\includegraphics[scale=0.9]{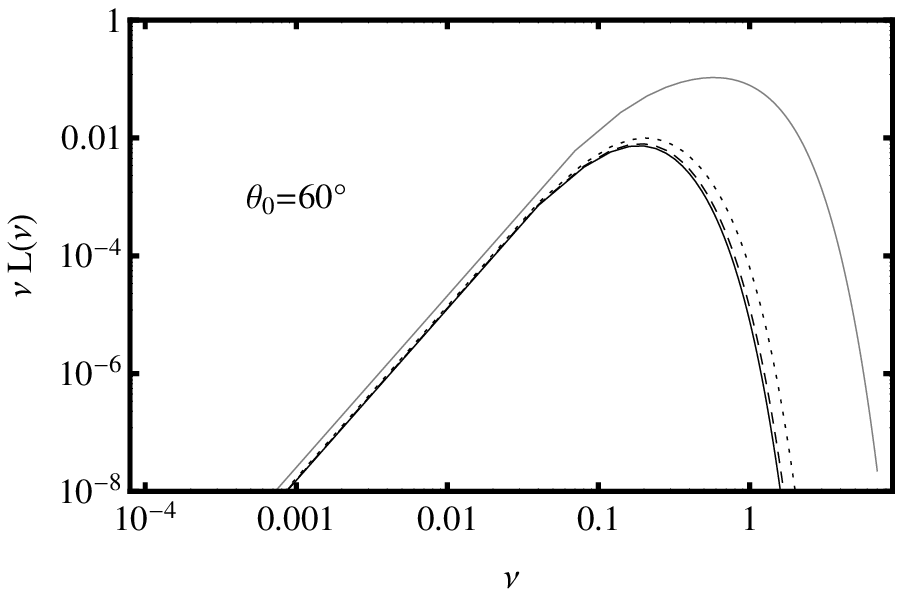}\\
\includegraphics[scale=0.9]{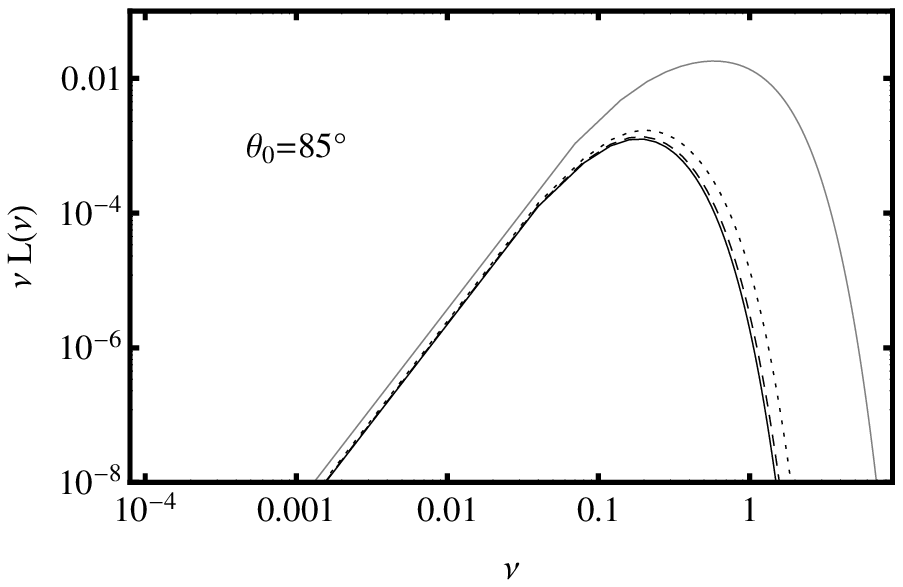}
\end{tabular}
\caption{The emission spectral continuum $\nu L(\nu)$, constructed for the outer (thermalized) Keplerian discs viewed under three representative values of the observer inclination angle ($30^\circ, 60^\circ, 85^\circ$, top to bottom) in the KS naked singularity and black hole spacetimes with four characteristic values of the Ho\v{r}ava parameter $\omega=0.1$ (gray), $0.3$ (black, dotted), $0.45$ (black, dashed), and $0.6$ (black).}
\end{center}
\end{figure}

\section{Profiled spectral lines}

We calculate the profile of a spectral (fluorescent) line, e.g., some of the Fe lines, generated by  the general relativistic effects -- the strong gravity focusing, the Doppler and gravitational frequency shift. We assume that the profiled spectral line is emitted from the innermost parts of the Keplerian disc due to some external irradiation that excites the fluorescent line radiation. For simplicity, we consider only the directly radiated photons that do not cross the equatorial plane where the radiating Keplerian disc is located. 

The observed flux is determined by $F_o = \int{I_o d\Pi}$ where $I_o$ is the observed intensity of the radiation flux and $d\Pi$ is the solid angle intended by the source on the observer sky. Since the emitted (e) and observed (o) radiation intensities are related by the Liouville theorem $\frac{I_o}{\nu^{3}_{o}} = \frac{I_e}{\nu^{3}_{e}} = const$, the observed flux can be given as $F_o =  \int I_e g^3 d\Pi$. For an observer at the distance $d_o$ from the source, the solid angle is expressed in terms of the coordinates $\alpha$ and $\beta$ on the observer plane by $d\Pi = \frac{1}{d^{2}_{o}} d\alpha d\beta$. The coordinates $\alpha$ and $\beta$ can be then expressed in terms of the motion constants $\lambda$ and $q$ or in terms of the radius $r_e$ of the orbiting matter and the redshift factor $g=\nu_{o}/\nu_{e}$ of the related photons. The corresponding Jacobian of the transformation $(\alpha,\beta) \rightarrow (r,g)$ then implies \cite{Sche-Stu:2013:JCAP:}
\begin{equation}
          \mathrm{d}\Pi=\frac{q}{d_o^2\sin\theta_o\sqrt{q-\lambda^2\cot^2\theta_o}}\left|\frac{\partial r_e}{\partial\lambda}\frac{\partial g}{\partial q}-\frac{\partial r_e}{\partial q}\frac{\partial g}{\partial \lambda}\right|^{-1}\mathrm{d} g\mathrm{d} r_e.
\end{equation}

In our simulations, we calculate the specific radiation flux taken by an observer at large distance from the disc using the formula
\begin{equation}
F_{\nu0}=\sum_{g}\sum_{r_{e}}g^{3}I_{e}(\nu_{e},r_{e})\Delta\Pi.
\end{equation}
The arriving photons are binned according to the particular frequency shift which is used to find the frequency of emitted photon $\nu_{e}=\nu_{o}/g$ for a particular observed frequency. We assume monochromatic emission; the emitter radiates locally at a narrow range of frequencies around $\nu_{0}$ with a narrow Gaussian shape of the local emission line. The amount of the emitted energy depends also on the radial extension of the radiating part of the Keplerian discs. The emissivity function is assumed in the form 
\begin{equation}
I_{e}(\nu_{e},r_{e})=I_{e0}\exp[-K(\nu-\nu_{0})^{2}]r^{-p}
\end{equation}
where parameter $p$ governs the power law determining the decrease of emissivity with increasing radius in the Keplerian disc. The profiled spectral line is constructed according to an actual form of the radiation intensity function $I_{e}$ of the radiating disc, namely, the polynomial radial dependence is assumed with factor $p=-2$.
 
\begin{figure}[H]
\begin{center}
\begin{tabular}{c}
\includegraphics[scale=0.9]{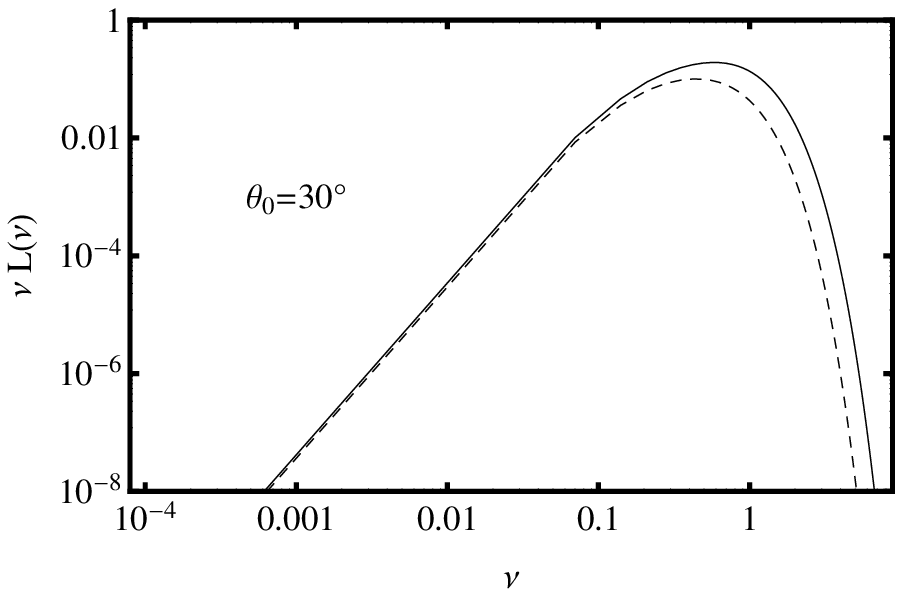}\\
\includegraphics[scale=0.9]{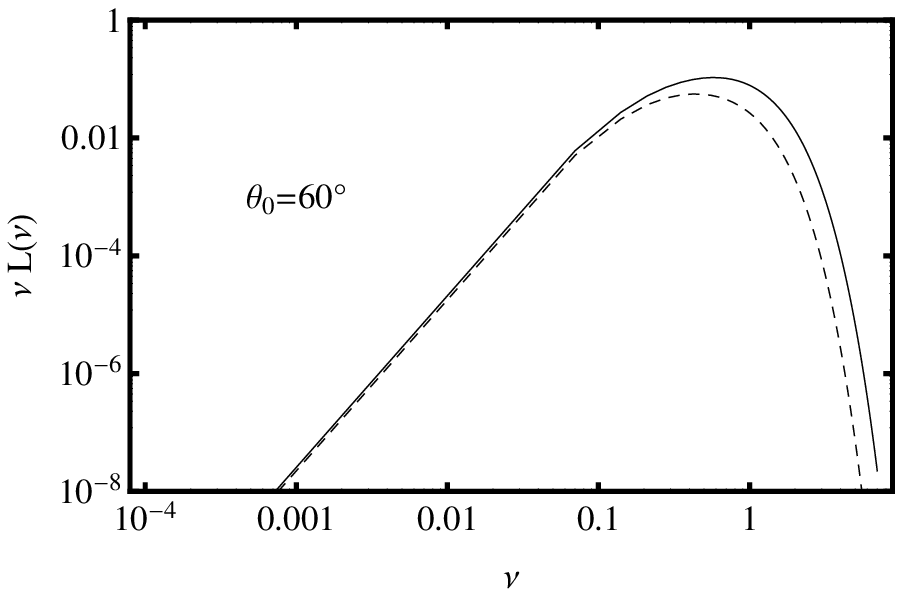}\\
\includegraphics[scale=0.9]{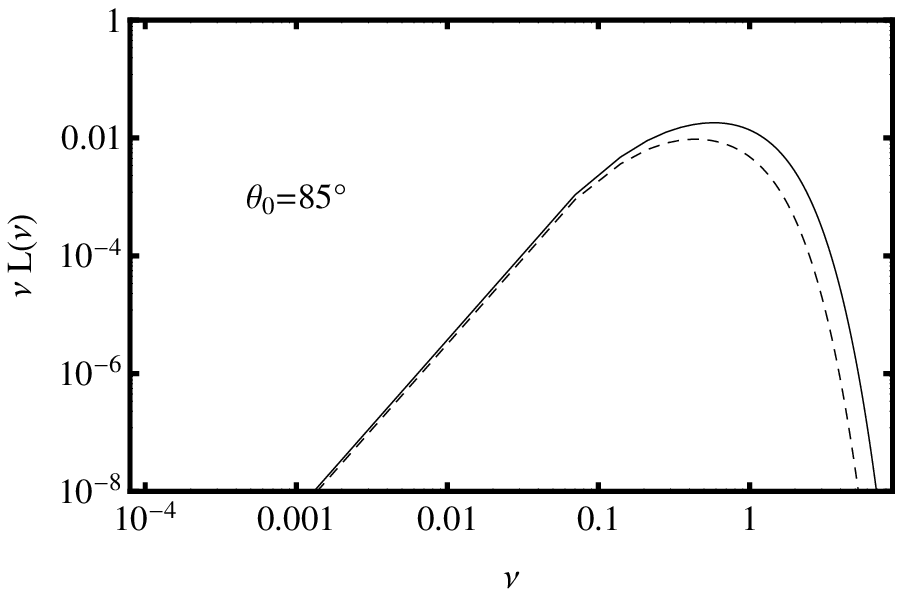}
\end{tabular}
\caption{The emission spectral continuum $\nu L(\nu)$ of the outer and the complete thermalized Keplerian discs orbiting a KS naked singularity with the parameter $\omega=0.1$, constructed for the three representative values of the observer inclination angle (top to bottom). The solid (dashed) line represents the spectral continuum generated in the Keplerian disc which spans from $r_{in}=r_{stat}(\omega)$ ($r_{in}=r_{\Omega max}(\omega)=(2/\omega)^{1/3}$).}
\end{center}
\end{figure}

\begin{figure}[H]
\begin{center}
\begin{tabular}{c}
\includegraphics[scale=0.8]{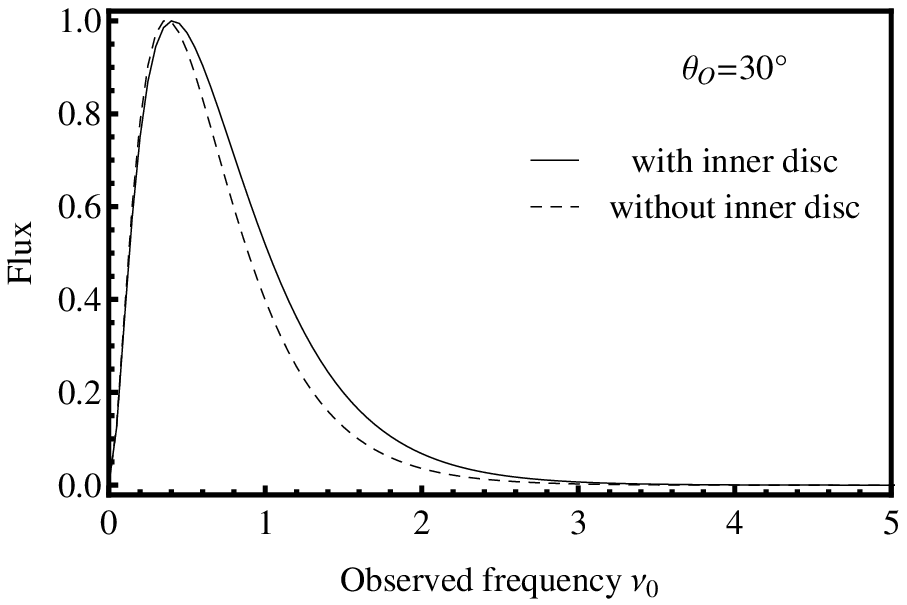}\\
\includegraphics[scale=0.8]{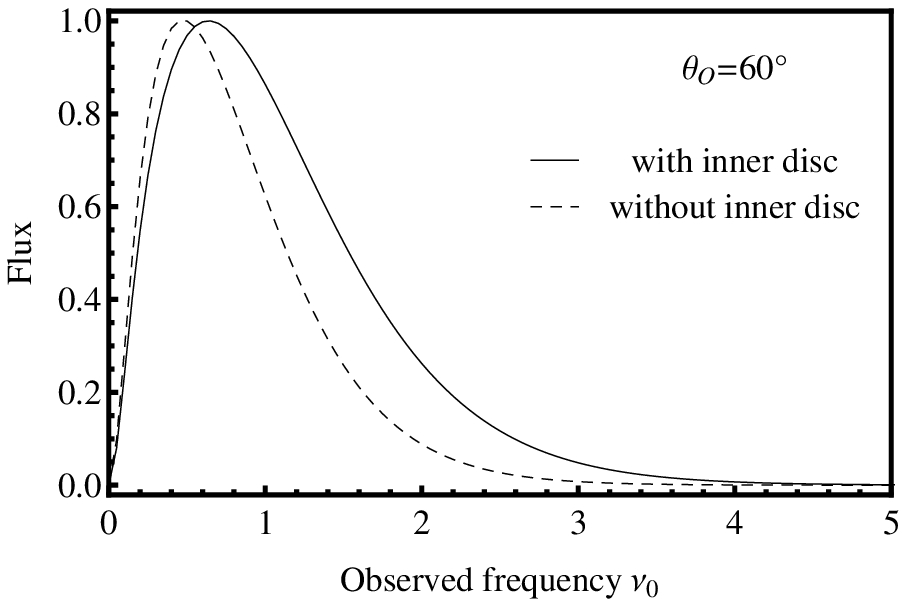}\\
\includegraphics[scale=0.8]{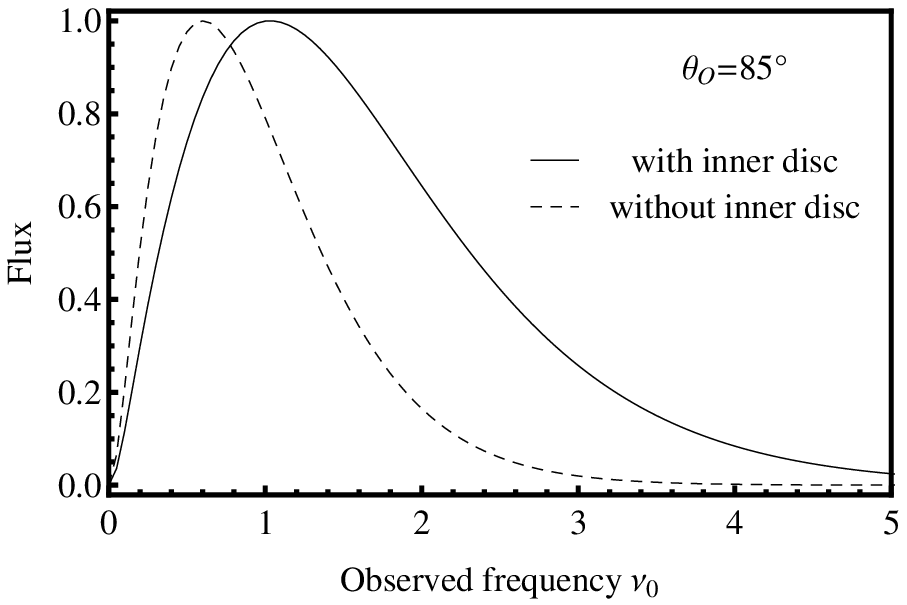}
\end{tabular}
\caption{Spectral distribution of the radiative flux $F(\nu)$ from the complete and the outer Keplerian discs orbiting a KS naked singularity characterized by the dimensionless parameter $\omega=0.1$. The Keplerian disc spans down to $r_{stat}$ (dashed line - complete disc), or to $r_{\Omega max}$ (solid line - outer disc). The spectral distribution is given for the three representative values of the observer inclination angle, $\theta_0=30^\circ$ (top), $60^\circ$ (middle) and $85^\circ$ (bottom).}
\end{center}
\end{figure}

We first give the profiled spectral lines for the outer Keplerian discs orbiting the KS naked singularity spacetimes with the three characteristic values of the parameter $\omega = 0.45, 0.3, 0.1$, and compare them to those constructed for the KS black hole spacetime with $\omega=0.6$, and the Schwarzschild spacetime ($\omega = \infty$) in Figure 21. The profiled spectral lines are constructed for three typical inclination angles to the distant observer $\theta_0=30^\circ$, $60^\circ$, $85^\circ$; we restrict the monochromatically radiating part of the outer Keplerian discs by the outer radius $r=20$ in all the considered cases, while the inner edge of the outer Keplerian disc is located at the ISCO in the spacetimes with $\omega > \omega_{ms}$, or at the radius of vanishing of the angular velocity gradient $r_{\Omega/max}$ in the case of the spacetimes with $\omega < \omega_{ms}$. In such a way, we give first the profiled lines generated in the Keplerian discs governed by the condition $\frac{d\Omega_{K}}{dr} < 0$ corresponding to the standard MRI viscosity mechanism.

Alternatively, we construct the (fluorescent) profiled spectral lines generated in addition to the outer Keplerian disc in the inner Keplerian disc at the region between $r_{stat}$ and $r_{E=E_{ISCO}}$ that we can assume in the KS naked singularity spacetimes with $1/2 > \omega > \omega_{ms}$ (Figure 22), and we add to the profiled spectral lines the contribution from the inner part of the Keplerian discs extending between the static radius $r_{stat}$ and $r_{\Omega/max}$ in the KS spacetimes with $\omega < \omega_{ms}$ (Figure 23). In the modelling of the fluorescent profiled spectral lines from the inner Keplerian discs in the KS naked singularity spacetimes, it is irrelevant to consider a viscosity mechanism related to the gradient of the angular velocity ($\frac{d\Omega_{K}}{dr} > 0$) in the inner discs, opposite to the angular velocity gradient in the outer disc that is related to the MRI viscosity mechanism, since the fluorescent spectral lines are generated by external irradiation of the orbiting matter. The fluorescent lines are not related to the viscosity mechanisms generating heat that can be radiated away, as in the case of the spectral continuum; it is quite enough to have a Keplerian disc governed by the radiatively inefficient processes, as gravitational radiation of the orbiting matter. 

Finally, we also test the role of the position of the outer edge of the inner Keplerian disc in modelling the profiled spectral lines, comparing the cases when the outer edge is at the radius corresponding to the energy $E=1$, to the cases when the outer edge of the inner Keplerian disc is located at the radius corresponding to the energy $E=E_{ISCO}$ (Figure 24). 

\begin{figure}[H]
\begin{center}
\begin{tabular}{c}
\includegraphics[scale=0.8]{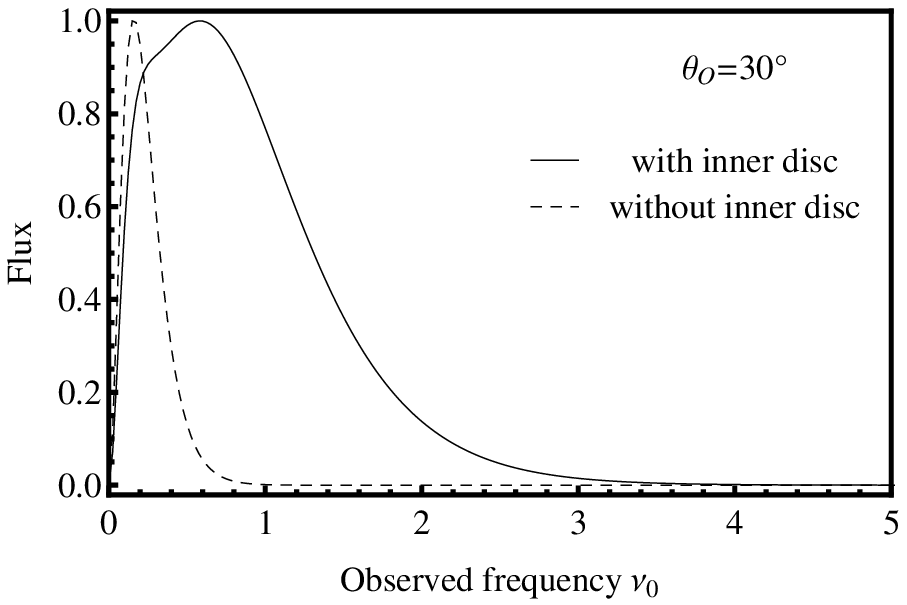}\\
\includegraphics[scale=0.8]{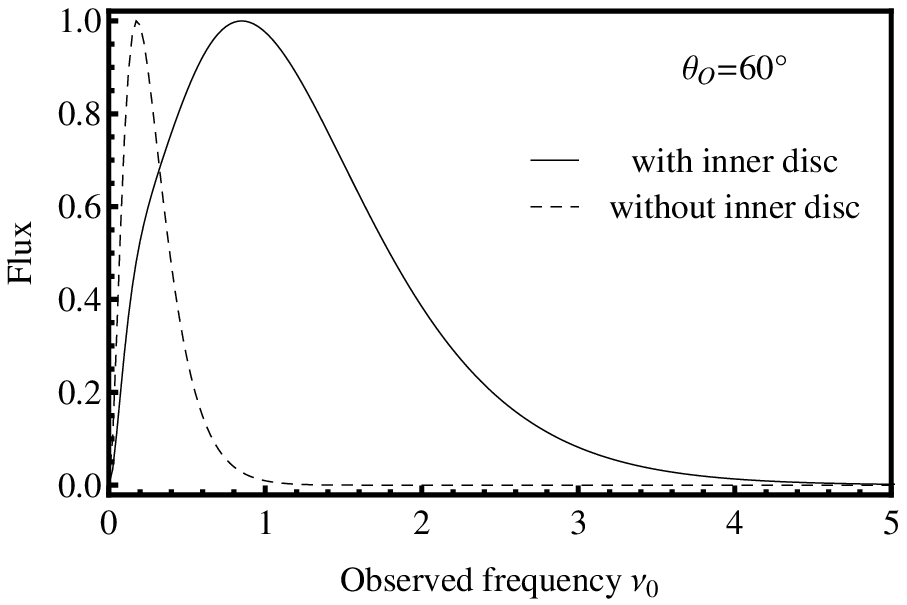}\\
\includegraphics[scale=0.8]{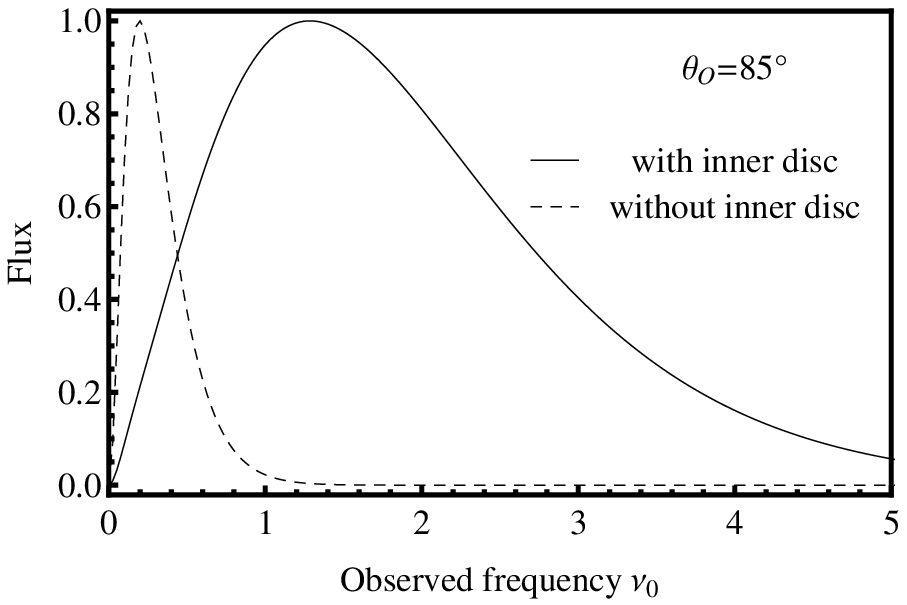}
\end{tabular}
\caption{Spectral distribution of the radiative flux $L(\nu)$ from the sole outer Keplerian disc and the combined inner and outer Keplerian discs orbiting a KS naked singularity characterized by the dimensionless parameter $\omega=0.282 > \omega_{ms}$. The dashed line corresponds to the radiation of the outer disc spanning between $r_{ISCO}$ and $r=20$. The full line corresponds to the radiation from the outer and inner discs - we assume that the inner disc spanning is between $r_{stat}$ and $r_K=r_K(E_{ISCO})$, and the outer disc is spanning between $r_{ISCO}$ and $r=20$. The three representative values of the observer inclination angle, $\theta_0=30^\circ$ (top), $60^\circ$ (middle) and $85^\circ$ (bottom), are considered.}
\end{center}
\end{figure}

\section{Discussion}

We have demonstrated that all the optical phenomena related to the Keplerian discs considered in the present paper are strongly dependent on the parameter $\omega$ of the KS naked singularity spacetimes and on the inclination angle of the discs relative to the distant observers. We can see that generally the $\omega$-dependence of the optical effects is demonstrated most profoundly for Keplerian discs viewed under large inclination angles as expected intuitively because of the strongest influence of the Doppler shift for large inclination angles. However, the optical phenomena give clear signatures of the naked singularity spacetimes for any inclination angle because of the inner Keplerian discs (or more complex internal structures) occuring in the innermost regions of the KS naked singularity spacetimes. \footnote{It should be stresses that some strong, qualitative phenomena related to the innermost accretion structures could be expected also in some other spherically symmetric naked singularity spacetimes due to the specific character of their spacetime structure allowing for existence of stable photon orbits and the existence of the inner Keplerian discs -- these can occur, e.g., in the spacetimes of the Reissner-Nordstrom type \cite{Stu-Hle:2002:ActaPhysSlov:,Kot-Stu-Tor:2008:CLAQG:,Stu-Kot:2009:GenRelGrav:,Pug-etal:2012:PHYSR4:}.}

We discuss separately properties of the three optical phenomena studied in our work for the Keplerian discs.

\subsection{Appearance of the Keplerian discs}

The phenomena related to the appearance of Keplerian discs can be observable for KS naked singularities located to the observer close enough in order to enable observations of the innermost parts of the Keplerian discs. The optical signatures of KS naked singularities can be relevant even for the outer Keplerian discs where the standard viscosity mechanism works -see Figs 8-10. The characteristics are of quantitative character in this case, the signature is given by the range of the frequency shift. The range decreases with the inclination angle decreasing due to the suppression of the role of the Doppler shift. Both the maximal and minimal frequency shift depend on the inclination angle. 

For small angles (Fig.8 for $\theta = 30^\circ$) the extremal shifts depend only slightly on the parameter $\omega$. The maximum of the shift is largest for the black hole and naked singularity spacetimes with large $\omega$ and slowly decrease with decreasing $\omega$, while the minimum of the  shift decreases for $\omega$ decreasing from the large values reaching a minimal value for a mediate value of the parameter $\omega$ and increases for decreasing small values of $\omega$.  

For mediate angles (Fig.9 for $\theta = 60^\circ$) the extremal shifts depend more steeply on the parameter $\omega$ in comparison with small inclination angles. The maximum is largest for the black hole spacetimes and slowly decreases for the naked singularity spacetimes with decreasing $\omega$, while the minimum decreases for $\omega$ decreasing from the large values and increases from a mediate value of $\omega$.  

For large angles (Fig.10 for $\theta = 85^\circ$) the extremal shifts depend more steeply on the parameter $\omega$ in comparison with mediate inclination angles. Now, the maximum is largest for the naked singularity spacetimes with a mediate value of $\omega$ and exceeds those of the black hole spacetimes. The maximum decreases more steeply for the naked singularity spacetimes with small $\omega$. The minimum again decreases for $\omega$ decreasing from the large values and increases from a mediate value of $\omega$, being largest for the black hole spacetimes.  

The inner Keplerian discs, if they exist, bring clear qualitative signatures of the presence of the naked singularities since no such structures can appear in the field of KS black holes, or standard black holes -- see Figs 11-13. The related quantitative features of their appearance, namely the frequency shift range, give the possibility to restrict the Ho\v{r}ava parameter $\omega$ of the KS naked singularity spacetimes. The constant frequency shift of radiation coming from the inner edge of the inner Keplerian discs (that occurs because matter on the static radius is stationary relative to distant observers), is one of the most relevant signatures of the KS naked singularities, or the other spherically symmetric naked singularities, e.g., of the Reissner-Nordstrom naked singularities. 

In Figure 11, it is clearly demonstrated that for the KS naked singularity spacetimes with $\omega > \omega_{ms}$, the frequency shift maximum of the inner and outer Keplerian discs is the same or slightly increased in comparison to those of the outer discs, while the frequency shift minimum is substantially lowered. The differences in both the blue and red end of the frequency range increase with increasing inclination angle and with increasing parameter $\omega$. The minimal frequency shift is lowered by one order or more in the inner discs as compared to the outer discs in the KS naked singularity spacetimes with $\omega > \omega_{ph}$, but it is of the same order in the KS spacetimes with $\omega_{ph} > \omega > \omega_{ms}$. We can see that in the KS spacetimes with $1/2 > \omega > \omega_{ms}$, the gravitational redshift effect in the innermost parts of the Keplerian discs  dominates the frequency shift related to the Doppler effect, and its role increases with increasing Ho\v{r}ava parameter $\omega$.

A dramatically different behaviour is observed in the KS naked singularity spacetimes with $\omega < \omega_{ms}$ -- see Figs 12 and 13. In this case, the frequency range of radiation coming from the total Keplerian discs and the outer Keplerian discs remains the same. On the other hand, with increasing of the inclination angle of the distant observers, the frequency minimum decreases, while the frequency maximum increases, as in the the case of Keplerian discs orbiting the KS naked singularity spacetimes with $1/2 > \omega > \omega_{ms}$. For the KS naked singularities with $\omega < \omega_{ms}$, the frequency range increases with increasing $\omega$ for all inclination angles.

\subsection{Spectral continuum}

We demonstrate in Figure 15 the radial profiles of the radiation flux of both the outer and inner Keplerian discs for one representative value of Ho\v{r}ava parameter $\omega = 0.45$, representing the KS spacetimes with two circular photon orbits and no critical point of the angular velocity profile (case $1/2 > \omega > \omega_{ph}$), and for two values of $\omega = 0.35, 0.3$, representing the KS spacetimes with the critical point of the angular velocity profile at the inner Keplerian disc (case $\omega_{ph} > \omega > \omega_{ms}$). The radiation flux of the outer Keplerian discs vanishes at the ISCO, representing the inner edge of the outer discs. The radiation flux from the inner Keplerian discs vanishes always at the static radius $r_{stat}$.

For the inner Keplerian discs in the KS spacetimes with $\omega_{ph} > \omega > \omega_{ms}$, the radiation flux vanishes also at the critical radius $r_{\Omega/max}$. Magnitude of the radiation flux from the inner disc at its maximum is by four orders larger in comparison to the flux maximum observed in the outer Keplerian disc, if the accretion flow is assumed the same. The inner Keplerian discs in the KS naked singularity spacetimes with $\omega > \omega_{ph}$, if thermalized due to some efficient viscosity mechanism, would be extremely hot and efficiently radiating due to the extremely steep decreasing of the radial profiles of energy, angular momentum and angular velocity; especially in the case of the near-extreme KS naked singularity spacetimes the gradients of these radial profiles are extremely large and the inner Keplerian discs would be extremely strongly radiating. If the accretion flow will be the same as those in the outer Keplerian discs, the radiation flux maximum from the inner discs exceeds very strongly, by many orders, the radiation flux maximum of the outer discs. The radiation flux maximum can diverge as the outer edge of the inner Keplerian disc approaches the stable photon circular orbit. 

Our results indicate that creation of thermalized inner Keplerian discs is astrophysically extremely unrealistic -- both because of the unrealistic conditions for creation of matter structures corresponding to the inner Keplerian discs, and the impossibility of existence of extremely efficient viscosity mechanism for the heating of the inner discs. In the KS spacetimes with the Ho\v{r}ava parameter $1/2 > \omega > \omega_{ms}$, only the outer thermalized Keplerian discs are physically realistic. On the other hand, existence of cold inner Keplerian discs with range limited by the outer edge at the radius corresponding to the circular geodesics with energy $E=1$ (or $E=E_{ISCO}$) cannot be excluded. However, the most probable is occurence of the thick toroidal accretion structures, or even spherical structures mixing matter and trapped radiation and orbiting around the static radius in the KS naked singularity spacetimes with $1/2 > \omega > \omega_{ms}$.

The radiation flux of the total Keplerian discs orbiting in the KS naked singularity spacetimes with $\omega < \omega_{ms}$ is illustrated in Figure 16 for two characteristic values of the parameter $\omega$. The radiation flux vanishes again at the critical radius where $\frac{d\Omega_{K}}{dr} = 0$ and at the static radius. Our results are normalized to the accretion flow, and we observe that the radiation flux maximum increases with increasing parameter $\omega$, reflecting thus the increasing  efficiency of the Keplerian accretion discs. The radiation flux from the inner part of the complete Keplerian discs extending below the critical radius (at $r < r_{\Omega/max}$) exceeds the radiation flux from the outer part of the Keplerian discs. However, it is not clear, if an efficient viscosity mechanism can work at the inner region, and we know from the discussion in the previous section that there are some difficulties in transporting the accreting matter accross the critical radius where no viscosity mechanism works.  

The temperature profiles of the outer Keplerian discs in the KS spacetimes with $1/2 > \omega > \omega_{ms}$ are compared to the temperature profile of a KS black hole and the temperature profile of the complete Keplerian disc around a KS naked singularity spacetime with $\omega < \omega_{ms}$ in Figure 14. We see immediately that the differences in the temperature profiles increase with decreasing $\omega$. The temperature increases due to the increasing efficiency of the Keplerian accretion in the outer discs. 

We demonstrate in Figure 17 a directly observable quantity, spectral continuum $\nu L(\nu)$, constructed for the outer Keplerian discs orbiting KS naked singularity spacetimes with typical values of the parameter $\omega$ and compare them to the one corresponding to the Keplerian disc around a KS black hole. These outer Keplerian accretion discs can be physically quite realistic, it is thus important that we can observe clear quantitative differences that could be observationally detected, especially for the KS naked singularity spacetimes with relatively low parameter $\omega$. The maximal magnitude of the spectral-continuum frequency profile depends on the inclination angle of the distant observers, being decreasing with increasing inclination. However, the frequency corresponding to this maximum is almost independent of the inclination angle. 

For the KS naked singularities with $\omega < \omega_{ms}$, we compare the spectral continuum of the complete Keplerian discs and the spectral profiles corresponding to the outer parts of the discs where $\frac{d\Omega_{K}}{dr} < 0$. It is demonstrated in Figure 18 that the spectral continuum is clearly different for all the inclination angles of the observers, in both the extension of the spectral profile in frequency, and the location and magnitude of the spectral-profile maximum.  

Finally, we present for completeness the radiation flux in dependence on the observed frequency of the emitted radiation ($F(\nu)$), and the inclination angle to the distant observers, for the complete Keplerian discs and their outer parts orbiting in the field of the KS spacetime with $\omega = 0.1 < \omega_{ms}$ in Figure 19. The flux dependences on the observed frequency, radiated from the inner and outer Keplerian discs, and the outer discs, orbiting in the field of the KS spacetime with $\omega = 0.282$ are compared in Figure 20; the inner disc has its outer edge at the radius corresponding to the energy at ISCO. In both cases we see strong differences in the frequency distribution -- the discs radiating also in their inner parts demonstrate clear shift of the flux maximum to the larger values of the observed frequency. 

\begin{figure}[H]
\begin{center}
\begin{tabular}{c}
\includegraphics[scale=0.8]{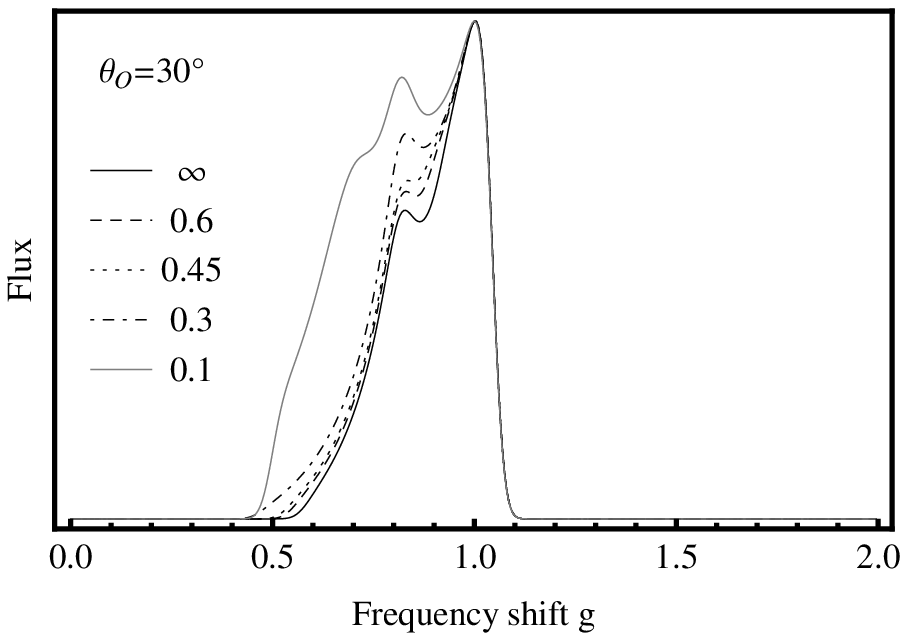}\\
\includegraphics[scale=0.8]{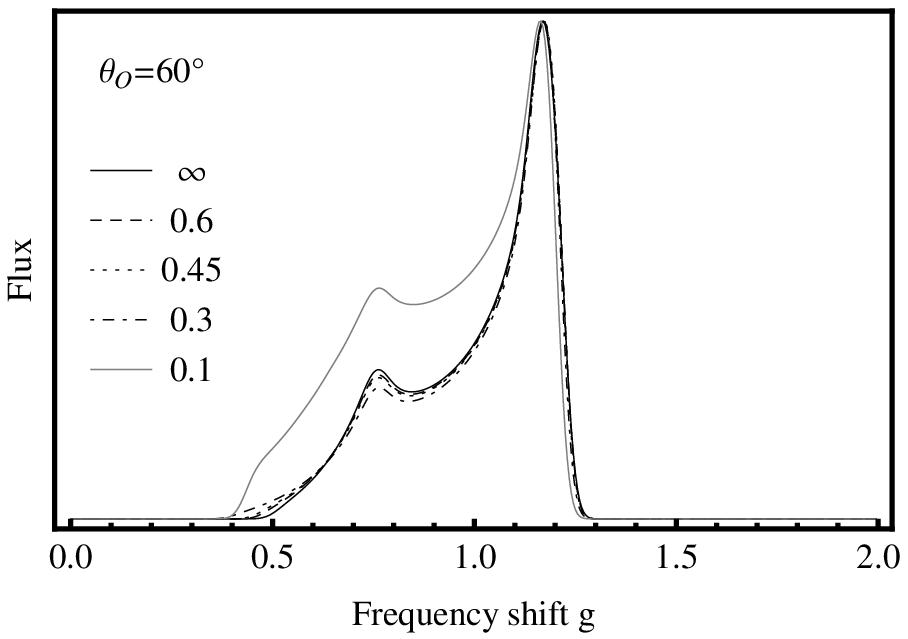}\\
\includegraphics[scale=0.8]{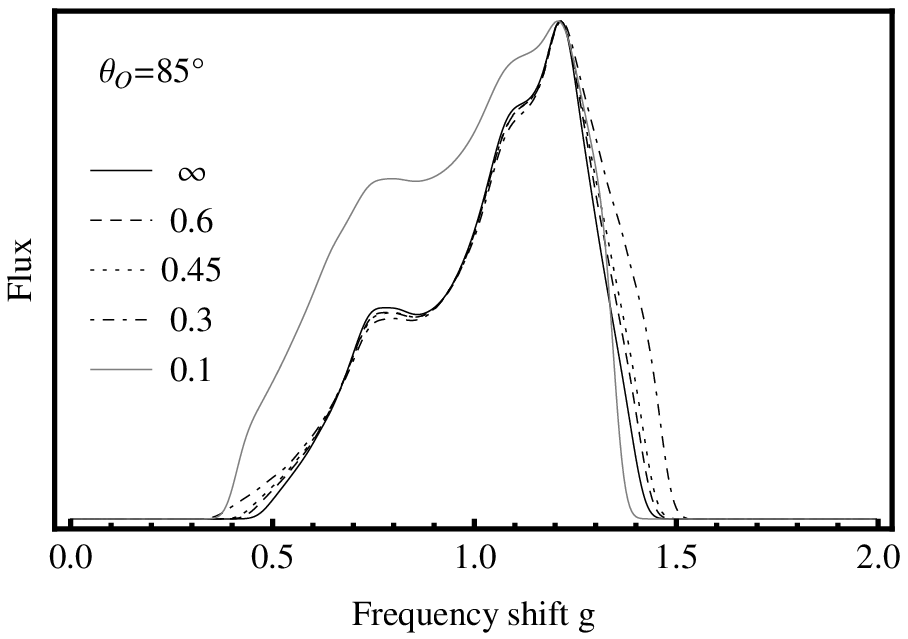}
\end{tabular}
\par\end{center}
\caption{The profiled spectral lines generated by monochromatic radiation from the outer Keplerin discs orbiting KS black hole or naked singularity spacetimes with five representative values of the Ho\v{r}ava parameter $\omega=\infty$ (black), $0.6$ (dashed), $0.45$ (dotted), $0.3$ (dot-dashed), and $0.1$ (gray). The profiled lines are constructed for the three representative values of the observer inclination angle, $\theta_{o}=30^/circ$(top), $60^\circ$ (middle) and $85^\circ$ (bottom).}
\end{figure}



\begin{figure}[H]
\begin{center}
\begin{tabular}{cc}
\includegraphics[scale=0.6]{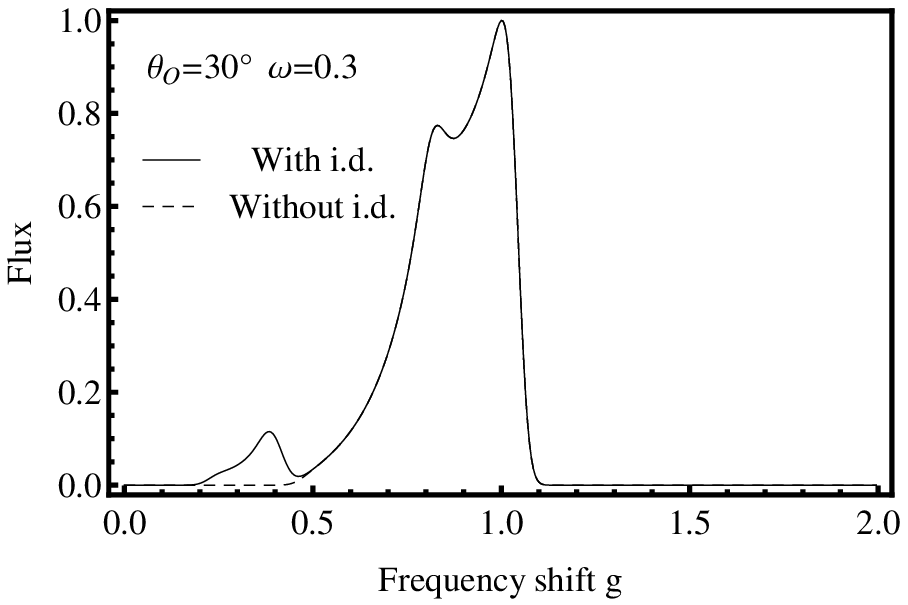}&\includegraphics[scale=0.6]{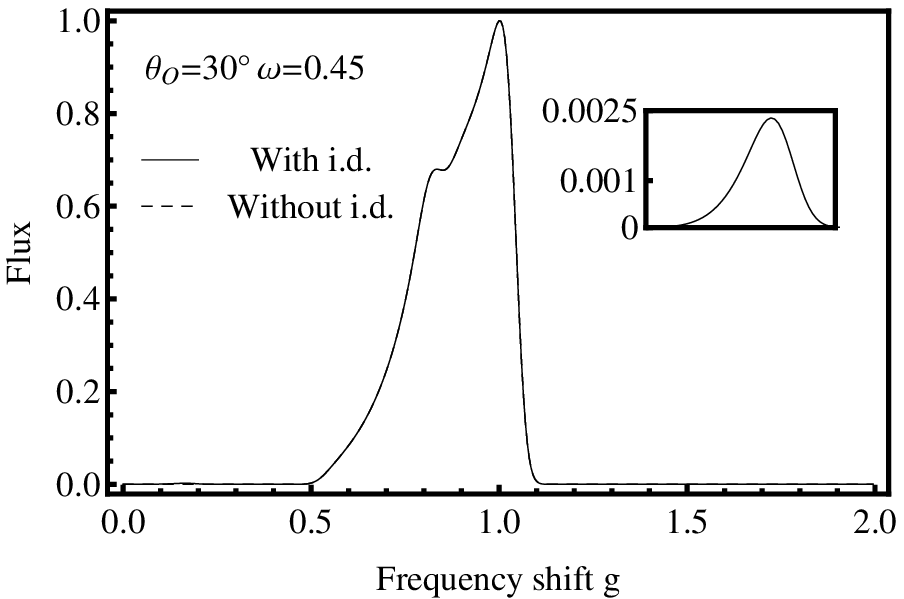}\\
\includegraphics[scale=0.6]{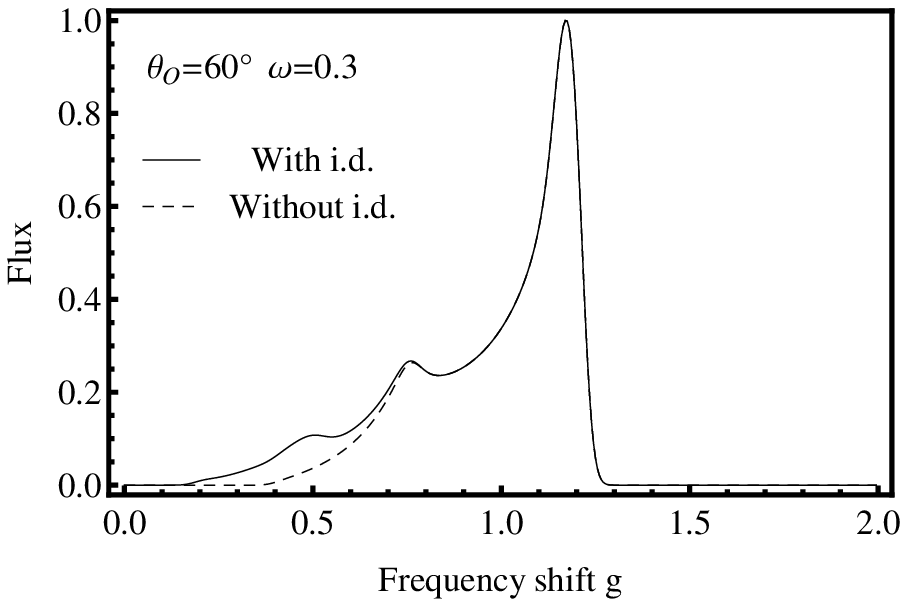}&\includegraphics[scale=0.6]{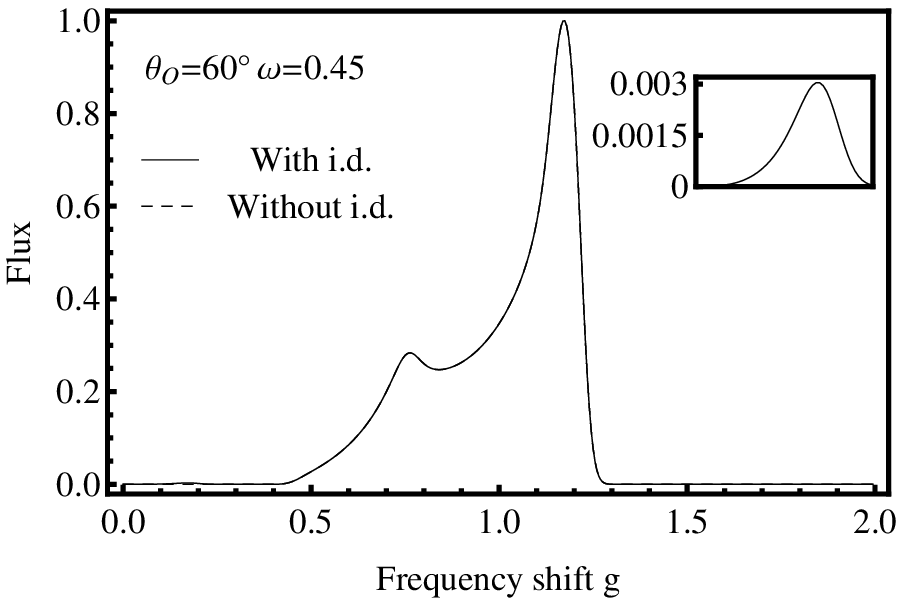}\\
\includegraphics[scale=0.6]{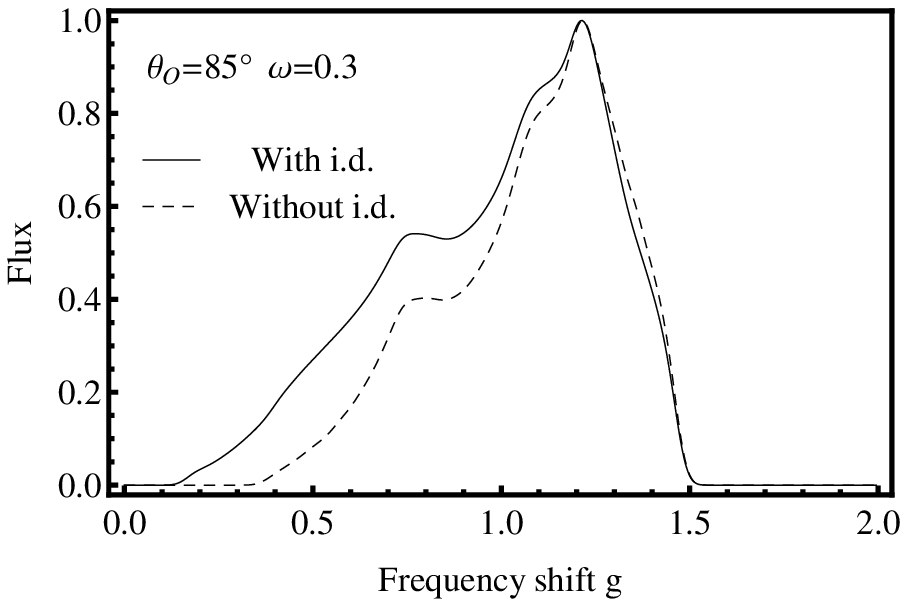}&\includegraphics[scale=0.6]{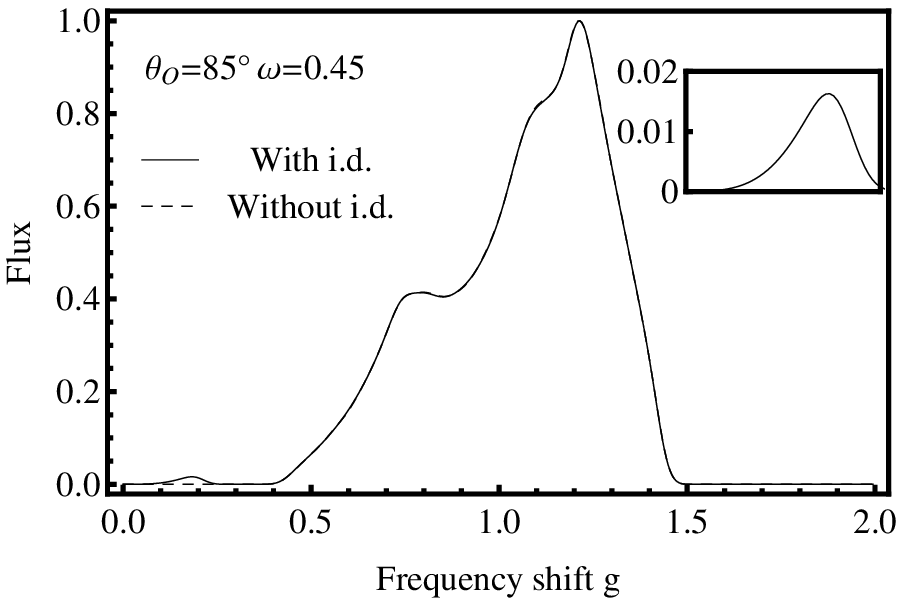}
\end{tabular}
\end{center}
\caption{The profiled spectral lines generated by monochromatic radiation from the Keplerin discs orbiting KS naked singularity spacetimes with two representative values of the Ho\v{r}ava parameter $\omega = 0.45, 0.3$ . The profiled lines are constructed for the three representative values of the observer inclination angle, $\theta_{o}=30^/circ$(top), $60^\circ$ (middle) and $85^\circ$ (bottom). The dashed curve represents the profiled spectral line generated by both the inner and the outer Keplerian discs, while the solid line represents the profiled spectral line generated only by the outer Keplerian disc.}
\end{figure}

\begin{figure}[H]
\begin{center}
\begin{tabular}{cc}
\includegraphics[scale=0.6]{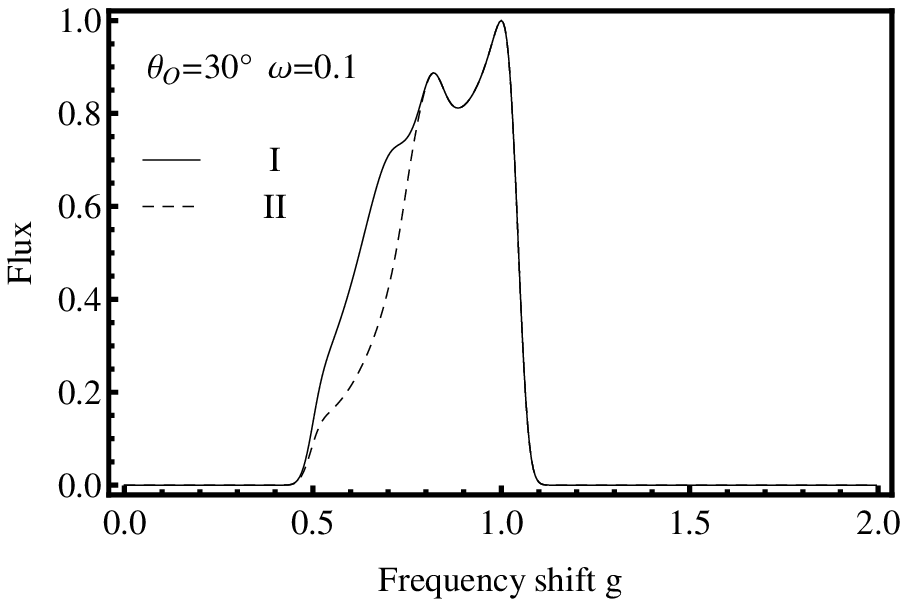}&\includegraphics[scale=0.6]{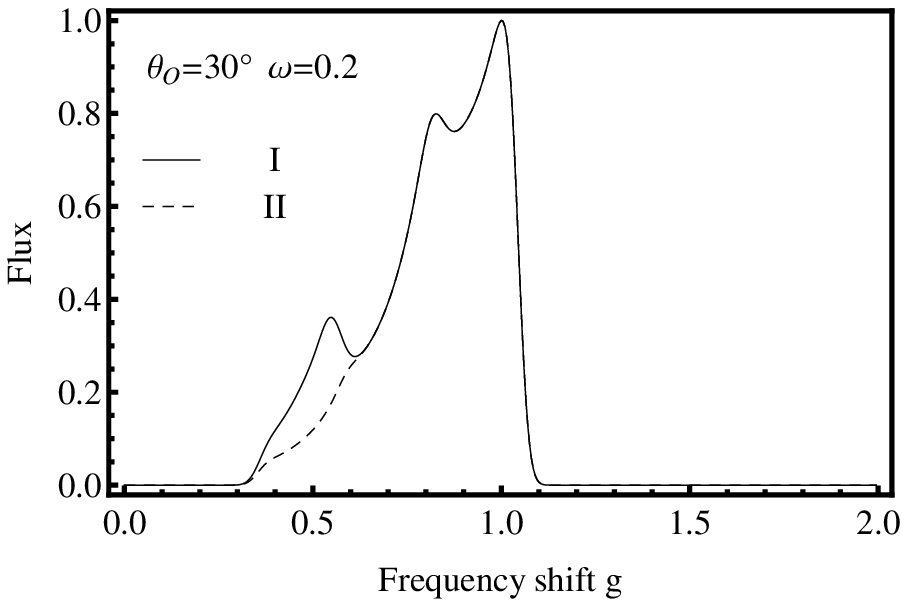}\\
\includegraphics[scale=0.6]{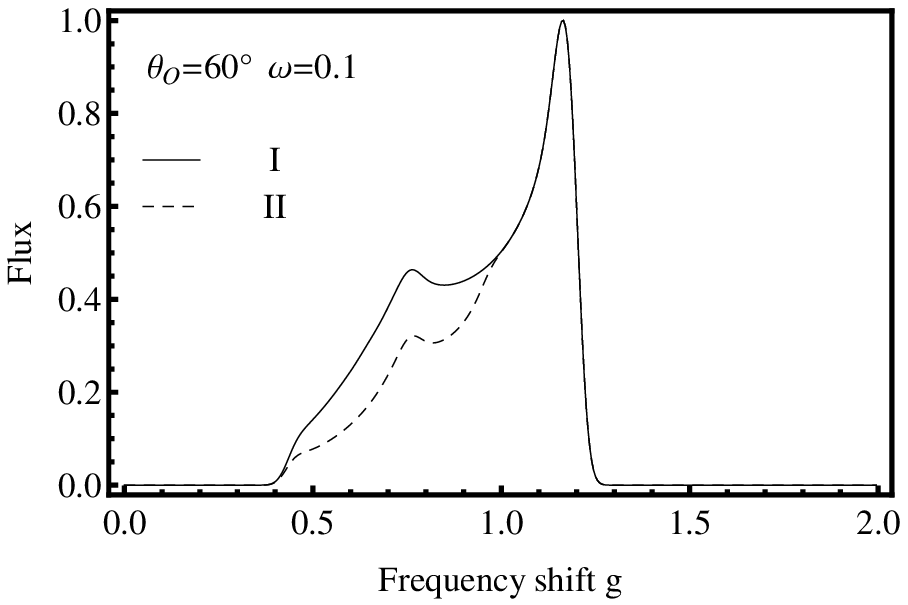}&\includegraphics[scale=0.6]{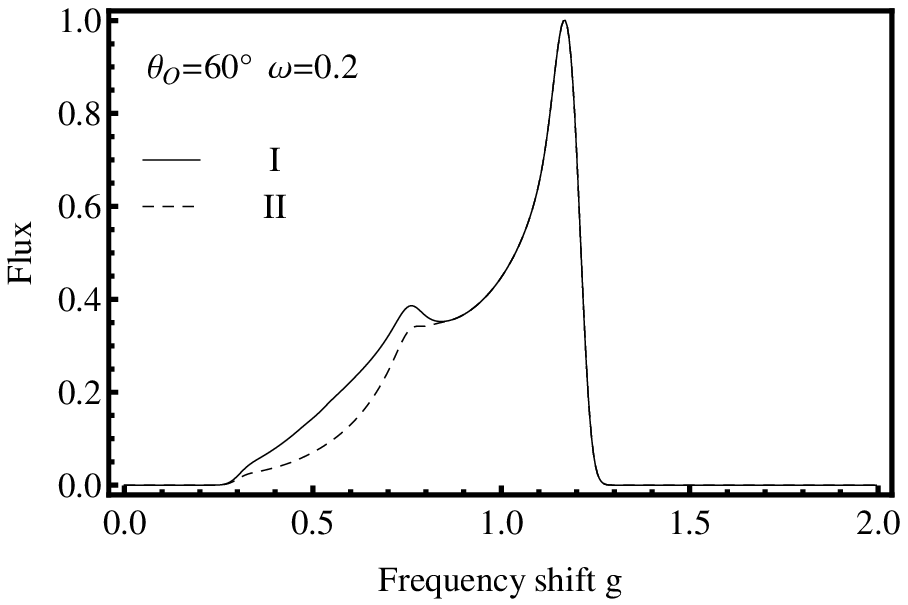}\\
\includegraphics[scale=0.6]{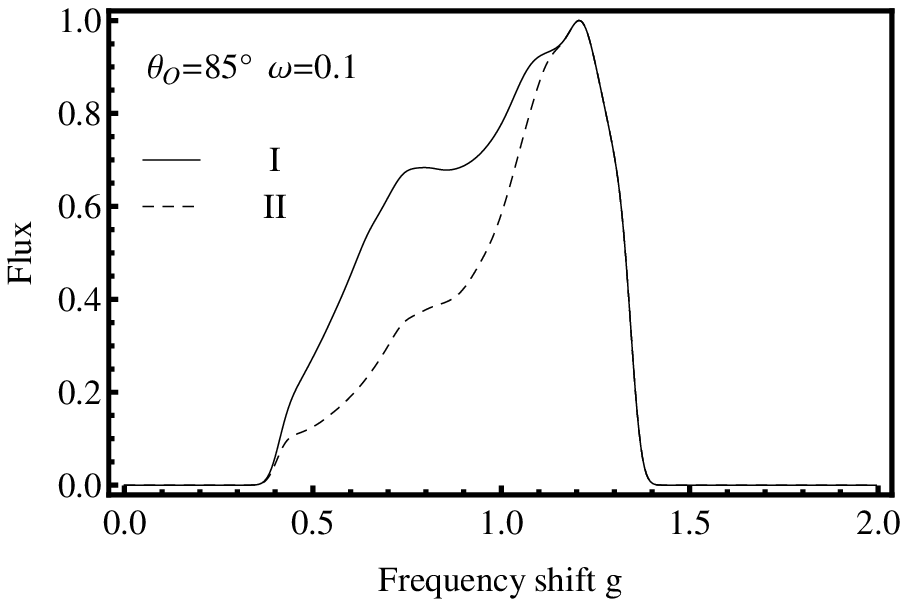}&\includegraphics[scale=0.6]{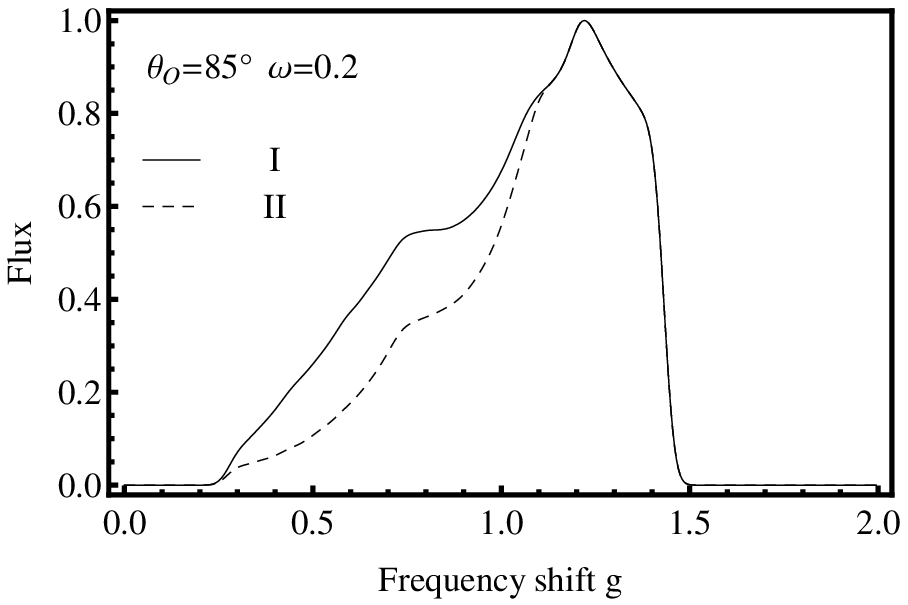}
\end{tabular}
\end{center}
\caption{The profiled spectral lines generated by monochromatic radiation from the Keplerin discs orbiting KS naked singularity spacetimes with two representative values of the Ho\v{r}ava parameter $\omega=0.1$ (left) and $\omega=0.2$ (right) . The profiled lines are constructed for the three representative values of the observer inclination angle, $\theta_{o}=30^/circ$(top), $60^\circ$ (middle) and $85^\circ$ (bottom). The dashed curve represents the profiled spectral line generated by both the inner and the outer Keplerian discs, while the solid line represents the profiled spectral line generated only by the outer Keplerian disc. In all the cases, the dashed curve represents the profiled line generated by the outer Keplerian disc extending to the inner edge at $r_{in}=r_{\Omega max}(\omega)$, while the solid curve represents the profiled line generated by the complete Keplerian disc extending down to the inner radius $r_{in}=r_{stat}(\omega)$.}
\end{figure}

\begin{figure}[H]
\begin{center}
\begin{tabular}{cc}
\includegraphics[scale=0.6]{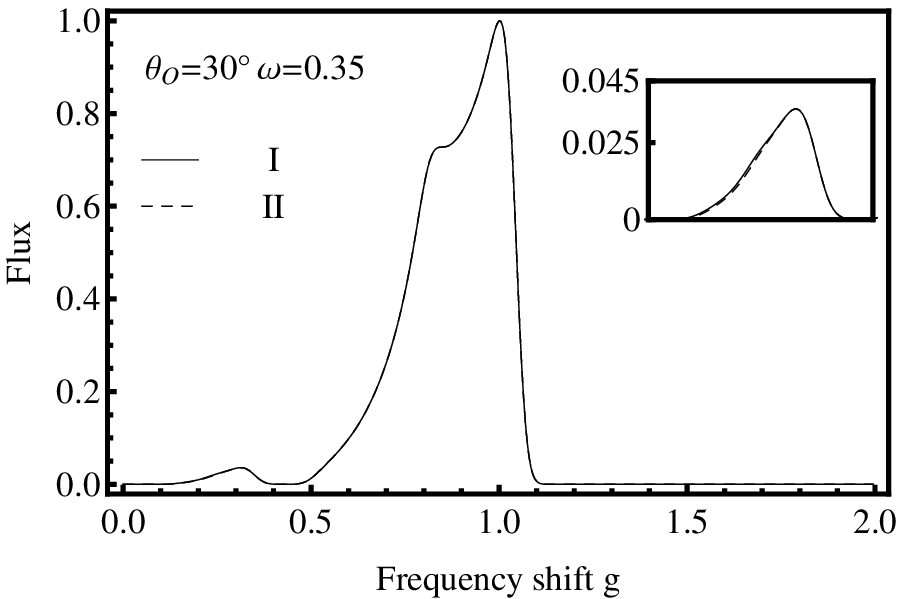}&\includegraphics[scale=0.6]{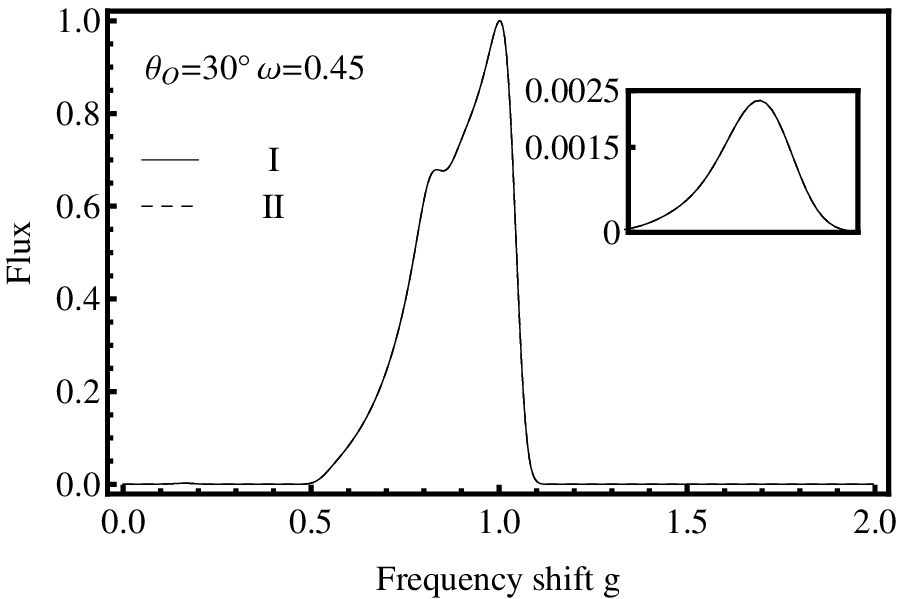}\\
\includegraphics[scale=0.6]{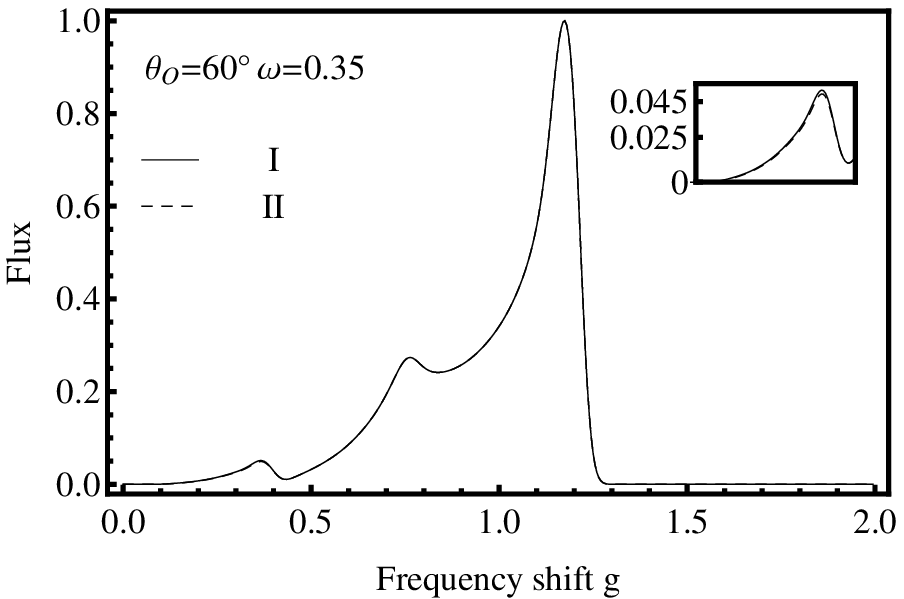}&\includegraphics[scale=0.6]{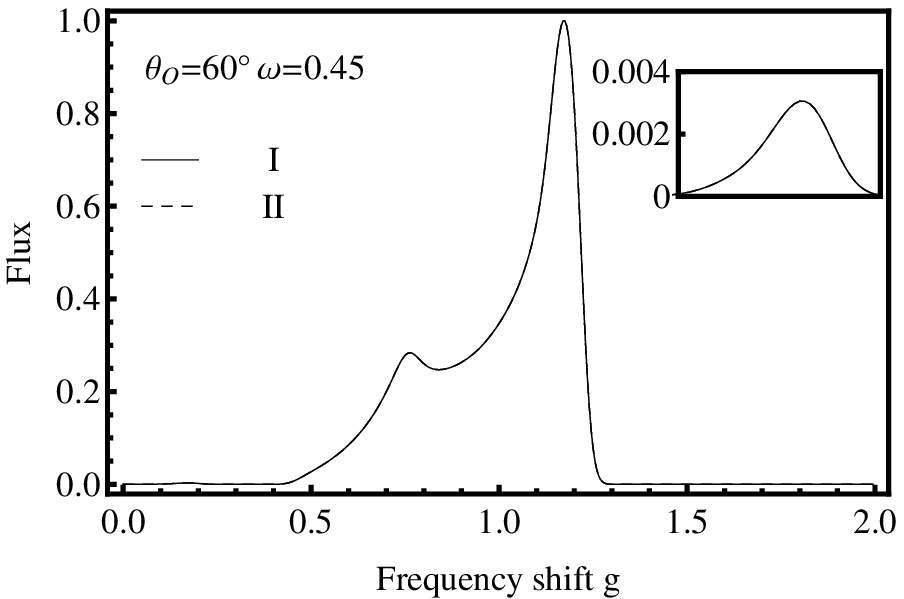}\\
\includegraphics[scale=0.6]{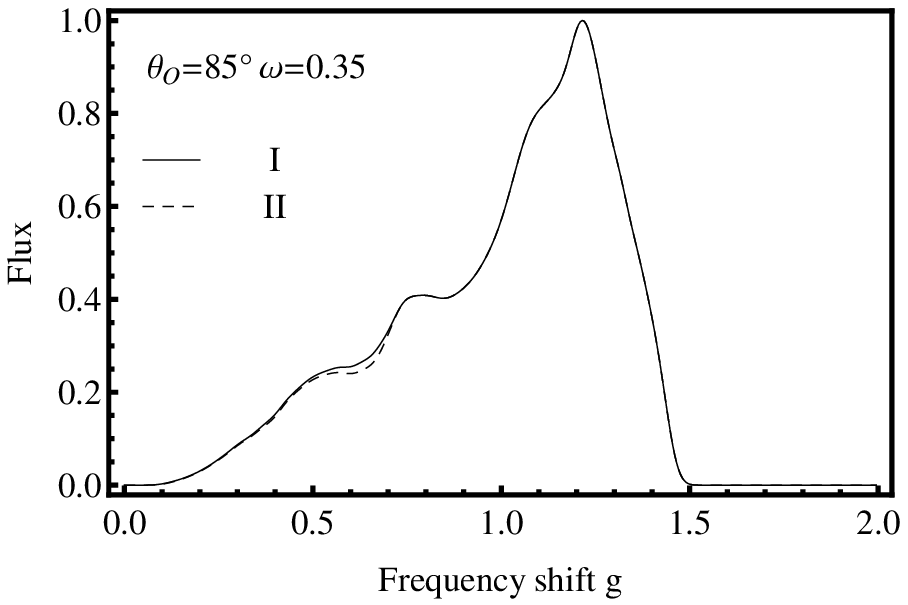}&\includegraphics[scale=0.6]{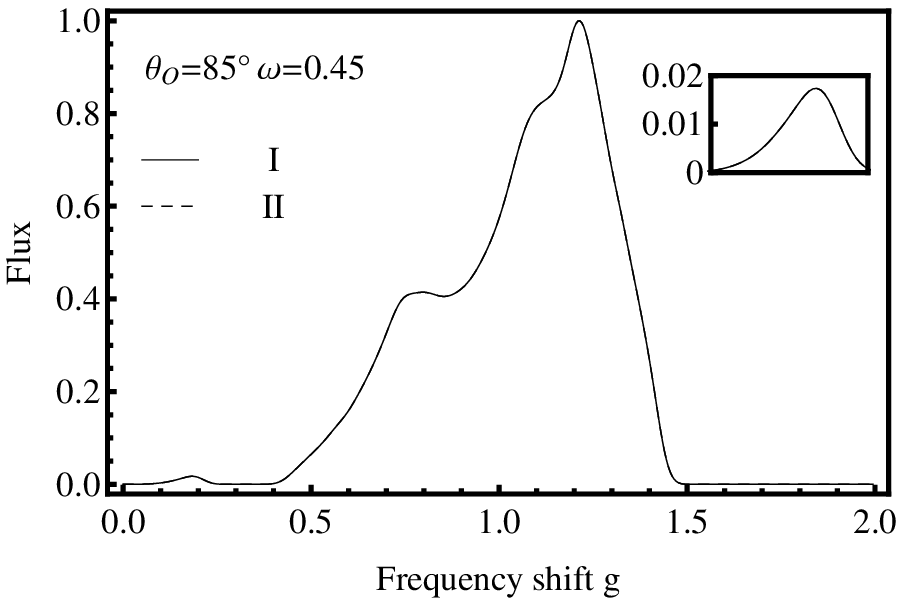}
\end{tabular}
\end{center}
\caption{The profiled spectral lines generated by monochromatic radiation from the inner and the outer Keplerin discs orbiting KS naked singularity spacetimes with two representative values of the Ho\v{r}ava parameter $\omega = 0.45, 0.3$ . The profiled lines are constructed for the three representative values of the observer inclination angle, $\theta_{o}=30^/circ$(top), $60^\circ$ (middle) and $85^\circ$ (bottom). The inner Keplerian disc is considered in two variants related to the location of its outer edge. The dashed curve represents the case of $r_o=r_{K}(E_{K}=1)$, while the solid curve represents the case of $r_o=r_{K}(E=E_{K}(r_{ISCO}))$. The values of the radius at the outer edge of the inner disc: for $\omega=0.35$, there is $r_o=1.762\,(E_{K}=1)$, $r_o=1.70492\,(E_{K}=E_{ISCO})$, and for $\omega=0.45$, there is $r_o=1.15575\,(E_{K}=1)$, $r_o=1.1549\,(E_{K}=E_{ISCO})$.}
\end{figure}

\subsection{Profiled spectral lines}

We consider the fluorescent spectral lines, generated by the Keplerian discs and profiled by the effects of the KS naked singularities; we compare them to those generated in the field of the KS black holes. Since the fluorescent lines are excited by an external irradiation of the Keplerian discs, it is not necessary to consider in this case the thermalization of the Keplerian discs -- also cold Keplerian discs can emit the fluorescent radiation. For this reason, we study in detail the role of the inner Keplerian discs in shaping of the fluorescent lines. The outer edge of the outer Keplerian discs is assumed at $r=20$, the inner edge at $r_{ISCO}$. The inner Keplerian discs have their inner edge at $r_{stat}$ and the outer edge at $r=r_{K}(E_{ISCO})$. In the KS spacetimes with $\omega < \omega_{ms}$, we consider the complete Keplerian discs in the range $r_{stat} - r=20$, and, alternatively, in the range $r_{\Omega/max} - r=20$. The profiled spectral lines are constructed in all the considered cases for three representative values of the observer inclination angle $\theta_{O}=\{30^\circ,\,60^\circ,\,85^\circ\}$. 

First, we have constructed the profiled spectral lines for the outer Keplerian discs orbiting KS naked singularities with three representative values of $\omega=\{0.1,\:0.35,\:0.45\}$ in Figure 21; each of them represents one particular case relative to the values of $\omega_{ms}$, $\omega_{ph}$ and $\omega_{h}$, discussed above. In the case of $\omega=0.1$, the complete Keplerian disc is assumed.  These profiled lines are compared to those constructed for the black hole spacetimes with $\omega=0.6$ and $\omega = \infty$ corresponding to the Schwarzschild black holes. The profiled spectral lines generated in spacetimes with different parameter $\omega$ can be clearly distinguished by the shape of the profiled line and its frequency extension. The shape differences are clearly demonstrated especially for small inclination angles, while the extension of the profile is most clearly reflected for large inclination angles. The differences increase with decreasing parameter $\omega$. 

The signatures of the inner Keplerian discs in the profiled spectral lines generated in the field of the KS naked singularities with $\omega > \omega_{ms}$ are demonstrated in Figure 22 for $\omega = 0.3, 0.45$. In both cases, the signature is represented by a small hump in the red edge of profiled spectral lines. However, the hump is unobservable for near-extreme KS naked singularities. 

For the KS naked singularity spacetimes with $\omega < \omega_{ms}$, the profiled spectral lines generated by the complete Keplerian discs and their outer parts (at $r > r_{\Omega/max}$) are compared in Figure 23 for two characteristic values of $\omega = 0.1, 0.2$. It is clearly demonstrated that the complete Keplerian disc gives a clear signature in the profile lines for all the inclination angles and both the values of $\omega$.

Finally, we have tested the role of the shift of the outer edge of the inner Keplerian discs on the profiled lines generated in the field of KS naked singularities with parameter $\omega = 0.35, 0.45 > \omega_{ms}$. It is demonstrated in Figure 24 that the shift of the outer edge from $r=r_{K}(E_{ISCO})$ to $r=r_{K}(E=1)$ gives quite negligible changes in the shape and extension of the profiled lines. 

We can summarize that the profiled spectral lines can give clear signatures of the KS naked singularity spacetimes. We can even obtain an interesting information related to the extension of the fluorescenting Keplerian discs in the shape of the fluorescent spectral lines. 

\section{Conclusions}

We have studied the optical phenomena related to the Keplerian accretion discs orbiting the KS naked singularity spacetimes. We have considered three optical phenomena: a) the appearance of the Keplerian discs demonstrating the influence of the strong gravity effects on the shape of the discs and the frequency shift of the radiation generated by them, b) the radiation flux of thermaly radiating (Page-Thorne) Keplerian discs, their temperature profile, and the corresponding spectral continuum observed at large distances, c) the spectral (fluorescent) lines generated by an outer irradiation of the Keplerian discs and profiled by the strong gravity effects of the spacetime. 

The KS naked singularity spacetimes are separated into three distinct classes according to the properties of the geodetical (Keplerian) circular orbits in dependence on the dimensionless Ho\v{r}ava parameter $\omega < 1/2$. For all three classes of the KS naked singularity spacetimes, the innermost circular orbits correspond to the stable equilibrium point with $L=0$ at the so called static radius where particles are at rest relative to distant observers. The other properties of the circular geodesics are quite different in those three classes of the KS naked singularities. 

For small values of the naked-singularity Ho\v{r}ava parameter, $\omega < \omega_{ms}$, only stable circular orbits can exist at such KS naked singularity spacetimes -- no unstable circular orbits and no circular photon orbits exist. For mediate values of the Ho\v{r}ava parameter, $\omega_{ms} < \omega < \omega_{ph}$, two regions of stable circular orbits exist, being separated by a region of unstable circular geodesics -- no photon circular orbits are possible in such spacetimes. For large values of the naked-singularity Ho\v{r}ava parameter, $\omega_{ph} < \omega < 1/2$, two regions of stable circular orbits exist again; the inner region is limited from above by a stable photon circular orbit; under the outer region of stable circular geodesic orbits a region of unstable circular geodesics extends being limited from below by an unstable photon circular orbit. No circular geodesics are allowed between the stable and unstable photon circular orbits. 

In all three classes of the KS naked singularity spacetimes, there are regions where the gradient of the angular velocity of the circular geodesics is oppositely oriented, while in the black hole spacetimes, there is always $\frac{d\Omega_{K}}{dr} < 0$ along the whole Keplerian discs allowing thus for the MRI viscosity mechanism to work in whole the region of stable circular geodesics. In the KS naked singularity spacetimes with  $\omega_{ph} < \omega < 1/2$, the regions with $\frac{d\Omega_{K}}{dr} > 0$ and $\frac{d\Omega_{K}}{dr} < 0$ are separated, while in the spacetimes with $\omega < \omega_{ph}$, there is a radius where $\frac{d\Omega_{K}}{dr} = 0$ and these regions are smoothly matched there, but the MRI viscosity mechanism governing the standard Keplerian accretion stops its functioning there. We call the discs where $\frac{d\Omega_{K}}{dr} < 0$, having the standard MRI viscosity mechanism at work, the outer Keplerian discs, while those where a region with $\frac{d\Omega_{K}}{dr} > 0$ occurs we call the inner Keplerian discs. In the inner Keplerian discs we assume occurence of a non-standard viscosity mechanism creating a thermalized disc, or we assume that the accretion is governed by the gravitational radiation of the orbiting matter and the disc has to be cold. It should be stressed that a non-Keplerian accretion structures (toroidal or even spherical) are more probable to occur in the regions related to the inner Keplerian discs, nevertheles, occurence of limited inner Keplerian discs cannot be excluded and that is the reason why we have studied such structures in our paper. 

All three considered optical phenomena, the appearance of the Keplerian discs, their spectral continuum and their profiled spectral lines demonstrate clear signatures of the KS naked singularity spacetimes, enabling us to distinguish them from the Keplerian discs orbiting the KS or Schwarzschild (Kerr) black holes. The strong, qualitative signatures of the KS naked singularity spacetimes are related mainly to their inner Keplerian discs (or inner accretion structures, if Keplerian discs are destroyed), and enable even strong restriction of the dimensionless parameter $\omega$ in order to distinguish the three clasess of the KS naked singularity spacetimes. Moreover, even the optical phenomena related only to the outer Keplerian discs where the standard accretion governed by the MRI mechanism occurs alow for quantitative estimates of the parameter $\omega$ and distinguishing of the KS naked singularities. 

The most profound quantitative signatures of the KS naked singularities are related to the profiled spectral lines generated in both the outer and inner Keplerian discs. However, the signatures are strongly dependent on the inclination of the discs to the distant observer that should be known apriori, in order to obtain a convincing restrictions on the KS spacetime parameter. For objects close enough, the details of the appearance of the inner Keplerian discs could be observed giving direct qualitative signatures of the presence of KS naked singularity spacetimes. 

Strong qualitative signatures of the KS naked singularities are expected in the case of the toroidal or spherical structures in the innermost parts of the near-extreme naked singularity spacetimes, being  influenced by the trapped photons. Of course, this is a theme for a future study. 

\section*{Acknowledgements}
We would like to express our gratitude to the~Czech grant GA\v{C}R~202/09/0772. The~authors further acknowledge the~project Supporting Integration with the~International Theoretical and Observational Research Network in Relativistic Astrophysics of Compact Objects, reg. no. CZ.1.07/2.3.00/20.0071, supported by Operational Programme \emph{Education for Competitiveness} funded by Structural Funds of the~European Union and state budget of the~Czech Republic.

\end{document}